\documentclass[12pt]{article}
\usepackage{amssymb, amsmath, epsfig, natbib, url, titling, pdfpages, wrapfig, framed, multirow, graphicx, doi, authblk, float}

\usepackage{hyperref}
\hypersetup{
    colorlinks=true,
    linkcolor=blue,
    citecolor=blue,
    urlcolor=cyan,
    }

\begin{document}

\bibliographystyle{aasjournal}
\title{NIAC project report: Solar system-scale VLBI to dramatically improve cosmological distance measurements}
\date{}
\author[1,2]{Matthew McQuinn\thanks{mcquinn@uw.edu}}
\author[2]{Miguel Morales}
\author[3,4,5]{Casey McGrath}
\author[1]{Alyssa Alvarez}
\author[6]{Katelyn Glasby}
\author[7]{T. Joseph W. Lazio}
\author[8]{Kiyoshi Masui}
\author[2]{Lyujia Pan}
\author[9]{Jonathan Pober}
\author[10,11]{Huangyu Xiao}

\affil[1]{Department of Astronomy, University of Washington, 3910 15th Ave NE, Seattle, WA 98195, USA}
\affil[2]{Dark Universe Science Center (DUSC); Department of Physics, University of Washington, 3910 15th Ave NE, Seattle, WA 98195, USA}
\affil[3]{Center for Space Sciences and Technology, University of Maryland, Baltimore County, Baltimore, MD 21250, USA}
\affil[4]{Center for Research and Exploration in Space Science and Technology II, NASA/GSFC, Greenbelt, MD 20771, USA}
\affil[5]{Gravitational Astrophysics Lab, NASA Goddard Space Flight Center, Greenbelt, MD 20771, USA}
\affil[6]{Department of Physics, Oregon State University, Corvallis, OR 97331,  USA}
\affil[7]{Jet Propulsion Laboratory, California Institute of Technology, 4800 Oak Grove Dr, Pasadena, CA 91009, USA; Independent Researcher}
\affil[8]{
MIT Kavli Institute for Astrophysics and Space Research, Massachusetts Institute of Technology, 77 Massachusetts Ave, Cambridge, MA 02139, USA;
Department of Physics, Massachusetts Institute of Technology, 77 Massachusetts Ave, Cambridge, MA 02139, USA}
\affil[9]{Department of Physics, Brown University, Providence, RI 02912, USA}
\affil[10]{Physics Department, Boston University, Boston, MA 02215, USA}
\affil[11]{Department of Physics, Harvard University, Cambridge, MA, 02138, USA}

\def\matt#1{{\bf\textcolor{magenta}{{[MM: #1]}}}}

\maketitle

\clearpage

\section*{Executive Summary}
\addcontentsline{toc}{section}{Executive Summary}

Precise measurements of cosmic expansion are central to modern cosmology. Current methods for measuring the Hubble constant using standard candles show significant tension with large-scale structure measurements that assume the $\Lambda$CDM  cosmological model to measure this parameter. Additionally, a variety of other tensions have emerged among cosmological data sets that probe the late-time cosmic expansion.  These raise the possibility of new physics beyond the $\Lambda$CDM cosmological model or unrecognized systematics in existing methods. This report evaluates a fundamentally new approach to measuring extragalactic distances that could provide more robust constraints on late-time cosmic expansion.  This report also evaluates secondary science applications that could be pursued alongside.

\textbf{Mission Concept:} This NIAC Phase I study evaluates the Cosmic Positioning System (CPS): a constellation of five radio telescopes distributed throughout the outer Solar System that measures cosmological distances by directly detecting electromagnetic wavefront curvature from fast radio bursts (FRBs). Operating on baselines of tens of astronomical units, CPS enables Global Positioning System (GPS)-like trilateration on cosmological scales, directly measuring distances to sources at cosmologically-interesting distances.

\textbf{Key Findings:} CPS could achieve percent-level distance measurements to individual FRBs at hundreds of megaparsecs, enabling sub-percent constraints on the Hubble constant and potentially the dark energy equation of state with tens of detected sources. Beyond cosmology, CPS enables high-impact secondary science. The differential gravitational time delays experienced by the FRB radiation traveling to the widely separated CPS spacecraft probe dark-matter structure on $\sim 100$ AU scales. The configuration is sensitive to gravitational waves in the $10^{-7}$--$10^{-4}$ Hz band, bridging the gap between pulsar timing arrays and LISA. CPS also offers a potentially powerful tool for mapping the mass distribution of the outer Solar System.

\textbf{Technical Feasibility:} Phase I analysis demonstrates technical feasibility using existing and near-term technologies. A constellation of five spacecraft equipped with $8$-$9$~m lightweight deployable antennas, receivers operating at $3$-$6$~GHz with tens-of-Kelvin system temperatures, and space-qualified atomic clocks similar to those on GPS satellites can meet the required timing and ranging precision. Spacecraft-to-spacecraft ranging at the centimeter level, clock stability at the tenth-of-a-nanosecond level over hours, and onboard data storage at the 10~TB scale are challenging but assessed to be achievable. The constellation enables self-calibration through mutual ranging, with a network of $3$--$4$ ground stations (equipped with $\sim$8~m antennas) providing FRB alerts and clock synchronization, with data downlink requiring several hours of Deep Space Network time monthly. Analysis of environmental accelerations and antenna torques is presented to understand positional calibration times and reaction wheel requirements. Whether these systems can conform to tight mass and power requirements for outer Solar System missions powered by radioisotope generators is discussed, but requires further study.

\textbf{Key Uncertainty:} The most significant viability issue concerns FRB properties in the targeted frequency band. This report developed models for distance and flux distributions of repeating FRBs, extrapolating from lower frequencies where they are better characterized. Community efforts to study repeating FRBs at several-GHz frequencies would close this critical knowledge gap.\\

CPS represents a transformative opportunity to establish a new geometric foundation for cosmology, while enabling other frontier measurements in astrophysics and fundamental physics.

\clearpage

\tableofcontents

\clearpage
\begin{figure}[h]
  \centering
  \includegraphics[width=0.95\textwidth]{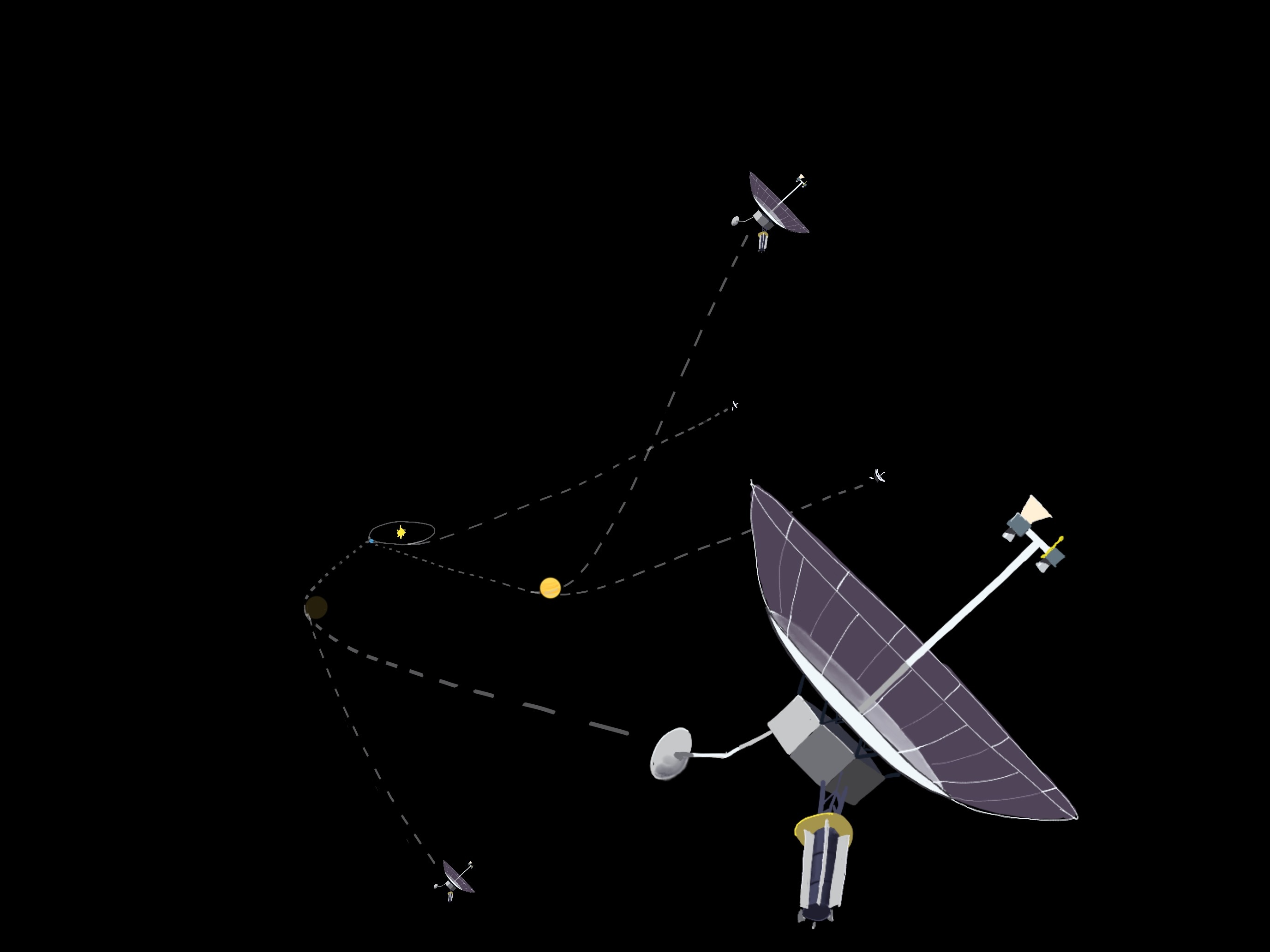}
  \label{fig:Miguels_artwork}
\end{figure}
\begin{center}
``Once you get to earth orbit, you’re halfway to anywhere in the solar system.''\\
-- Robert Heinlein 
\vspace{.5cm}\\
``It is no argument against any project to say `The idea’s fantastic!’ Most of the things that have happened in the last fifty years have been fantastic, and it is only by assuming that they will continue to be so that we have any hope of anticipating the future.''\\
--Arthur C. Clarke
\end{center}

\clearpage

\section{The Cosmic Positioning System (CPS)}
\label{ss:intro}

Measurements of both distance and recessional velocity to extragalactic sources provide crucial constraints on the Universe's expansion history. Over the past two decades, these observations have revealed an accelerating cosmic expansion,  requiring 70\% of the Universe's energy content to be in `dark energy' -- a negative-pressure substance that has profound implications for fundamental physics. Intriguingly, recent precision measurements hint at further cosmological surprises: a 10\% discrepancy exists between cosmic distances determined by the two most established measurement techniques, currently a tension with a statistical significance of $\approx 7~\sigma$ \citep{2022ApJ...934L...7R, 2025arXiv251023823H}, although there is an active debate about whether the errors are underestimated \citep[e.g.][]{2025ApJ...985..203F}. The leading model for explaining this discrepancy suggests an earlier cosmic epoch dominated by a different form of dark energy \citep{2021CQGra..38o3001D}; a result that, if true, would upend cosmology.  Recently favored models for dark energy by the Dark Energy Spectroscopic Instrument (DESI), in which the late-time dark energy density evolves, exacerbate this tension \citep{2025arXiv250314738D}. However, the most direct method for measuring expansion relies on a complex series of calibration steps, establishing a ``distance ladder'' to progressively more luminous astrophysical sources until distances can be measured to objects far enough away to precisely probe cosmic expansion. The field would benefit tremendously from a more straightforward and precise method for measuring the cosmic expansion. A reliable measurement would constrain at least one free parameter in the late-time expansion history, improving constraints on dark energy and even the masses of neutrinos \citep{2024JCAP...12..048L}.

\citet{2022arXiv221007159B} proposed a purely geometric method for measuring distances to cosmological sources, thereby determining the Universe's expansion history directly without the need for the distance ladder. This approach operates on principles similar to GPS geolocation. By placing antennas throughout the outer Solar System, one could directly measure the curvature of electromagnetic wavefronts from distant sources, allowing for precise distance calculations through trilateration. The initial analysis by \citet{2022arXiv221007159B} suggests that if antennas are separated by a few tens of astronomical units, this method could yield cosmological distance constraints that are potentially an order of magnitude more precise than current methods. This report investigates the technical viability and scientific potential of a NASA mission that uses this method, a mission we call {\bf the Cosmic Positioning System (CPS)}.

Beyond cosmological expansion measurements, the NIAC Phase I study summarized by this report has identified three additional major scientific applications for antennas in the outer Solar System. First, the most motivated models for the dark matter -- weakly interacting massive particles and axions -- predict clumping on scales smaller than a parsec. \citet{xiao24} showed that gravitational time delays would alter the arrival times between CPS antennas of light from astrophysical sources.  This effect could be used to probe the small-scale structure of the dark matter, potentially revealing the nature of this elusive particle that constitutes 85\% of all matter. Second, \citet{2024arXiv241115072M} showed that CPS could potentially be sensitive to interesting levels of gravitational waves in the microHertz frequency band -- a band where we may lack the technology to detect gravitational waves using more traditional interferometric methods. Third, CPS could precisely map the mass distribution in the outer Solar System to better constrain the Kuiper belt and search for other planetary bodies, such as the conjectured Planet 9. This report summarizes each science case and evaluates the specific instrumental requirements and tolerance levels needed to achieve sensitivity benchmarks for these applications.


In what follows, we briefly outline these science drivers and mention a few others that CPS would enable.  The remainder of this report aims to evaluate the feasibility of this mission.

\subsection{Geometric distance measurements with time delays}
\label{ss:geometricdistaanceintro}

\begin{figure}
\vspace{-1cm}
\begin{center}
\includegraphics[angle=0, scale=.44]{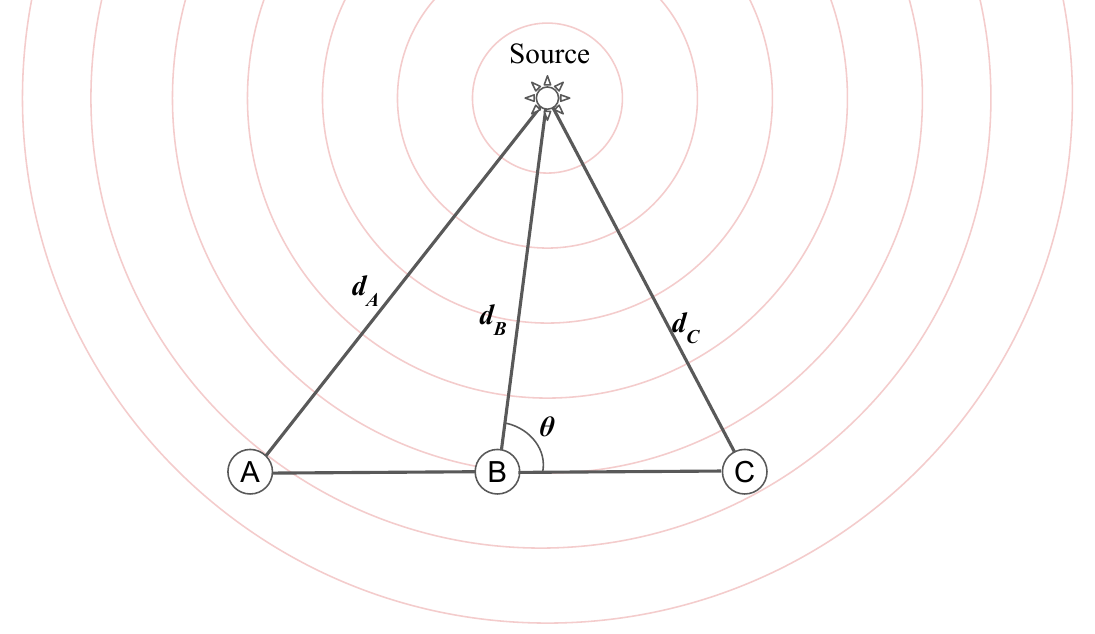}
\includegraphics[angle=0, scale=.44]{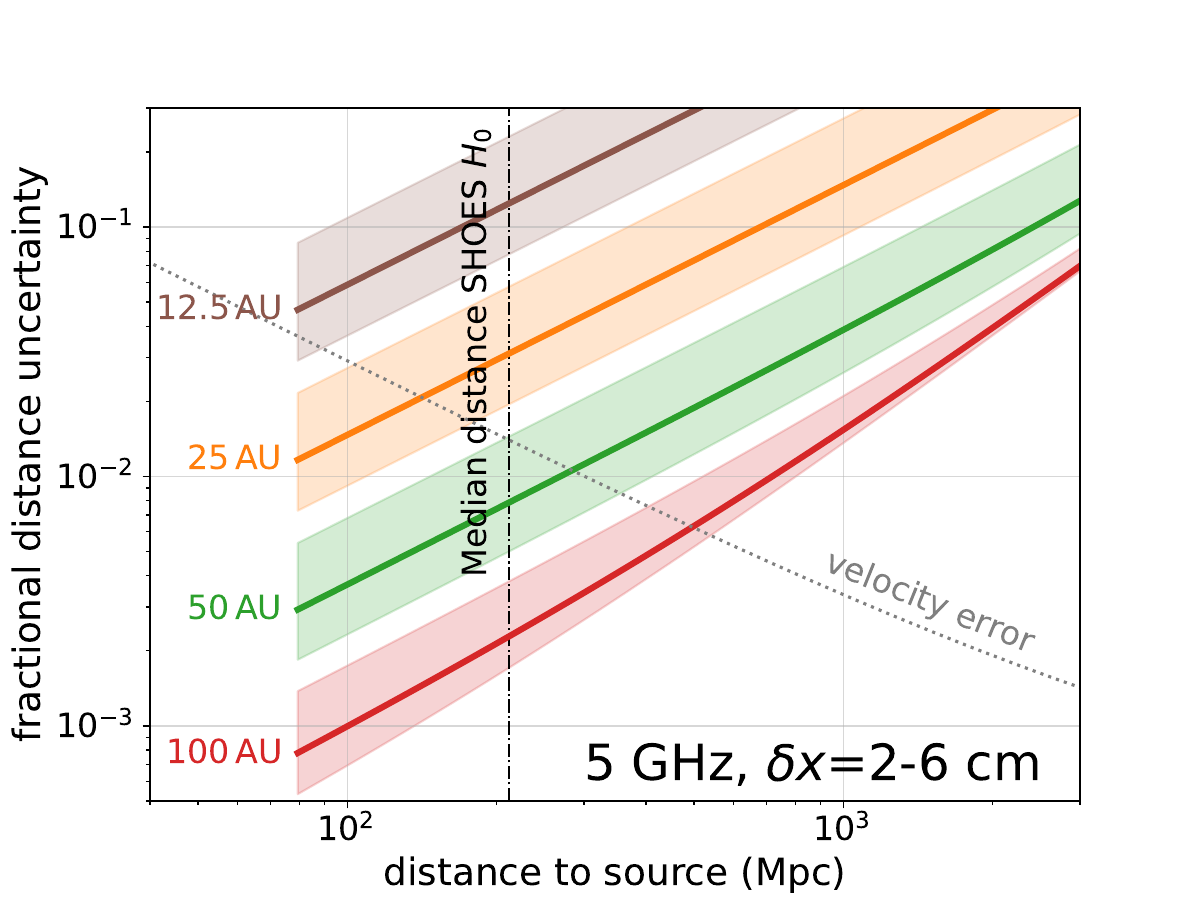}
\end{center}
\vspace{-0.5cm}
\caption{\small {\bf Left:} Illustration of the wavefront curvature distance measurement: The signal arrives at detector B before it arrives at detectors A and C. By comparing the arrival times at the three detectors (four detectors in 3D), the distance to the source can be inferred via the wavefront curvature. {\bf Right:} Estimated fractional distance error, $\sigma_d/d$, to a \emph{single} FRB as a function of the distance to the FRB, with distances of $>100$ Mpc being interesting for measuring the cosmic expansion and the median distance of supernova for the SHOES $H_0$ measurement annotated.  The typical baseline length of detectors in the constellation is annotated on the left of the curves, and the highlighted band assumes detector positions errors of $\delta x = 2-6\,$cm. 
{\bf CPS aims for much more precise distance constraints, which in turn would result in the most precise direct constraints on the late-time cosmic expansion.}
\label{fig:curvature_overview}
\label{fig:distance_uncertainty}}
\end{figure}

A primary scientific objective of CPS is to measure the expansion history of the Cosmos geometrically by detecting the curvature of the electromagnetic wavefront from distant astrophysical sources. This principle is illustrated in the left panel of Figure~\ref{fig:curvature_overview}. Light from an extragalactic point source reaches detector B before arriving at detectors A or C due to the spherical nature of the propagating wavefront.\footnote{Note that, in this illustration, if the light propagated as a plane wavefront from the source, it would arrive at A, B, and C in this figure at the same time if $\theta = 90^\circ$.} By precisely measuring the differential arrival times across three detectors for this 2D illustration (four detectors are actually required), the distance to the source can be inferred through simple geometric principles. 

An estimate for the fractional uncertainty in distance measurements using wavefront-curvature detection is given by \citep{2022arXiv221007159B}
\begin{align}
    \frac{\sigma_d}{d} \approx 2 \%
    \left( \frac{x}{40~\text{AU}} \right)^{-2}
    \left( \frac{d}{200~\text{Mpc}} \right)
    \left( \frac{5~\text{GHz}}{\nu} \right) \left( \frac{\sigma_t }{\nu^{-1}} \right),
    \label{eqn:error}
\end{align}
where $x$ represents the typical separation between detectors (an AU is the Earth-Sun distance), $\nu$ is the radio frequency that is observed, $d$ is the distance to the source, and $\sigma_t$ denotes the timing precision on the wavefront arrival. For meaningful cosmic expansion constraints, distance measurements must target sources at $d\gtrsim 50~$Mpc, as only beyond this distance do velocities from local accelerations contribute errors of $\lesssim 10$\% to the total velocity from cosmic expansion. 

Equation~(\ref{eqn:error}) is formulated to suggest that achieving $\sigma_t \sim \nu^{-1}$ is feasible. Attaining time delay measurements with precision $\sigma_t \sim \nu^{-1}$ requires both controlling non-geometric time delays to at least this level and modeling relative detector positions to an accuracy of $\sim c/\nu$, where $c$ is the speed of light.  For CPS's target frequency of $\nu=5~$GHz, this corresponds to $c/\nu = 6$~cm. The analysis by \citet{2022arXiv221007159B} demonstrated that various potential complications -- including wavefront timing errors, delays from propagation through intervening plasma, delays/smearing from diffractive and refractive propagation in the Milky Way, and gravitational time delays -- are all sufficiently small that a timing precision of $\sigma_t \approx (0.1-1)\times \nu^{-1}$ appears achievable for $\nu \gtrsim 3~$GHz.  We further investigate the refractive time delay in this report, as this delay is potentially the most concerning.

Fast radio bursts (FRBs) -- millisecond radio transients for which there are $10^4$ on the sky per day \citep[e.g.][]{petroff19} -- are the only known source class that is unresolved by AU-scale baselines and, hence, point-like, which is ideal for this experiment  \citep{2022arXiv221007159B}.  Their large electromagnetic fluxes, with many reaching tens of Janskys over their millisecond duration, enable this experiment without launching extremely large antennas or requiring huge data transfer rates. Some FRBs repeat on day to week timescales, meaning that CPS spacecraft do not need to survey the sky to find FRBs and can instead focus on the less computationally expensive task of detecting FRBs from known sources. Furthermore, the $\sim 10^2-10^3~$Mpc distances of repeating FRBs are ideal for precise distance measurements with CPS.  Owing to their decreasing flux and rate at frequencies higher than $1~$GHz, FRB properties also favor targeting frequencies that are not too large \citep{2022arXiv221007159B}.  Thus, FRB and timing precision considerations set the preferred frequency band for CPS to be $3 \lesssim \nu \lesssim 6~$GHz.   

Equation~(\ref{eqn:error}) shows that once baselines reach tens of Earth-Sun distances, the distance to a \emph{single} 200~Mpc FRB can be directly measured at the percent level.  A distance of 200 Mpc is the median of the supernovae used to achieve the leading error on the cosmic expansion rate \citep{2022ApJ...934L...7R}.  The right panel in Figure \ref{fig:distance_uncertainty} further quantifies the potential sensitivity to distance for observations of a single FRB. 
The typical baseline length of detectors in the constellation is annotated on the left of the curves.\footnote{Later in this report, we perform more detailed calculations and find that the length most relevant is the second shortest baseline for our nominal five detector configuration.} The highlighted band around each curve assumes $\delta x = 2-6\,$cm for the detector positional uncertainty.  
By $x\approx25~$AU, Figure \ref{fig:distance_uncertainty} suggests that $\sim 1\%$ measurements of a single FRB are possible out to a cosmologically interesting distance of $100~$Mpc.   By $x\approx 100~$AU, it suggests that $1\%$ measurements are possible up to $1000~$Mpc. Furthermore, the error can be reduced by observing more bursts; in this report, we forecast that tens to hundreds of FRB sources would be detected with CPS. 

Currently, hundreds of supernovae are used to place a 1-3\% constraint on the cosmic expansion rate \citep[also known as the Hubble constant;][]{Riess2019, 2025ApJ...985..203F, 2025arXiv251023823H}.  For supernova measurements, the power output (`luminosity') of each supernova must be known; a complex `distance ladder' in which each rung calibrates the next is used to ultimately calibrate the luminosity of a certain type of supernova.  Inevitably, there are questions about the rungs in the ladder and even whether there is evolution in the properties of these supernovae over time that would lead to systematic errors (as different distances originate from different cosmic times).  
The purely geometric distance measurements of CPS would circumvent the distance ladder completely, while potentially achieving superior precision.

For CPS to be a viable candidate for a future cosmological mission, it must be able to measure the Hubble constant so accurately that it eliminates this significant source of uncertainty in cosmological analyses that determine other cosmological parameters.  We show in this report that an accurate sub-percent measurement of the Hubble constant would substantially decrease the uncertainty in measurements of the properties of the dark energy.  A second goal of CPS would be to directly constrain the time evolution of the dark energy by comparing the expansion rate at different distances.  There has recently arisen a several-$\sigma$ tension with the expectation that dark energy is a cosmological constant, as indicated by the Dark Energy Spectroscopic Instrument (DESI; \citealt{2025arXiv250314738D}).  We show that CPS may be able to test this by directly measuring the dark energy equation of state.  



\subsection{The clumpiness of the dark matter}
\label{ss:darkmatter_preview}

The clumpiness of the dark matter has a direct bearing on its particle nature. In cold dark matter models with inflation -- the leading cosmological paradigm -- dark matter is anticipated to form structures down to extremely small scales. For example, in models where dark matter consists of weakly interacting massive particles (WIMPs), collapsed structures are expected with sizes as small as $100~$AU \citep{PhysRevLett.97.031301}. Structures on such scales are also predicted if the dark matter is an axion -- a hypothetical particle whose existence is motivated in part because it would also solve the strong CP problem of Quantum Chromodynamics \citep{HOGAN1988228}.

In certain models, the dark matter distribution should be even clumpier if additional sources of density fluctuations exist beyond the standard adiabatic inflationary fluctuations. This enhanced clumpiness occurs in several scenarios. For example, (1) for the QCD axion if the Peccei-Quinn symmetry breaking that results in axion production occurs after inflation \citep{HOGAN1988228}, (2) in models where the Universe experiences an early matter-dominated era, or (3) in models in which some of the dark matter consists of ``primordial'' black holes created in the early universe \citep{2019JCAP...08..031M}.

\begin{figure}[h!]
    \centering
\includegraphics[width=8cm]{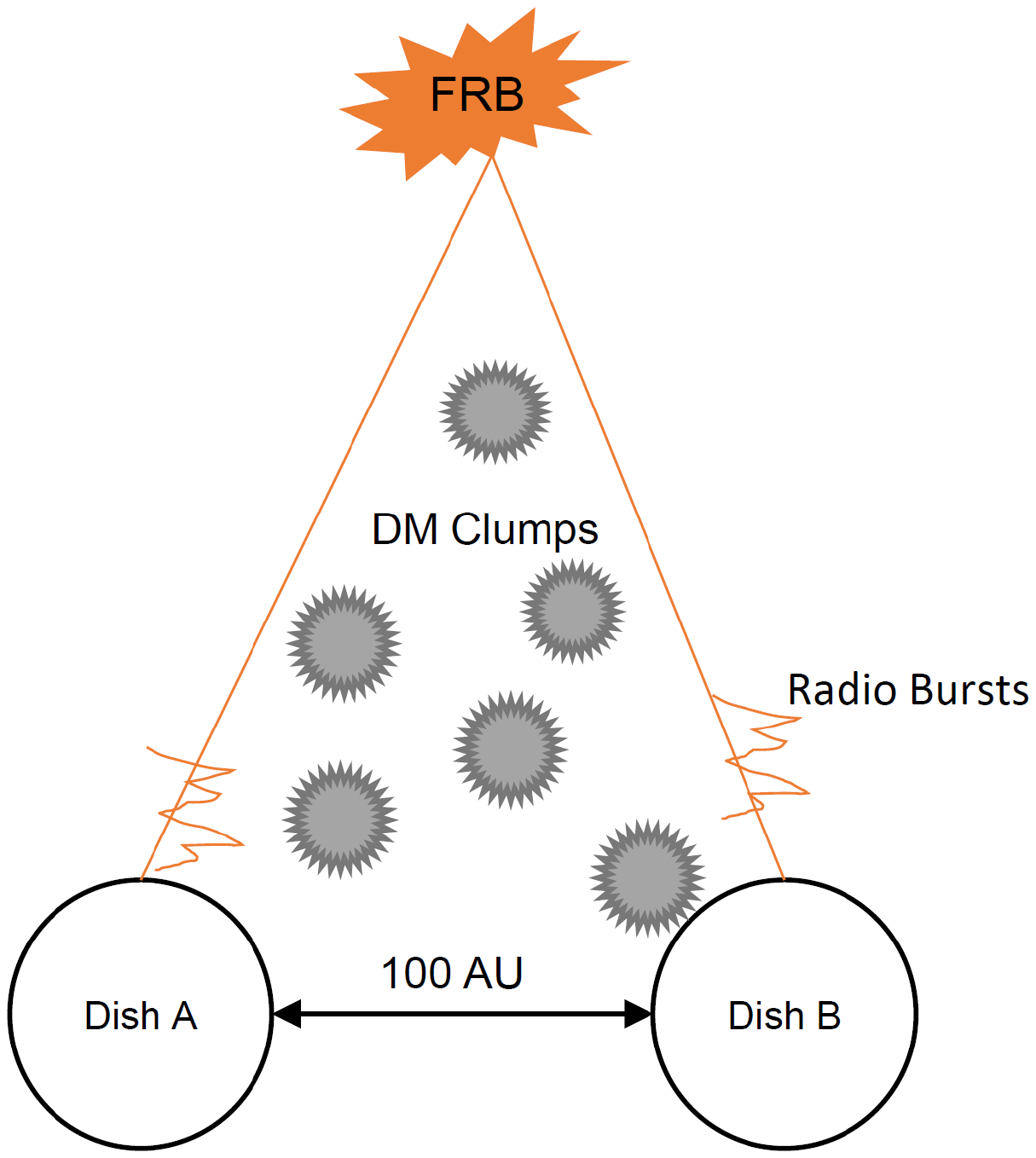}
\vspace{-1cm}
\caption{\small
Illustration of configuration that would constrain the clumpiness of the dark matter, from \citet{xiao24}. Shown are two radio antennas separated by 100 AU observing the same FRB source but along different sightlines. Each sightline experiences slightly different Shapiro time delays caused by intervening dark matter clumps. As a result, the FRB voltage-field time-series detected by CPS (illustrated by the orange jagged lines) will have different arrival times at each detector (shown here, the same signal arrives at Dish A before arriving at Dish B).  Such a measurement is sensitive to diffuse dark matter structures with smaller sizes than probed by other methods.
}
\label{fig:cartoonShapiro}
\end{figure}

The small-scale clumpiness of dark matter will impart subtle gravitational time delays that differ between CPS detectors.  The time variations and signatures of delays from dark matter will be much different from other (generally more massive and much less abundant) structures such as stars and, thus, likely can be distinguished \citep{xiao24}.  The effect of gravitational time delays from passing dark matter clumps is illustrated in Figure~\ref{fig:cartoonShapiro}. CPS can constrain these differential time delays by monitoring a repeating FRB over many cycles -- the apparent angular position of the FRB will wobble on the sky as dark matter clumps pass between the CPS telescopes and the source. \citet{xiao24} found that baselines of $10~$AU may be sufficient to probe the enhanced clumpiness predicted in the post-inflation QCD axion scenario or in models featuring an early matter-dominated era extending almost to Big Bang Nucleosynthesis. With $100~$AU baselines, CPS could potentially be sensitive to the standard dark matter clumpiness driven solely by inflationary fluctuations \citep{xiao24}. 

\subsection{$\mu$Hz gravitational waves}
\label{ss:gw}
\begin{figure}[htbp]
\centering
\includegraphics[width=.7\textwidth]{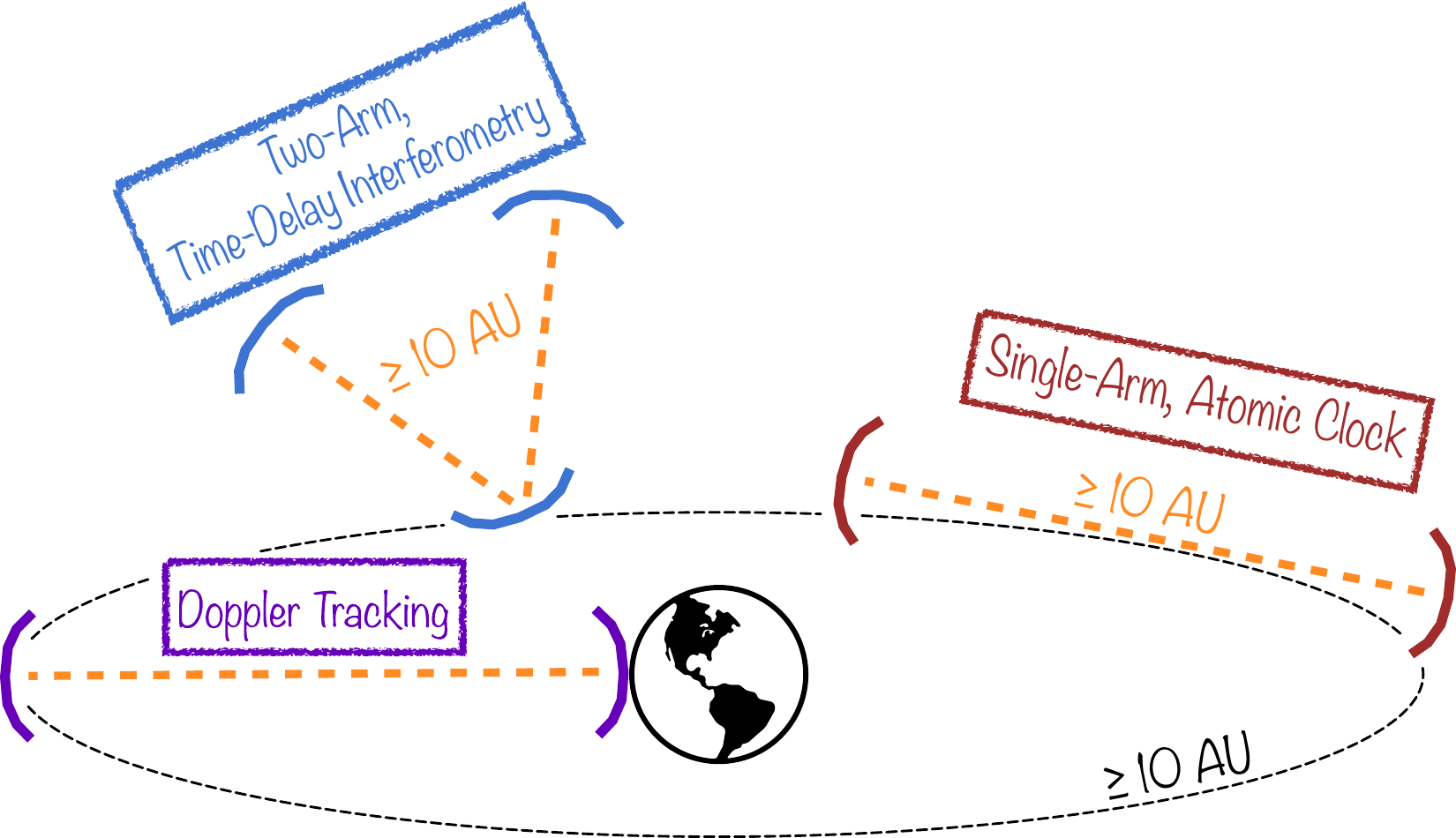}
\caption{\small 
Potential configurations for CPS to detect gravitational waves in the frontier $10^{-7}$--$10^{-4}$~Hz frequency band, from \citet{2024arXiv241115072M}. Three configurations are shown: (1) traditional Doppler tracking using an Earth station, (2) a single-arm configuration leveraging an onboard atomic clock, and (3) a two-arm time-delay interferometry setup requiring two radio antennas for at least one spacecraft. The latter configuration, while requiring more onboard instrumentation for at least one spacecraft, offers substantially higher sensitivity by eliminating onboard clock noise (which likely dominates the error without significant advances in space clock technology).}
\label{fig:configurations}
\end{figure}

The $\mu$Hz frequency band, defined here as $10^{-7}$--$10^{-4}$ Hz, represents a critical yet unexplored window for gravitational wave astronomy \citep{2021ExA....51.1333S}. This band falls below the $10^{-4} < f < 1~$Hz range targeted by the Laser Interferometer Space Antenna (LISA; \citealt{2017arXiv170200786A}) and at higher frequencies than the $f \sim 10^{-8}$ Hz probed by pulsar timing arrays \citep{Dahal_2020}. Gravitational waves in the $\mu$Hz band would originate from the early inspiral phase of intermediate-mass black hole binaries (with masses of $\sim 10^5$--$10^{7} M_\odot$) that the LISA mission aims to observe closer to merger, as well as from the inspiral and merger of supermassive black hole binaries ($\sim 10^{7}$--$10^{10} M_\odot$).

While space-interferometry methods that use test masses to correct for non-inertial motion, exemplified by LISA, would require development to provide sufficient acceleration control to detect anticipated astrophysical events in this band, CPS offers a novel approach by leveraging the naturally low-acceleration environment of the outer Solar System.  Additionally, such a system with tens of AU arms offers the potential for better source localization than one with shorter arms, as the instantaneous localization error scales inversely with the baseline length \citep{2024arXiv241115072M}.

The CPS gravitational wave experiment builds upon the heritage of Doppler tracking experiments conducted with outer Solar System spacecraft, but it offers key advantages. Traditional Doppler tracking, which uses Earth as one node, is limited by atmospheric effects and mechanical distortions of the terrestrial station \citep{dopplerranging}.  While CPS spacecraft could be used as a precision node for Doppler tracking owing to their large antenna and precise radio instrumentation, CPS can forgo having an Earth station entirely and have two spacecraft as the nodes. The sensitivity of a spacecraft-spacecraft arm is then limited by the precision of the onboard clock on the CPS spacecraft.  Additionally, with some augmentation of one spacecraft so that it can receive weak radio signals from two directions, clock noise can be eliminated by having that spacecraft send and receive transmissions simultaneously with two spacecraft to synthetically create a Michelson-like interferometer. Figure~\ref{fig:configurations} diagrams the different configurations where CPS could potentially contribute to leading constraints on $\mu$Hz gravitational waves.  This report discusses the viability of the different approaches.

\subsection{The mass distribution in the outer Solar System}
\label{ss:solarsystemmass}
CPS has the potential to constrain the mass profile of the outer Solar System as well as search for additional major bodies in the far reaches of the Solar System. The precise positional calibration system of CPS potentially improves upon the constraints that are possible with Earth-spacecraft ranging methods, such as those used with the Cassini spacecraft to place constraints on the existence of the hypothesized `Planet 9' \citep{Holman_2016}. Earth-spacecraft ranging suffers from uncertainty related to Earth's ephemeris and requires the spacecraft to orbit a massive Solar System body (Saturn in the case of Cassini) to mitigate uncertainties from various sources of acceleration on the spacecraft (such as thrusts to dissipate spacecraft angular momentum, spacecraft reflections of solar light, and spacecraft emissions). In contrast, CPS can potentially model these sources of acceleration better using its spacecraft-spacecraft ranging capability. By avoiding the need to be bound to a planet, CPS can provide constraints much further out in the Solar System,  enhancing the sensitivity to distant bodies.

CPS enables this science through precise measurements of spacecraft separation and GPS-like positioning of individual spacecraft. To estimate the potential sensitivity, we consider a spacecraft at a distance $r$ from the Sun. The proposed system could constrain the mass interior to the spacecraft's orbit with an accuracy of:
\begin{equation}
\delta M \sim \frac{r^2 \delta x}{G t^2} = 1.5 \times 10^{-3} \left(\frac{\delta x}{30 ~{\rm m}}\right) \left(\frac{r}{30 ~\rm AU}\right)^{2} \left(\frac{t}{{\rm 1 ~yr}}\right)^{-2} ~M_{\oplus},
\label{eqn:massperturbations}
\end{equation}
where $M_{\oplus}$ indicates an Earth mass, $\delta x$ is the positional uncertainty, and $t$ is the observation time. We find in \S~\ref{ss:accelerations} that the non-gravitational displacement on a year timescale at $r = 30~$AU is $\delta x \sim 30~$cm for $t= 1~$yr, though achieving this precision would also require fantastic monitoring of the accelerations from the thrusters by the ranging system, as well as modeling the mean acceleration from solar radiation.  We suspect that the latter are likely to set $\delta x$ (\S~\ref{ss:accelerations}). If the positional uncertainty of $\delta x \lesssim 300~$m can be achieved for spacecraft at $30$~AU over a year, equation~(\ref{eqn:massperturbations}) suggests that CPS would be sensitive to the mass distribution in the Kuiper belt (at $\approx 30$-$50~$AU) and in trans-Neptunian objects, both of which have masses estimated to be in the range $10^{-2}$-$0.1~M_\oplus$ \citep{2018AstL...44..554P}.  Furthermore, there is evidence for substantial mass outside the Kuiper belt `cliff' at 50~AU \citep{2024PSJ.....5..227F}, a hypothesis that could be directly tested by CPS, especially with an extended mission lifetime beyond the nominal ten years.

The system would be sensitive to detecting a potential Planet 9, which would have a tidal effect on CPS spacecraft. This tidal effect on two spacecraft antipodal to the Sun increases the $\delta M$ by a factor of $d^3/(2r^3)$ relative to equation~(\ref{eqn:massperturbations}), where $r$ is the spacecraft's distance from the Sun and $d$ is the distance to Planet 9. Based on current models, Planet 9 would produce a relative displacement in a year of $10^2-10^4~$m for spacecraft at $r=30~$AU, assuming a mass of $1$-$10~M_\oplus$ and a distance of $300$-$700~$AU as suggested by the models in \citet{Brown_2022}.  


\subsection{Other science}
\label{ss:sience}

Other scientific advances and tools that we think CPS would be able to contribute include:
\begin{description}
\item[Solar System-scale navigation system:]  
The proposed system could potentially provide precise positioning for interplanetary spacecraft by having the CPS telescopes direct ranging signals to spacecraft traveling to destinations such as Mars, thereby enhancing navigation capabilities for future NASA missions, including the Artemis Program. Additionally, CPS could serve as a resilient backup to GPS. In scenarios where GPS becomes inoperative, the CPS network could be configured to broadcast ranging signals to Earth. While the baseline five-spacecraft constellation would not typically provide the four simultaneous (unobstructed) links necessary for complete space-time trilateration, when combined with ground-based reference stations similar to the US's LORAN system, this configuration could potentially achieve positioning accuracies comparable to GPS. This capability would be most applicable to larger vehicles, ships, and aircraft equipped with several meter radio dishes capable of receiving the weak ranging signals of the CPS spacecraft.

\item[Solar system barycenter determination:]  
CPS spacecraft could provide precise distance measurements between Earth and the outer Solar System, enabling a significantly improved determination of Earth's orbital motion. Currently, the Solar System barycenter is known only to approximately 100 meters. With its precise ranging system providing tight constraints on the Earth's position relative to the spacecraft, CPS would substantially reduce this uncertainty. Such improvement would be particularly valuable for pulsar timing arrays, where ephemeris errors can both mimic gravitational wave signals and limit sensitivity to actual gravitational waves \citep{2020ApJ...893..112V}.  This could therefore benefit ultra-low gravitational wave sensitivity and science discovery.

\item[Peculiar velocities of extragalactic sources:]  
The proper motion of FRB host galaxies will be directly measurable with CPS for sources detected multiple times. Cosmological objects move approximately $\sim 1~(d/100 \text{ Mpc})^{-1}~\text{micro-arcsecond yr}^{-1}$, a motion that CPS could detect at a precision of $\sim 10^{-3}$ for a point source at $d=100~\text{Mpc}$ \citep{2022arXiv221007159B}. While some component of this motion will be due to the FRB source's movement relative to its host galaxy, much of it reflects the influence of large-scale cosmic structure, providing constraints on cosmic flows.

\item[Interstellar medium turbulence and scattering:]
Differential dispersion delays between different spacecraft observing the same FRB would be readily detectable with CPS \citep{2022arXiv221007159B}. Such measurements would constrain the spectrum of interstellar medium density fluctuations at Solar System scales (tens to hundreds of AU), complementing existing measurements at both smaller and larger scales. This would provide valuable insights into the spectrum of interstellar turbulence \citep{1981Natur.291..561A} and test models of electron inhomogeneities from $\sim 1~\text{AU}$ current sheets that have been proposed to explain various aspects of scattering and scintillation \citep{2014MNRAS.442.3338P}.
\end{description}

\section{Heritage}
CPS benefits from a heritage of space telecommunications, radio science space missions, outer Solar System spacecraft, and global satellite navigation systems. Below we highlight research, systems, and precursor missions that demonstrate much of the technology required for CPS.

\paragraph{Large-diameter high-gain space antenna:}
Many large-diameter antennas have been successfully deployed in space for telecommunications purposes \citep{Imbriale2012SpaceAntenna}. These include NASA's ATS-6 satellite ($D=9.1$~m; 1974), Asia Cellular Satellite's Garuda 1 satellite ($D=12$~m; 2000), ICO Global Communications' ICO  satellite ($D=12$~m; 2008), TerreStar Corporation's Terrestar satellite ($D=18$~m; 2009), and LightSquared's SkyTerra 1  satellite ($D=22$~m; 2010). Notably, RadioAstron (also known as Spektr-R) of the Russian space program, launched in 2011 for radio astronomy, featured a $D=10$~m fold-out dish \citep{2013ARep...57..153K}. RadioAstron had an orbit with a 50 Earth radius apogee, approximately the Earth-Moon distance. It was equipped with two hydrogen maser frequency standards, plus a phased link to even more precise ground-based clocks. RadioAstron operated over 300~MHz--25~GHz, achieving a system temperature of $T_{\rm sys} = 43$~K with passive cooling for its L-band receiver. The CPS spacecraft would employ a similar passive cooling scheme to achieve comparable receiver temperatures. A second notable mission is NASA's Soil Moisture Active Passive (SMAP) satellite.  It was launched in 2015 and equipped with a 56~kg fold-out $D=6$~m dish equipped with an L-band receiver \citep{smap}.  The dish's wire mesh makes it mostly transmissive to optical wavelengths.   CPS requires a similarly lightweight, optically transmissive radio dish design to minimize launch mass, maintain sufficient sensitivity, and minimize torques from solar radiation pressure.

\paragraph{NASA missions to the outer Solar System:}  
NASA has a history of success with outer Solar System missions that inform CPS trajectory planning and spacecraft design. These missions include the Galileo, Pioneer, Voyager, and Cassini probes. Voyager 1 and 2 have famously reached beyond 140~AU \citep{Bell2015}. During their journeys to these distances, the Voyager spacecraft scattered gravitationally off Jupiter and other gas giants to boost their velocity, resulting in trajectories that are inclined 35 and 48 degrees relative to the ecliptic plane. In our nominal mission, two of the CPS spacecraft would also utilize Jupiter gravity assists to achieve trajectories off the ecliptic, which is critical for localizing the spacecraft via spacecraft-spacecraft trilaterations in the dimension out of the ecliptic.

In 2006, NASA launched the New Horizons spacecraft to explore Pluto. New Horizons has a dry mass of 400~kg, only 30~kg of which was the science system, and it was launched with an additional 80~kg of propellant. New Horizons was powered by a 250~Watt radioisotope power source (from 10~kg of plutonium-238), with the science system drawing less than 30~W of power \citep{2008SSRv..140...23F}. The CPS spacecraft likely would have to operate with a comparable power budget. New Horizons was launched on an Atlas V 551 rocket and achieved an Earth escape velocity squared (C3) of 157~km$^2$~s$^{-2}$ (the characteristic energy per unit mass out of Earth's orbit). New Horizons was boosted from 19~km~s$^{-1}$ to 23~km~s$^{-1}$ via a gravity assist from Jupiter. 
It took 9.5~years to reach Pluto at 34~AU. For the CPS spacecraft to reach $\gtrsim 20$~AU within a decade (as we find is needed to achieve competitive cosmic expansion constraints), the C3 for the CPS spacecraft must be at least comparable to that of New Horizons (\S~\ref{ss:trajectories}).

\paragraph{Global satellite navigation systems, space clocks, and ranging:}
CPS consists of a network of spacecraft with precise ephemeris constraints, similar to Global Navigation Satellite Systems (GNSSs; \citealt{misraenge}), as pioneered by the United States' Global Positioning System (GPS).  CPS would use trilateration to establish the spacecraft positions, just as GNSS uses trilateration to provide precise geolocations. The underlying methods of GNSSs scale up to the Solar System scale of the CPS network, with the additional advantage of the lower acceleration environment of the outer Solar System. GPS satellite orbital positions are known with a few-centimeter precision and can be used to trilaterate terrestrial positions to a similar precision \citep{igs_precise}. The ranging and clock technology on GPS satellites provides a proven foundation for the CPS system. The hydrogen maser clocks on the Galileo GNSS satellites drift by 0.5~ns per day \citep{leonardo_hydrogen_maser}, demonstrating stability similar to that of the clocks required on the CPS spacecraft.

NASA has developed space clocks for future interplanetary missions. Their Deep Space Atomic Clock, a Mercury trapped ion clock, was launched in 2019 and operated for a year in orbit \citep{2021Natur.595...43B}. With performance in space comparable to that of the Galileo satellites (and superior performance in laboratory conditions), it is also somewhat lighter and requires less power than the Galileo hydrogen maser clocks.  Additionally, there has been research to miniaturize the Deep Space Atomic Clock technology \citep{m2picclock}, as well as to develop even more precise optical atomic space-clocks \citep{Tomio_2024}.

NASA itself has extensive experience with precise ranging to interplanetary spacecraft. Cassini ranging at 10~AU achieved roughly 1~meter precision, limited primarily by the solar plasma (which could be eliminated using dual-frequency ranging) and internal delays within the DSN stations \citep{2009IPNPR.177C...1N}. Additionally, while not measuring absolute distance, Cassini phase tracking has demonstrated the ability to measure changes in distance with precisions of $\sim 0.1$~cm for gravitational wave constraints \citep{dopplerranging}. Laser ranging has also been performed with an accuracy of approximately 10~cm over 24 million kilometers with the Mercury probe MESSENGER \citep{Sun2012MESSENGER}.

\paragraph{Very long baseline interferometry (VLBI):}  
VLBI correlates the electric field received at one radio receiver with that received at another. With receivers often separated by continental distances, this technique allows for milli-arcsecond source localizations \citep{SCHUH201268}. VLBI has successfully been applied to some fast radio bursts (e.g., \citealt{VLBIFRB}), the principal source for CPS distance measurements. The longest baselines for VLBI have been the 50 Earth radii (equivalently $2\times10^{-3}$~AU) between terrestrial telescopes and the RadioAstron satellite \citep{2013ARep...57..153K}. CPS would expand the longest baseline achieved by a factor of approximately $10^4$, creating unprecedented angular resolution of $\sim 10^{-4}$micro-arseconds \citep{2022arXiv221007159B}. This improves by a factor of $10^5$ over the state-of-the-art $\sim 10~$micro-arcsecond angular resolution achieved with the Gaia spacecraft and the Event Horizon Telescope. Section~\ref{ss:signalacquisition} discusses the additional challenges that come with detecting correlations over these longer baselines.

\section{Mission architecture and instrumentation}

\begin{figure}[h]
  \centering
  \includegraphics[width=0.9\textwidth]{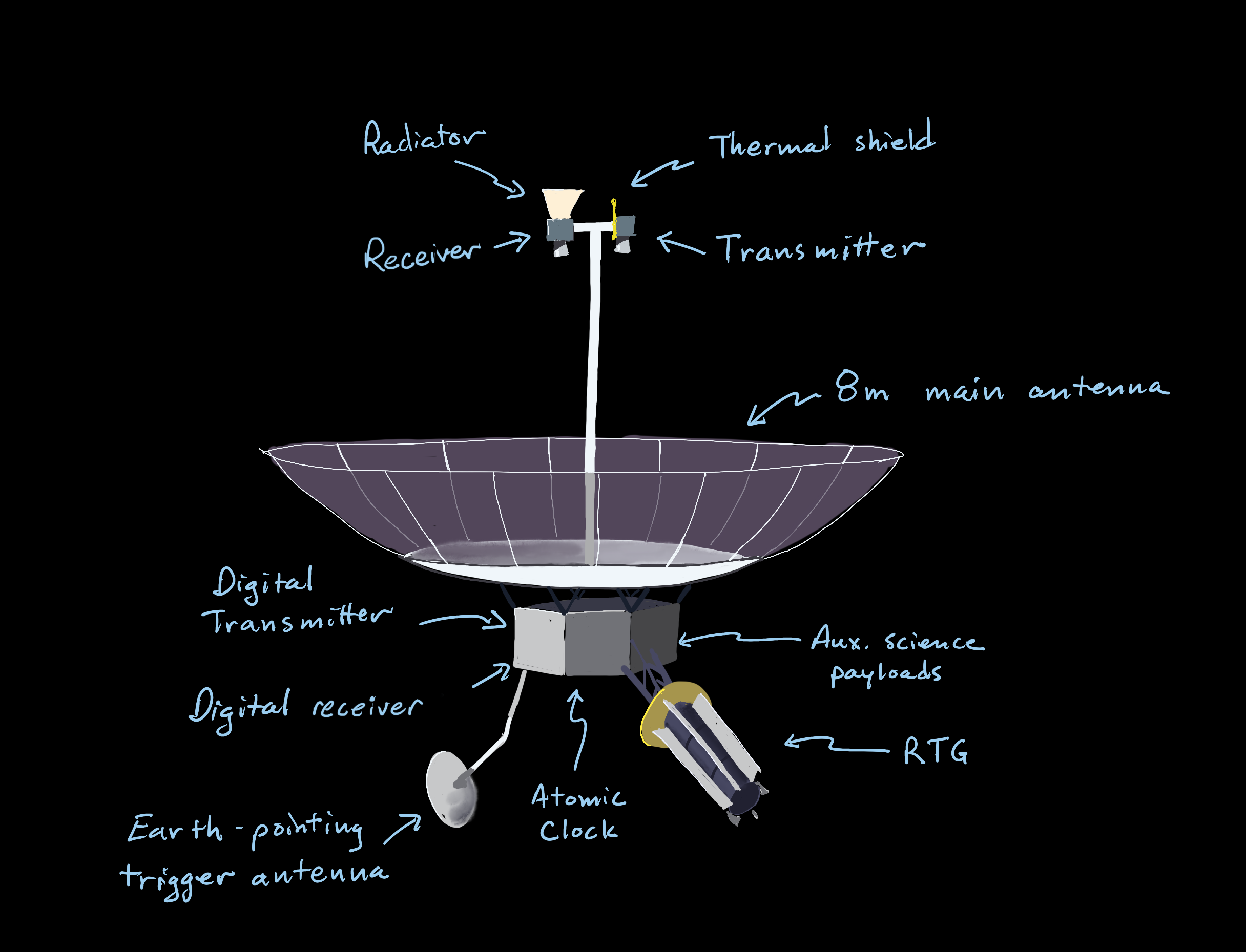}
  \caption{\small Conceptual design of a CPS spacecraft. The primary $D\approx 8~$m dish dominates the structure, with the receiver mounted on a passively cooled cold plate separated from the transmitter to minimize thermal coupling. The spacecraft bus below the main dish houses the mission-critical payload, including the digital receiver, digital transmitter, and atomic clock. A steerable smaller antenna for receiving terrestrial alerts is positioned at the bottom of the spacecraft. Power is supplied by a radioisotope thermoelectric generator (RTG) mounted on an extended boom to minimize thermal and radiation effects on sensitive instrumentation.}
  \label{fig:Miguels_satellite}
\end{figure}

\begin{table}[h]
\centering
\small
\begin{tabular}{|l|l|l|}
\hline
\textbf{Parameter} & \textbf{Requirement} & \textbf{Comments} \\ \hline
10 year distance & $r \geq 20$~AU for each spacecraft &  \S~\ref{ss:trajectories}; required for sufficient sensitivity on $H_0$ (\S~\ref{sec:science})  \\ \hline
number of spacecraft & 5&  scientific returns similar with 4 spacecraft (\S~\ref{ss:groundstations})\\ \hline
spacecraft C3 & $>150~$km$^2$/s$^2$ & requires $M<1500~$kg or $< 750~$kg paired \\ 
& to achieve minimum $r$ & for SLS Block 1B with advanced $3^{\rm rd}$ stage  (\S~\ref{ss:trajectories})\\ 
\hline
spacecraft spacetime& 10 cm when $r<30~$AU & requiring clock drift of $< 0.3~$ns in 4 hours (\S~\ref{ss:atomic_clocks}); \\ 
position  error ($\delta x$)& (most relevant for shortest arms) & trilateration errors of $<5~$cm in each dim (\S~\ref{ss:ssranging}) \\ \hline
spacecraft flux limit & 20 Jy for 1~ms observation& set by requirement of detecting $\gtrsim 100$ FRBs (\S~\ref{ss:rates}); \\ 
& integration to achieve SNR=10 & requires $\left(\frac{D_{\rm eff}}{8 ~{\rm m}}\right)^2\left(\frac{T_{\rm sys}}{20 \; {\rm K}} \right)^{-1}  \left(\frac{B}{500 \;{\rm MHz}} \right) \gtrsim 1$ (\S~\ref{ss:radioastronsystem})\\ \hline
pointing accuracy & 3 arcminute & $\delta \theta < 0.2 \lambda/D$ for $D=8~$m for X-band (\S~\ref{ss:attitude})\\ \hline
frequency & 3-6 GHz & set by scattering and FRB flux considerations (\S~\ref{sec:science});\\ 
& & science enhanced with L \& Ka bands (\S~\ref{sss:hga})\\ \hline
bandwidth & $\geq1$ GHz & necessary to detect sufficient FRBs (\S~\ref{sss:bandwidth}, \ref{ss:rates}) \\ \hline
slew speed & full turn in less than $1~$day & required to not lose significant time on sky (\S~\ref{ss:attitude})\\ \hline
storage & $> 10$ TB, $>1\,$yr continuous write & needed for terrestrial follow-up; \S~\ref{sss:bandwidth}\\ \hline
off-gassing/& $<100$ nano-Newton & non-modeled component (\S~\ref{ss:accelerations}) \\ 
re-radiation & & GW science potentially more stringent; \\ 
& &  based off spacecraft mass of $10^3~$Kg  \\ \hline
ground support & triggering, clock calibration, &  three or four $D= 8~$m dishes (\S~\ref{ss:groundstations});\\
& and downlink &$\sim 5\,$hr month$^{-1}$ for DSN downlink (\S~\ref{ss:signalacquisition})\\ \hline 
triggering & to detect trigger alert from ground &  $D= 1\;$m with Ka band; $\sim 3 \pi$ angular coverage;
 \\
antennas & &  offset by several meter from main dish (\S~\ref{ss:groundstations})\\ \hline
radiation exposure & 100~kRad & for electronics to survive mission lifetime (\S~\ref{ss:trajectories})\\ 
\hline 
\end{tabular}
\caption{\small Mission requirements for the Cosmic Positioning System.}
\label{table:instrument_requirements}
\end{table}

\begin{table}[h!]
\centering
\footnotesize
\begin{tabular}{|p{2.2cm}|p{3cm}|p{3cm}|c|p{1cm}|p{1cm}|p{4cm}|}
\hline

\textbf{Component} & \textbf{Use/Requirement} & \textbf{Example} & \textbf{\#} & {\textbf{Weight} \textbf{(kg)}} & {\textbf{Power} \textbf{(W)}} & \textbf{Example's specs/comments} \\ \hline

Analog to Digital converter (ADC) & $>0.5$~GHz bandwidth digitization & Teledyne EV12AQ600   & 1 & & 6.6 & 3.2 GigaSamples per second, dual channel; allows $B=1.6$~GHz; TRL9\\ \hline

{Low noise amplifiers (LNAs) + receiver electronics} & processing prior to ADC; low $T_{\rm sys}$ & specialized equipment. & -- & -- & $\sim $10 & dual channel LNA over science band; X-band LNA for ranging \\ \hline

{Board for processing voltages} & {FRB flagging and dedispersion} & Xilinx XQR UltraScale+ (Radiation-Hardened version) & 1 & $0.05$ & $5-10$ & can achieve $>30\,$Gbps; FPGAs; power consumption estimated for $10~$Gbps; TRL9 \\ \hline

{Trigger antenna receiver} & {receive trigger alert} & New Horizons receiver & 1 & 0.5 & 3 & operating in Ka Band ideal (rather than X-band like New Horizons); TRL9 \\ \hline

{Data storage} &  {$>10~$TB with enough write cycles to last mission lifetime} & Phison’s space-qualified 8~TB M.2 2280 SSD (6500 MB/s) & 4 & $4\times 2$ & 5-9 & weight reduced if all SSD cards put in single enclosure; power is for write to single card; TRL6 \\ \hline

Clock & to keep accurate time with $\sigma_y(\text{1\,s}) < 10^{-12}$ & Deep Space Atomic Clock (DSAC) & 1 & 17 & 44 & DSAC lab performance: $\sigma_y(\text{1\,s}) =(1-2)\times 10^{-13}$; TRL$8\to9$  \\ \hline \hline

Star tracker & $\lesssim 3$ arc-minute pointing & Jena-Optronik ASTRO APS Star Tracker & 1 & 1.4 & 5 & TRL9 \\ \hline

{Reaction wheels} & for turning and stabilization & Blue Canyon Technologies RW8 & 4 & {$4\times 4.4$} & {$4\times 5.6$} & maximum angular momentum $8$~N~m~s; max torque $0.25~$N~m; allows full turn in a few hours for nominal antenna specs; quoted power at zero angular momentum; TRL9  \\ \hline

Batteries & To strengthen ranging signal + other power redistribution & GomSpace NanoPower BPX & 3 & $3\times 0.3$ & -- & 38~Watt-hr; discharged weekly to boost power during $\sim 10$~min. ranging transmissions; TRL9 \\ \hline
\end{tabular}
\caption{\small {\bf Representative space-qualified components for each CPS spacecraft:} Certain components may draw power only a fraction of the time.  The minimum number of each component is listed under `\#' -- this number does not account for additional redundancy that may be advantageous. The \href{https://www.nasa.gov/directorates/somd/space-communications-navigation-program/technology-readiness-levels/}{technology readiness level (TRL)} of the representative component is listed under the comments. 
}
\label{table:representativecomponents}
\end{table}

Each CPS spacecraft consists of a 10~m-scale antenna and its receiver system, an atomic clock for precise time keeping, and a meter-scale triggering antenna for receiving FRB alerts. A conceptual design is given in Figure \ref{fig:Miguels_satellite}. System requirements are summarized in Table~\ref{table:instrument_requirements}, and representative components that may satisfy these requirements are given in Table~\ref{table:representativecomponents}.

\subsection{Radio astronomy system}
\label{ss:radioastronsystem}
The primary radio science objective for CPS involves timing the arrival of light from fast radio bursts (FRBs) to constrain the expansion history of the Cosmos and the clumpiness of the dark matter. The flux sensitivity of our system scales as $\propto D_{\rm eff} ^2/({\rm min}[B, \Delta \nu_{\rm FRB}] \cdot T_{\rm sys})$, where $D_{\rm eff}$ is the effective diameter of the radio dish, $B$ is the receiver bandwidth, $\Delta \nu_{\rm FRB}$ is the intrinsic spectral width of the FRB, and $T_{\rm sys}$ is the system temperature. To achieve sufficient sensitivity to detect an FRB with $B < \Delta \nu_{\rm FRB}$, we find that the main antenna and receiver on each spacecraft must satisfy:
\begin{equation}
\left(\frac{D_{\rm eff}}{8 ~{\rm m}}\right)^2\left(\frac{T_{\rm sys}}{20 \; {\rm K}} \right)^{-1}  \left(\frac{B}{500 \;{\rm MHz}} \right) \gtrsim 1.
\label{eqn:baselinespecs}
\end{equation}

Parameter configurations that result in the left-hand side of equation~(\ref{eqn:baselinespecs}) significantly below unity would risk detecting an insufficient sample of FRBs (as detailed in \S~\ref{ss:rates}). It is important to note that many observed FRBs have spectral widths smaller than or comparable to 500~MHz. When an FRB's spectral width is much smaller than the system bandwidth such that $\Delta \nu_{\rm FRB} < B$, a larger bandwidth still provides benefits by increasing the probability that brighter burst components fall within the observed frequency range.  Additionally, CPS requires ${\rm min}[B, \Delta \nu_{\rm FRB}] \gtrsim 0.1 \nu$ to remove the light travel delays from plasma dispersion without incurring a significant loss of timing accuracy \citep{2022arXiv221007159B}.

In the subsequent analysis, we assume a baseline configuration where each parameter meets or exceeds the threshold values in equation~(\ref{eqn:baselinespecs}). Specifically, our nominal values for the CPS mission are $D_{\rm eff} = 8 ~{\rm m}$, $T_{\rm sys} = 20 \; {\rm K}$, and $B = 1.6 \;{\rm GHz}$. We assess the technical feasibility of achieving these parameters within the mission constraints.

\subsubsection{High-gain antenna and feeds}
\label{sss:hga}
Here we assess the feasibility of deploying antenna elements with effective diameters of at least $D_{\rm eff} = 8~$m on each spacecraft. Achieving this value requires a geometric dish size of $D=8.5$-$9~$m with feed-horn illumination efficiency of 80-90\%, which is higher than the $70\%$ achieved in many terrestrial systems, such as the Green Bank Telescope. This illumination efficiency can be realized when the feedhorn radiation pattern extends beyond the dish's reflecting surface, taking advantage of the low sky temperature behind a dish situated in the outer solar system.\footnote{The Sun at 3 GHz -- the bottom of the nominal CPS band -- has a flux density of $\sim 1\times 10^6$ Jy \citep{SWPC_F107}, which at $20~$AU would be reduced to $\sim 2500$~Jy. A flux of 2500 Jy contributes $30 {\cal B}(\theta)$ K to the system temperature of a $D=8~$m dish, where ${\cal B}(\theta)$ represents the response of the feedhorn at angle $\theta$ off its axis, with ${\cal B}(\theta \approx 0) \approx 1$. As long as ${\cal B}$ is somewhat less than unity for angles where the Sun might be located, CPS could pursue observations using such an over-illuminated feedhorn without significantly increasing $T_{\rm sys}$ above our nominal value of $20~$K. As the frequency increases above $3~$GHz, the Sun's flux declines substantially, making this consideration even less relevant.}

The CPS radio dish should be as lightweight as possible, as it likely dominates the spacecraft moment of inertia. Reducing dish mass, therefore, decreases reaction wheel requirements and enables faster slew speeds for target acquisition. We base our mass estimate on the Soil Moisture Active Passive (SMAP) satellite, which deployed a 56~kg fold-out antenna with $D=6~$m, to estimate an achievable dish mass. While SMAP operated at $1.4~$GHz (L-band), its mesh had a spacing of 0.7~mm and a surface error of 1.5~mm \citep{smap}, suggesting that similar antenna designs could accommodate the higher frequencies required by CPS (3-8 GHz). Since CPS does not require high-fidelity imaging as SMAP did, it can tolerate greater surface imperfections.  Taking dish mass to scale with collecting area (e.g. $M \propto D^2$), the implied mass for a 9~m SMAP-like dish would be $\approx 126$~kg. 

Additionally, to reduce angular momentum buildup, it is ideal that the dish be largely transmissive to solar radiation (\S~\ref{ss:attitude}).  This is the case for the SMAP antenna, owing to its mesh design.

The high-gain antenna must operate in the 3-6 GHz range for the FRB science, as described in \S~\ref{ss:geometricdistaanceintro}. However, ranging operations (and gravitational wave science) would benefit from higher frequencies, with NASA heritage in spacecraft ranging at 8 GHz (X-band) and 30 GHz (Ka-band). 
For our nominal design, we specify an operational frequency range of 3-8 GHz, with ranging conducted at 8 GHz and FRB science in the 3-6 GHz band.

We envision that most science operations will utilize a wideband feedhorn sensitive over the 3-6 GHz spectral range. Since the system temperature can be made lower in a receive-only configuration, our fiducial design incorporates a separate 8 GHz feedhorn dedicated to ranging and telemetry functions. This dual-feedhorn configuration is illustrated in Figure~\ref{fig:Miguels_satellite}, which also shows a thermal shield separating the two horns to minimize electromagnetic and thermal coupling. Additionally, during ranging operations where spectral overlap exists between the two horns, it may be possible to use the low-temperature radio science feedhorn for the reception of the ranging signal, while the transmit feedhorn functions solely for transmission, as simultaneous transmission and reception are not required.

\subsubsection{Bandwidth, analog to digital conversion, and data storage}
\label{sss:bandwidth}
The radio receiver requires wide bandwidths satisfying $B /\nu \gtrsim 0.1$ in order to avoid losing significant sensitivity when correcting for timing errors due to differential dispersion between the telescopes \citep{2022arXiv221007159B} and to achieve adequate flux sensitivities (\S~\ref{ss:rates}). However, the wider the bandwidth, the higher the bit rates required for on-board processing. Fortunately, space-qualified analog-to-digital converters capable of the gigabit processing speeds required by our bandwidths are available. As a representative component, we use the Teledyne EV12AQ600 space-qualified analog-to-digital converter. It can process 6.4 giga-samples per second, which for two polarizations would allow for a bandwidth of $B = 1.6~$GHz.

The voltage timeseries also needs to be stored, especially in cases where the CPS spacecraft cannot self-trigger on the FRB and, therefore, must instead store a continuous buffer over at least one light travel time between the Earth and the spacecraft in order not to lose the data before it is notified that a burst has occurred. To estimate the data storage requirements, we assume 2 polarizations and 2-bit quantization.\footnote{Two-bit quantization leads to a 12\% loss in flux sensitivity from quantization \citep{2017isra.book.....T}. We incorporate this loss in subsequent sensitivity estimates.} The required minimum storage capacity for one light travel time for a spacecraft at a distance $r$ is:
\begin{equation}
\text{\#~Bytes} = 15~ \text{TB} \left( \frac{B}{1.6~{\rm GHz}} \right) \left( \frac{r}{20~{\rm AU}} \right),
\label{eqn:disk}
\end{equation}
where $B=1.6~$GHz is the nominal bandwidth (Table~\ref{table:representativecomponents}) and $r=20~$AU is a typical spacecraft distance after ten years for our nominal trajectories (\S~\ref{ss:trajectories}).
Equation~(\ref{eqn:disk}) indicates that a minimum of $\sim 10$~TB of data storage is needed on each spacecraft if a signal is to be sent shortly after an FRB has occurred at the maximum distance when using the full sampling rate of the Teledyne EV12AQ600. 

While 10~TB exceeds the storage capacity previously deployed on NASA spacecraft, it would be used to store the voltage timeseries rather than mission-critical operational data. Therefore, moderate bit corruption would not compromise the scientific objectives. Terabyte space-qualified storage options are becoming available.  We use Phison's 8~TB M.2 2280 SSD flash drive (TRL 6; \citealt{phison_8tb_ssd}) as our representative component. 

Each CPS spacecraft would be required to store data for at least 10\% of the time, when CPS is coordinating observations with terrestrial observatories.  (This concern is not relevant when the CPS spacecraft self trigger or for gravitational wave detection, where these modes do not require the full voltage timeseries to be stored;  \S~\ref{sec:science}.) However, SSD flash storage typically fails after $N_{\rm cycles}\sim 3000$ write cycles, which results in the storage having a lifetime of
\begin{equation}
\tau_{\rm write} = 1.9~\text{yr}~ \left(\frac{N_{\rm cycles}}{3000} \right)\left( \frac{B}{1.6~{\rm GHz}} \right)^{-1} \left( \frac{\text{\#~Bytes}}{\rm 32~TB} \right),
\end{equation}
where $\tau_{\rm write}$ is the number of years of continuous write before storage failure.
This lifetime must be $\gtrsim 1$ years over the ten-year nominal mission to provide sufficient operational time for joint observations with ground-based telescopes; though several years would be ideal.  

To meet the light travel time and lifetime requirements, our nominal design uses four 8~TB Phison drives per CPS spacecraft.  Even more storage might be ideal, as in our nominal mission many spacecraft reach 2-3$\times$ the $r=20~$AU light travel delay that motivates this amount of storage after ten years (\S~\ref{ss:trajectories}).

In conclusion, the required data processing and storage capabilities are challenging but potentially feasible. We further estimate that these processing and storage systems would require approximately 10-20~Watts (cf. Table~\ref{table:representativecomponents}).  Half of this power is the analog path and its digitization. The other half is for the digital processing.  




\subsubsection{System temperature}
\label{sss:Tsys}
The system temperature, $T_{\rm sys}$, is principally determined by the thermal noise entering the receiver, the noise contribution from the low-noise amplifier (LNA), and various signal losses in the system. In the primary $3-6~$GHz frequency band used by CPS, $T_{\rm sys}$ can be significantly lower than the ambient temperature even when the LNA is operating at room temperature.  Space-qualified receivers operating at room temperature have demonstrated system temperatures of approximately 50~K at $3$-$8~$GHz \citep{nardamiteq_lna_catalog}.\footnote{In fact, \citet{weinreb2021low} found an LNA temperature of $7$ (3.3)~K for an LNA operating at $\nu = 1.4~$GHz and at a temperature of $T=310~(230)~$K.} Further reductions in system temperature can be achieved by mounting the analog receiver onto a cold plate that is isolated to maintain significantly lower temperatures than the spacecraft internal temperature. 

The RadioAstron satellite provides a relevant example. It was designed to achieve $T_{\rm sys} = 33~\mathrm{K}$ and ultimately reached $T_{\rm sys} = 43~\mathrm{K}$ in the L-band by attaching electronics to a passively cooled cold plate at approximately $T \approx 130~\mathrm{K}$ (see Table 6 in \citealt{RadioAstronUserHandbook2019}). Its 5~GHz C-band receiver was designed to reach $T_{\rm sys} = 66~\mathrm{K}$, being located in a less thermally isolated region of the cold plate than its L-band counterpart. However, this specification was not met, and $T_{\rm sys} = 130~\mathrm{K}$ was achieved instead \citep{2013ARep...57..153K}. The higher-than-expected $T_{\rm sys}$ were attributed to increased losses in the antenna feed assembly, impedance mismatches between polarization channels and low-noise amplifiers, and reduced isolation leading to self-excitation when both polarization channels operated simultaneously \citep{2013ARep...57..153K}. CPS would benefit from the substantial improvement in LNA technology (particularly in indium phosphide and metamorphic high electron mobility transistors) over the ensuing decades since the early 1990s when RadioAstron was engineered.  

One promising approach to achieve lower $T_{\rm sys}$ without employing cryogenics -- which would likely be prohibitively resource-intensive for CPS -- is cooling using light emitting diodes (LEDs). This technique uses specialized LEDs that convert phonons (vibrational energy within a material) into photons, which are then radiated away, thereby maintaining equipment at lower temperatures. While well-established in laboratory settings, this technology has not yet been deployed in space. NASA is investigating this technology for maintaining liquid hydrogen at 20~K and reducing boil-off to facilitate Mars exploration missions \citep{nasa_zero_boil_off}.

We conclude that achieving our fiducial specification of $T_{\rm sys} = 20~\mathrm{K}$ for the wideband receiver would be challenging but potentially feasible. 




\subsubsection{Attitude control and pointing}
\label{ss:attitude}
The CPS mission requires reaction wheels for controlling spacecraft orientation (at least between when an FRB occurs and when ranging operations can be performed) because using thrusters would likely compromise the precise positional calibration needed between ranging operations.\footnote{It is possible thrusters could be configured to exert a torque without much net force, in which case thursters alone would be sufficient.} To determine appropriate reaction wheel specifications, we first consider the spacecrafts' moment of inertia, which will be dominated by the radio dish. For a uniform on-axis dish of mass $M$, the moment of inertia for rotations along the dish axis is $I = M D^2/8$. Given our estimated dish mass of $\approx 100~$kg (\S~\ref{sss:hga}) and diameter $D= 9$ meters, this yields $I \approx 1000~$kg~m$^2$.

We have identified several commercial reaction wheels with maximum torque of $\tau_{\rm max} \sim 0.1~$N\,m and maximum angular momentum capacity of $H_{\rm max} = 10~$N\,m\,s. These typically consume between 4 and 10 Watts during normal operation and weigh several kilograms. 
Reaction wheels with $H_{\rm max} = 10~$N\,m\,s meet our requirement of one full rotation within a day. With $\tau_{\rm max}=0.1~$N~m and $H_{\rm max}=10~$N~m~s applied to a spacecraft with $I = 1000~$kg~m$^2$ , the maximum angular acceleration and angular velocity would be $a_{\rm max} = \tau_{\rm max}/I = 1300~$radian~hr$^{-2}$ and $\omega_{\rm max} = H_{\rm max}/I = 36~$radian~hr$^{-1}$, respectively. This maximum rate exceeds our requirement of one full turn per day (Table~\ref{table:instrument_requirements}) by more than an order of magnitude.

Reaction wheels will accumulate angular momentum over time, necessitating periodic momentum unloading via thrusters. For CPS, thruster firings are acceptable only if the interval between firings is longer than the time after an FRB occurs for the system to establish spacecraft spacetime positions via trilaterations (hours to days). For long-term spacecraft tracking to constrain the Solar System mass distribution, even less frequent thruster firings would be beneficial (\S~\ref{ss:solarsystemmass}). 

The outer Solar System environment provides a significant advantage with minimal environmental torques. For a spacecraft with effective surface area $A_{\rm eff}$ with the center of solar radiation force offset by $\delta b$ from the center of mass, the angular momentum buildup would be
\begin{equation}
H \sim A_{\rm eff} \, \delta b \, \frac{F_{\odot}}{c} \, t = 
30 
\left(\frac{A_{\rm eff}}{10~{\rm m}^2}\right) \left(\frac{\delta b}{1~{\rm m}}\right) 
\left(\frac{r}{30~{\rm AU}}\right)^{-2} 
\left(\frac{I}{1000~{\rm kg~m}^2}\right) 
\left(\frac{t}{1~{\rm week}}\right) 
\text{~N~m~s},
\end{equation}
where $F_{\odot}$ is the solar radiative flux.  A value of $A_{\rm eff} = 10~{\rm m}^2$ would result from a $D=9~$m mesh dish oriented towards the Sun absorbing $20\%$ of the incident solar radiation.  This calculation suggests that momentum unloading would be required on timescales of days to weeks at $10-30~$AU from the Sun.  The reaction wheels that control the spacecraft angular momentum perpendicular to the dish axis would experience the greatest load.  At shorter distances from the Sun, more frequent discharge would be necessary.

Additional design elements could potentially be incorporated to extend the time required for momentum unloading via thrusts. Possible solutions include solar radiation flaps offset from the spacecraft center of mass to actively reduce angular momentum buildup (a reflective 1~m$^2$ flap at the edge of the dish would be able to compensate for $ A_{\rm eff} \delta b \sim 10\;$m$^3$), magnetorquers utilizing the small $\mu$Gauss interplanetary magnetic field (although we find that this requires ones capable of generating a magnetic moment of 100 A m$^2$ and likely drawing a challenging tens of Watts of power), or implementing observation schedules that strategically reorient the spacecraft toward different FRBs in a manner that reduces angular momentum buildup.

For attitude determination, a star tracker is required. The pointing precision needed is approximately $\delta \theta = 0.2 \lambda/D$, where the $0.2$ factor ensures that the target FRB or spacecraft remains well within the antenna beam. This expression translates to $\delta \theta = 3$~arc-min~$(D/8~\text{m})^{-1}\times(8\text{ GHz}/\nu_{\text{max}})$. Given the minimal environmental torques in the outer Solar System, the pointing precision will likely be limited by the angular accuracy of the star tracker itself. Commercial options such as the Jena-Optronik ASTRO APS Star Tracker (2~kg and 5~Watts) can achieve sub-arcsecond pointing precision \citep{jena_optronik_astro_aps}, easily meeting the CPS requirement.



\subsection{Spacecraft-spacecraft and terrestrial station-spacecraft ranging system}
\label{ss:ssranging}

CPS needs to constrain $n_{\rm pos} = 3N_{\rm sat} - 3 - 2$ spacecraft spatial position parameters to accurately measure wavefront curvature, where the subtraction of 3 accounts for insensitivity to overall translations and the subtraction of 2 accounts for insensitivity to overall rotations. With spacecraft-spacecraft ranging, there are $N_{\rm sat}(N_{\rm sat} - 1)/2$ constraints available. Equating these expressions shows that a minimum of 5 spacecraft are required to fully constrain their spatial positions via self-ranging alone. However, telemetry with ground stations can provide four additional constraints through distance measurements using one-way ranging, allowing CPS to operate with 4 spacecraft.\footnote{Given that the Solar System barycenter is currently uncertain to $\sim 100~$m, ranging from Earth to the different CPS spacecraft must be performed within a window of several hours to ensure that the relative uncertainty in Earth's position remains below a few centimeters.} Additionally, two-way ranging (where the spacecraft rebroadcast in phase the received signal) must be performed between each spacecraft and a ground station to calibrate the onboard clocks.

Radio ranging constitutes the only direct communication required between CPS spacecraft, with all other communication occurring with terrestrial stations.\footnote{Given advances in the frequency stability of resonant cavities, it may now be possible to achieve stable delay locks with $\sim 10~$Watt lasers and $\sim 10~$cm diameter mirrors over tens of astronomical unit distances.  This could allow lasers to be used for CPS ranging rather than radio broadcasts, further avoiding the need to slew and reorient dishes.  See \citet{2024arXiv241115072M} for discussion.  However, such an approach would require equipping the CPS spacecraft with lasers, mirrors, and extremely stable resonators.} This configuration is advantageous because the signal-to-noise ratio for ranging operations (where the transmitted waveform is known) scales inversely with separation $x$. The carrier-to-noise density ratio ($C/N_0$) for spacecraft-spacecraft ranging can be expressed as:
\begin{eqnarray}
C/N_0 &=& 31 \text{ dB-Hz} + 40 \log_{10}\left(\frac{D_{\text{eff}}}{8~\text{m}}\right) - 10 \log_{10}\left(\frac{T_{\text{sys}}}{50 \;\text{K}}\right) \nonumber \\ 
&& - 20\log_{10}\left(\frac{x}{50 \text{ AU}}\right) + 20 \log_{10}\left(\frac{\nu}{8 \text{ GHz}}\right) + 10 \log_{10}\left(\frac{P_{\rm em}}{30 \text{ Watt}}\right),
\label{eqn:C/N}
\end{eqnarray}
where we have evaluated at plausible specifications and note that batteries could be used for a temporary power boost over the minutes during which ranging occurs.
 The $C/N_0$ for terrestrial station-spacecraft ranging would likely be substantially higher due to the larger dishes available on Earth. Furthermore, if the spacecraft can transmit and receive in the Ka band ($30~$GHz), $C/N_0$ would increase by an additional 11.5~dB-Hz. For comparison, the X-band downlink from Cassini at $10~$AU to a 34\,m Deep Space Network antenna achieved $[C/N_0]_{\rm dB-Hz} \approx 40$-$50$ \citep{Wang2005}, while the Voyager spacecraft at $\gtrsim 140$~AU still maintain communications at $[C/N_0]_{\rm dB-Hz} \approx 30$ with the 70\,m Deep Space Network Goldstone antenna \citep{spacecommunications}.

Despite the relatively low $C/N_0$ for spacecraft-spacecraft ranging, there is substantial precedent for successful ranging with weak signals from Global Navigation Satellite Systems. A carrier-to-noise density ratio of 35 dB-Hz is the typical ``acquisition threshold'' for Global Positioning System ranging code -- the threshold at which a delay-lock loop can acquire the frequency and delay of a signal through systematic search over a grid of delays and frequencies, with each frequency-delay combination in the search integrating for $1~$ms \citep{misraenge}. Since the CPS spacecraft' velocities and positions would be extremely well constrained and integration times could extend beyond $1~$ms, the threshold for acquisition would likely be considerably lower. Indeed, the Deep Space Network routinely achieves signal lock for transmissions with $[C/N_0]_{\rm dB-Hz}\approx 15$ \citep{Folkner2011}.

Once a delay lock is established, the expected ranging error is \citep{misraenge}:
\begin{equation}
\delta x = 1.6 \times 10^{\frac{C/N_0 - 31}{20}} \times \left(\frac{T_c}{10^{-7} \text{ s}}\right) \times \left(\frac{\epsilon}{0.1} \times \frac{1 \text{ min}}{t}\right)^{\frac{1}{2}} \text{ cm},
\label{eqn:error_ranging}
\end{equation}
where $T_c$ is the duration of a code chip in the pseudorandom code, $t$ is the integration time, and $\epsilon$ is the correlator spacing of the delay lock loop circuit. Values of $T_c \sim 10^{-6}$~s and $T_c \sim 10^{-7}$~s are used for the civilian and military GPS coarse acquisition (C/A) pseudorandom codes, respectively. Values of $\epsilon \sim 0.1$ are typical for GPS receivers, primarily due to radio frequency interference considerations, though smaller values would likely be suitable for our application in the relatively radio-quiet outer Solar System \citep{misraenge, 2022arXiv221007159B}. Equation~(\ref{eqn:error_ranging}) indicates that with our fiducial specifications yielding $C/N_0 \approx 31$ dB-Hz over a $50~$AU baseline (Eqn.~\ref{eqn:C/N}), centimeter precision is achievable in one minute of integration using a code chip of $10^{-7}$s.

Thus, thermal noise contributions to ranging error are likely to fall below our requirement of $\delta x_{1\rm d} = 5$~cm in each spatial dimension.  (This requirement on $\delta x_{1\rm d}$ is most relevant for the second shortest arm for $N=5$ [the shortest for $N=4$]; \S~\ref{sss:trajectories_beam}.) Internal delays in the analog signal path must also be controlled at this level. 
Internal calibration is discussed further in \S~\ref{ss:internalcal}.

For X-band communications that either target Earth or pass through the inner Solar System, phase delays due to the interplanetary plasma and Earth's atmosphere can exceed our accuracy requirement of $\delta x_{1\rm d} = 5$~cm \citep{2022arXiv221007159B}. Consequently, CPS must implement dual-frequency ranging, similar to Global Navigation Satellite Systems, to eliminate plasma dispersion effects. If the two frequencies are separated by factors of 1.5 or 2, the optimal bias-free method for eliminating dispersion increases the error from ranging by a factor of 2 and 1.4, respectively, relative to equation~(\ref{eqn:error_ranging}) \citep{misraenge}. 

Earth-based ranging also encounters non-dispersive tropospheric delay, which cannot be eliminated with two-way ranging. At zenith, the dry component can be predicted to an accuracy of a few millimeters, while the wet delay at zenith can be modeled to an accuracy of 1-2~cm \citep{misraenge}, somewhat smaller than our $5~$cm error tolerance. Furthermore, the wet delay error can be substantially reduced by using water vapor radiometers, which have been installed at some DSN antenna sites \citep{dopplerranging}.

\subsubsection{Acceleration control}
\label{ss:accelerations}
\begin{figure}[htbp]
\centering
\includegraphics[width=\textwidth]{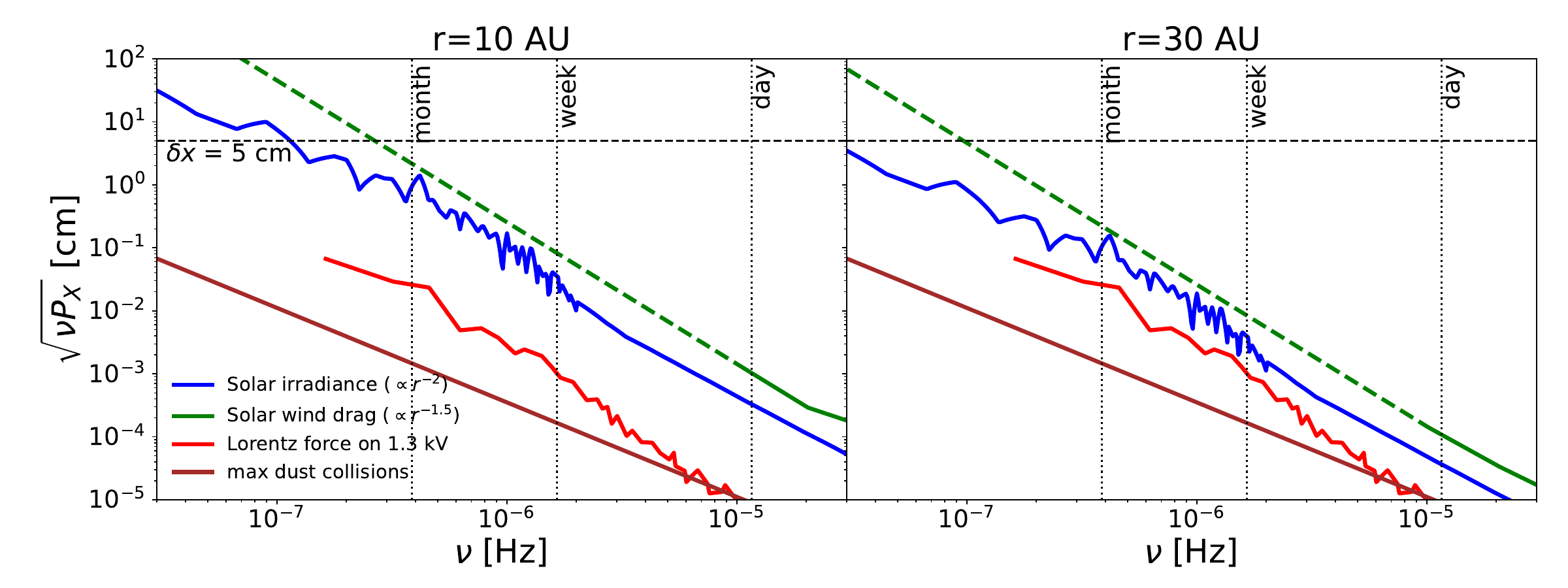}
\vspace{-0.5cm}
\caption{\small The square root of the dimensionless displacement power spectrum (multiplied by frequency) for the most significant stochastic sources of spacecraft displacement in the outer Solar System. This value, when evaluated at frequencies whose period corresponds to the day to possibly even month cadence of positional calibration, approximates the spacecraft displacement due to stochastic accelerations between calibrations. These calculations assume a spacecraft with mass $M_{\rm sat}=10^3$~kg and effective area $A_{\rm eff}=10~$m$^2$ at solar distances of $r=10$~AU (left panel) and $r=30$~AU (right panel). Note that displacements from solar irradiance variations and the solar wind act primarily in the radial direction, whereas dust collisions would be predominantly aligned with the interstellar dust flow. }
\label{fig:stochasticdisp}
\end{figure}

Beyond the trilateration error requirements, the CPS spacecrafts' positional calibration must be performed often enough to maintain knowledge of spacecraft space-time positions to within tolerance. This can be accomplished through regular spacecraft-spacecraft and spacecraft-Earth ranging. Additionally, after an FRB detection, trilateration can be performed immediately to provide positional constraints as temporally close as possible to the event.

The required frequency of calibration operations might be determined by the magnitude of stochastic accelerations in the outer Solar System environment. Figure~\ref{fig:stochasticdisp} shows the dimensionless power spectrum from various acceleration sources at $r=10$~AU and $r=30$~AU, including the accelerations from solar irradiance variations,\footnote{This ignores the mean radiation pressure from the Sun.} the solar wind, interplanetary magnetic fields acting on a \emph{maximally} charged spacecraft, and dust collisions. These estimates assume a spacecraft mass of $M_{\rm sat}=10^3$~kg and an effective area of $A_{\rm eff}=10~$m$^2$ (which would be double the projected geometric area for a perfectly reflective surface), and are detailed in \citet{2024arXiv241115072M} in the context of accelerations, noting the conversion from acceleration estimates to displacements is $-\omega^2 \widetilde{x} = \widetilde{a}$.  
The combined effect of these curves -- dominated by solar irradiance variations and solar wind drag -- gives the total amplitude of stochastic displacements. The figure annotates wave periods corresponding to a day, a week, and a month, demonstrating that the displacement remains under the $\delta x_{1\rm d} =5$~cm threshold per dimension for periods exceeding one month.

The remarkably small stochastic displacements in the outer Solar System environment necessitate a detailed understanding of accelerations from reflections, thermal emissions, outgassing, and thruster leakage. For FRB science, accelerations from reflections and re-emissions of solar radiation must be modeled for the interval between an FRB detection and subsequent calibration. The time for unmodeled mean acceleration from these effects to exceed our $\delta x_{1\rm d}$ tolerance can be expressed as:
\begin{equation}
\tau_{\odot} = 1.6 {\rm ~days}~ \left( \frac{\delta x_{1\rm d}}{5 ~\text{cm}} \right)^{1/2} \left(\frac{f_{\rm unmodel}}{0.1}\right)^{-1/2}  \left(\frac{A_{\rm eff}}{10~{\rm m}^2} \right)^{-1/2} \left(\frac{M_{\rm sat}}{10^3~{\rm kg}}\right)^{1/2}
\left(\frac{r}{30~{\rm AU}}\right)
\label{equation:acceleration}
\end{equation}
where $f_{\rm unmodel}$ represents the fraction of reflected and re-emitted radiation (relative to the solar radiation flux) that remains unmodeled. If post-detection calibration requires approximately one day, this suggests that re-emissions need to be modeled to a precision of $f_{\rm unmodel} \sim 0.1$ for plausible spacecraft specifications.  Otherwise, more frequent calibrations would be required. Achieving better modeling precision than $f_{\rm unmodel} \sim 0.1$ may be feasible, particularly since these parameters can be empirically calibrated by a period of focused ranging operations between CPS spacecraft.

Equation~(\ref{equation:acceleration}) neglects thermal emission from the spacecraft itself, which actually dominates over solar radiation effects in the outer Solar System. At $r = 30~$AU, the absorbed solar radiation is only $10~ (A_{\rm eff}/30~{\rm m}^2)$~Watts, whereas the spacecraft operates on $\sim 500$~Watts and the RTG dissipates more than ten times its electrical output as waste heat. One advantage of this thermal emission is that, to lowest order, it does not depend on spacecraft orientation with respect to the Sun. Furthermore, most of it emanates from the RTG.  Both advantages should enable more precise calibration. However, if we assume $100$~Watts of unmodeled thermal emission in a single direction, the timescale over which the resulting acceleration can persist before exceeding tolerance drops to $\tau_{\odot} = 4~$hr for $\delta x_{1\rm d} = 5~$cm and $M=1000~$kg. This approaches the minimum time required for triggering and ranging after an FRB detection, which must be longer than at least the Earth-spacecraft light travel time (\S~\ref{ss:ssranging}).

To constrain the distribution of mass in the outer Solar System, which requires spacecraft positions modulo accelerations to be tracked over extended periods, the mean acceleration from solar radiation is likely the limiting factor, making the precision of reflection and emission modeling particularly important for this science case.


Emissions from thrusters and off-gassing must also be carefully controlled to maintain positional accuracy. For a spacecraft with mass $M_{\rm sat}=10^3$~kg, a continuous thruster leak of just 10 nano-newtons would result in a displacement of approximately 10~cm over one week. Similarly, off-gassing in one direction produces a force of $F_{\rm off-gas} \sim  (\Delta m/ 10^{-2} {\rm g}) (t/ 1 ~\text{week})^{-1}~$nano-Newton if venting $\Delta m$ of $280$~K gas over time $t$.  If ranging calibration can be performed within one day after an FRB detection, this allows for a maximum continuous force from thruster leaks or off-gassing of approximately one micro-newton. The constraints become more demanding for the gravitational wave (as described in \citealt{2024arXiv241115072M}) and Solar System mass distribution sciences.  To address these challenges, precise ventilation systems would likely be required to control the direction and timing of any necessary emissions. 

\subsubsection{Spacecraft atomic clock}
\label{ss:atomic_clocks}
Time synchronization across the CPS constellation is critical for accurate wavefront timing. This can be achieved either through terrestrial beacons sending regular pulses to the spacecraft to calibrate their clocks or by transmitting a signal to the spacecraft immediately after an FRB detection that establishes the current time. For the latter, even with immediate signal transmission, an on-board clock must maintain stability such that it drifts less than $\delta t < 0.16~(\delta x_{1\rm d}/5~{\rm cm})$~ns before follow-up calibration can be performed.  At a minimum, the time for follow-up calibration is the light travel time to the spacecraft or $2.6~{\rm hr}~(r/20~{\rm AU})$. Assuming white frequency noise scaling, we can express these requirements as a constraint on the $\tau = 1$~s Allan deviation of the atomic clock:
\begin{equation}
\sigma_y(1~{\rm s}) \leq 1.3\times10^{-12} \left(\frac{\delta x_{1\rm d}}{5~{\rm cm}} \right) \left(\frac{t_{\rm follow}}{4~{\rm hr}} \right)^{-1/2}.
\label{eqn:atomicclockspec}
\end{equation}

The hydrogen maser clocks deployed on Europe's Galileo GNSS satellites achieve an Allan deviation of $\sigma_y(1~{\rm s}) = 7\times10^{-13}$, with performance following white frequency noise scaling for several hours \citep{leonardo_hydrogen_maser}.  The Deep Space Atomic Clock (DSAC) is a space-qualified atomic clock that was developed by NASA. It uses mercury trapped ion technology and has demonstrated similar Allan deviations to the Galileo clocks during a year-long low-Earth orbit mission, while exhibiting white noise scaling to $10^5$~seconds \citep{2021Natur.595...43B}. In laboratory conditions, DSAC achieved an Allan deviation of $\sigma_y(1~{\rm s}) = (1$-$2)\times10^{-13}$, exceeding the requirements of equation~(\ref{eqn:atomicclockspec}) even for $t_{\rm follow} \approx 1~$week. The outer Solar System environment would eliminate Earth magnetic field variations and the 9$^\circ$ temperature fluctuations experienced by DSAC in Earth orbit, potentially providing an environment in which DSAC achieves timing drifts closer to its laboratory performance.

However, both the Galileo hydrogen maser and DSAC have substantial resource requirements that present challenges for outer Solar System missions. DSAC has a volume of 7.5 liters, a mass of 17~kg, and it uses a mammoth 44~W of power, with hydrogen maser clocks on Galileo satellites having even somewhat higher resource demands. A micro-mercury trapped ion clock (M2TIC) was developed by \citet{m2picclock} that achieves $\sigma_y(1~{\rm s}) \approx 5\times 10^{-12}$ with white noise scaling to $\sim 10^5$~seconds, while requiring only 1.1~L of volume, 1.2~kg of mass, and less than 6~W of power. Although its Allan deviation is somewhat higher than the requirement specified in equation~(\ref{eqn:atomicclockspec}) for $t_{\rm follow} = 4$~hr, such miniaturization represents an important step toward CPS viability.

NASA is also developing more precise optical atomic clocks for space missions, particularly in connection with very long baseline interferometry (VLBI) imaging of black holes. Clocks based on optical frequency combs show particular promise for reduced size and power consumption. \citet{Tomio_2024} demonstrated laboratory performance of $\sigma_y(1~{\rm s}) \approx 5\times 10^{-14}$, significantly surpassing the mission requirement in equation~(\ref{eqn:atomicclockspec}). Such high-performance clocks would allow the calibration of the CPS clocks to be conducted on a weekly or even monthly cadence.  A precise onboard clock also benefits the CPS gravitational wave science, as the strain sensitivity of CPS scales inversely with the $\sigma_y$ of the onboard clock (\S~\ref{ss:science_grav_waves}).

\subsubsection{Ground station support}
\label{ss:groundstations}

In addition to periodic DSN communication for data downlink, CPS requires a network of $3-4$ radio antennas scattered across the Earth and equipped with X or Ka band receivers.  As shown below, these require effective diameters of $D_{\rm gr} \gtrsim 8~$m. These dishes would be used for notifying the CPS antennas that an FRB has occurred (when in the mode of coordinating with ground based observatories; \S~\ref{sec:science}), clock calibration, and reception of two-way ranging.  For an atomic clock with $\sigma_y(1~{\rm s}) \sim 10^{-12}$, like hydrogen masers and the in-space performance of DSAC, the clock calibration would need to be performed within several hours of an FRB detection to meet the $5~$cm temporal accuracy specification. Additionally, rapid notifications from the ground array reduces the onboard data storage requirements when observing in coordination with large radio telescopes on Earth.  

The calibration requirements differ depending on the number of spacecraft in the constellation:

\paragraph{Five spacecraft:} This number is sufficient for the constellation itself to calibrate its geometry.  The ground network serves primarily as a timing beacon for the CPS spacecraft, sending a signal soon after an FRB has occurred, without requiring precise knowledge of spacecraft distances, as an overall rotation of the array relative to the ground does not affect the wavefront curvature measurement. All spacecraft must receive the timing signal within several hours to ensure that uncertainties in Earth's ephemeris do not contribute significant errors to the relative clock calibrations.

\paragraph{Four spacecraft:} The system is one constraint short of being able to calibrate all the relevant spatial degrees of freedom of the spacecraft. Thus, two-way ranging (where the CPS spacecraft receives and rebroadcasts the signal back) with antennas on Earth is needed to provide additional constraints to calibrate distances.\\

\noindent Our nominal mission assumes five spacecraft, but if any spacecraft fails or is temporarily unavailable, then the second mode would be used. \\


The strength of a signal from a dish in the ground network to a CPS spacecraft is:
\begin{eqnarray}
C/N_0 &=& 40 \text{ dB-Hz} + 20 \log_{10}\left(\frac{D_{\text{trig}}}{1 \text{ meter}}\right) +20 \log_{10}\left(\frac{D_{\text{gr}}}{8~ \text{m}}\right) - 10 \log_{10}\left(\frac{T_{\text{sys}}}{150 \, \text{K}}\right) \nonumber \\
&& - 20\log_{10}\left(\frac{x}{30 \text{ AU}}\right) 
+ 20 \log_{10}\left(\frac{\nu}{30 \text{ GHz}}\right) + 10 \log_{10}\left(\frac{P_{\rm em, gr}}{10^3 \text{ Watt}}\right),\label{eqn:C/Ntrigger}
\end{eqnarray}
where we have evaluated this assuming 30~GHz Ka-band broadcasts to the trigger antenna, and the triggering antenna operates at a much higher system temperature than the main dish, nominally taken to be $T_{\text{sys}} = 150$ (although $8~$GHz X-band broadcasts would also be possible with a smaller  $T_{\text{sys}}$ or larger diameter antenna). As discussed earlier in this subsection, $C/N_0\gtrsim 30$ should be sufficient to achieve a delay lock within seconds after the interception of a C/A code transmitted from Earth with GPS-like specifications.  This code can then be used to calibrate the onboard clocks.  Thus, antennas with effective diameters of $D_{\text{gr}}\approx 8~$m capable of kilowatt transmissions, combined with $D_{\text{trig}} \approx 1~$m triggering antennas on the CPS spacecraft, should be sufficiently sensitive for notifying the CPS spacecraft that an FRB has occurred and then achieving a delay lock for clock and possibly distance calibration. The receiver for the trigger antenna would also need to periodically activate and search for an alert signal. For $D_{\rm trig} \approx 1$~m, the antenna would likely be too small to send a ranging signal back, requiring the CPS spacecraft to reorient its main dish to execute two-way ranging with the ground network (which is only necessary in the four spacecraft mode).

Not all CPS spacecraft would be able to direct their triggering antenna towards Earth at all times. Assuming the triggering antenna is located on the dish axis and half a dish diameter below the center of the main dish, its observation of polar angles less than $\approx 45^\circ$ (15\% of the sky) would be obscured by the main dish. This suggests that much of the time, one of the five spacecraft would not be able to direct its trigger antenna towards Earth while pointing towards an FRB source. When operating in conjunction with a terrestrial radio telescope, the spacecraft that cannot point their trigger antenna directly at Earth would need to periodically reorient by up to $\approx 45^\circ$ to check for FRB alerts. 

\subsubsection{Internal delay calibration}
\label{ss:internalcal}
GPS satellites achieve sub-nanosecond calibration of delays internal to the spacecraft electronics \citep{misraenge}.  To meet the $\delta x <5~$cm spacecraft localization requirement, CPS must calibrate internal delays to a similar precision.  If ranging is done in one direction between every spacecraft pair, plus one- and two-way ranging with the Earth dishes, this would provide five additional calibration constraints beyond the minimum $4 \, N_{\rm sat}$ constraints needed to calibrate the CPS spacetime positions.  (This becomes eight additional constraints with six spacecraft and twelve with seven.) Five additional constraints are sufficient to calibrate each spacecraft's internal delay for receiving, if the internal delay at transmission is perfectly known.  However, each spacecraft pair can also do ranging in the reverse direction, providing another $N_{\rm sat} (N_{\rm sat} - 1)/2$ constraints, more than enough to measure the instrumental delays at both transmission and reception.  

Additionally, with five spacecraft, there is an additional constraint from timing the light from an extragalactic source (like an FRB) beyond those required to solve for the wavefront curvature, CPS's observable for cosmic expansion science. This constraint can be used to refine the system calibration.\footnote{Additionally, FRBs at distances that are farther than those where CPS can measure their distance precisely provide two more constraints rather than one since CPS does not need to solve for the wavefront curvature.}  FRB sources from which CPS observes multiple FRBs over the mission lifetime would provide additional constraints that could be used to calibrate the array.

In conclusion, CPS should have enough constraints from self-ranging and FRB sources to calibrate internal delays in each spacecraft.


\subsection{Spacecraft trajectories}
\label{ss:trajectories}
\begin{figure}[H]
  \centering
    \includegraphics[width=\textwidth]{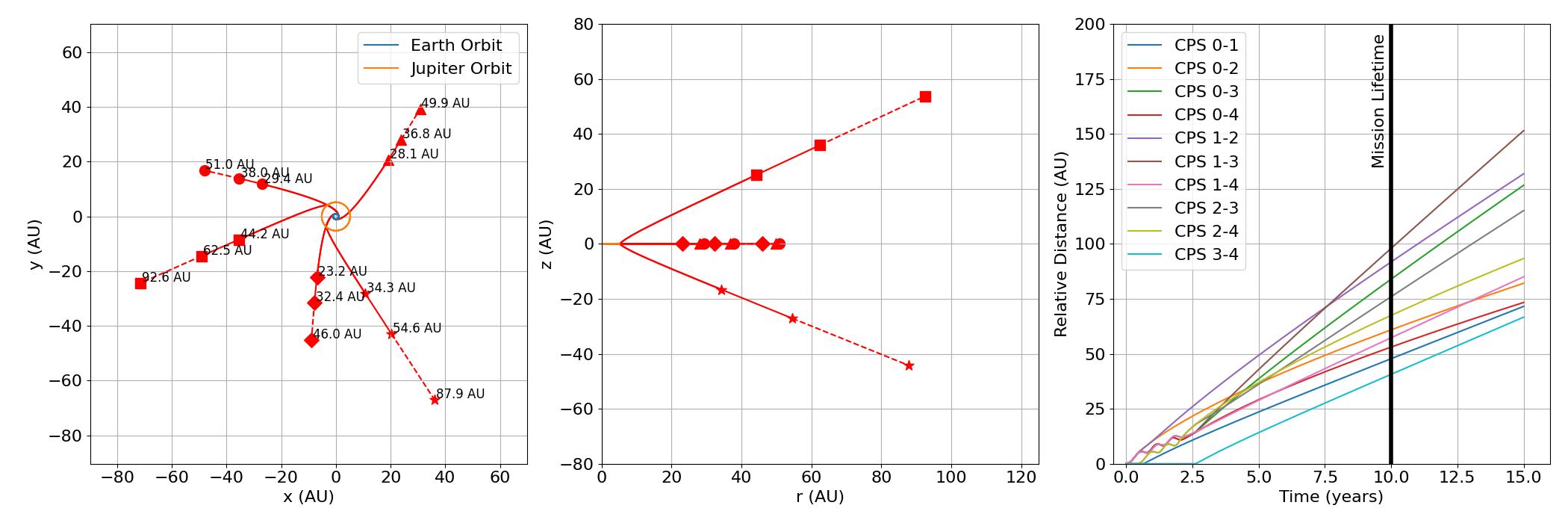}
    \includegraphics[width=\textwidth]{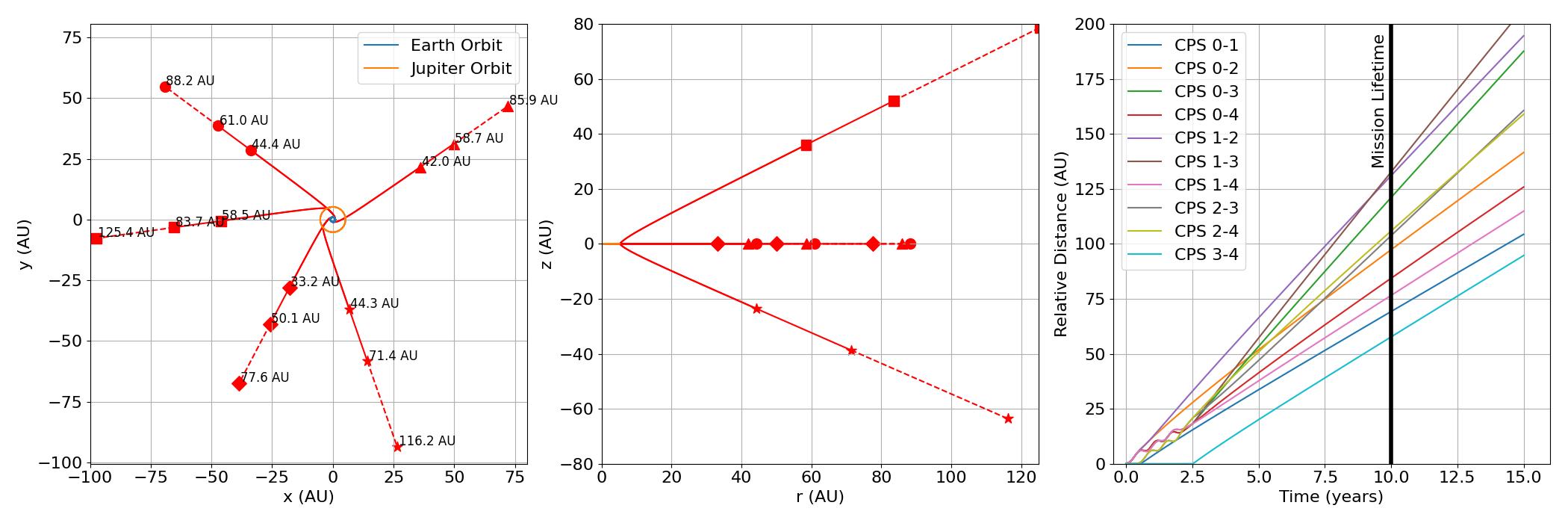}
  \caption{\small {\bf Nominal trajectories for the CPS spacecraft:} The top panels assume the same characteristic energy (C3) as the New Horizons spacecraft of $157~$km$^2$~s$^{-2}$, while the bottom panels assume $220~$km$^2$~s$^{-2}$, as may be achievable with next-generation third and fourth stage boosters. The left and middle panels show the orbits from a bird's-eye view of the ecliptic plane and a side view, respectively, with markers indicating distances from the Sun at $7$, $10$, and $15~$years and $r \equiv \sqrt{x^2+y^2}$.  The nominal 10-year mission lifetime is shown with solid lines, while a potential 5-year mission extension is represented by the dashed trajectory continuations.  The orbits of Earth and Jupiter are shown as circles in the left panels.  The right panels display the resulting separations between spacecraft pairs over time.}
  \label{fig:trajectories}
\end{figure}

For cosmic distance and dark matter science, the optimal deployment strategy involves launching the CPS spacecraft to the furthest separations in the shortest time possible, with at least one spacecraft's trajectory at a significant angle out of the ecliptic plane, as a three-dimensional constellation is necessary for precise spacecraft localizations with spacecraft-spacecraft trilaterations. The achievable spacecraft separations depend on both the initial launch velocity and subsequent planetary gravitational assists.  We only consider gravitational assists from Jupiter. Jupiter's substantial mass and location in the Solar System make it the optimal body for gravitational scattering and velocity assists to redirect spacecraft out of the ecliptic plane while boosting their velocities. However, it is possible that the mission benefits from using gravitational assists from other Solar System bodies.

The characteristic energy C3 -- defined as the spacecraft's velocity squared at infinity from Earth (excluding other Solar System masses) -- is the standard metric to quantify the kinetic energy of spacecraft after launch. The 500~kg New Horizons spacecraft, launched on an Atlas V rocket, achieved C3 = $157~$km$^2$~s$^{-2}$. With anticipated enhancements to NASA's Space Launch System (SLS), particularly improved third and fourth stage boosters on Block 1B rockets, New Horizons could potentially now be launched with a C3 of $ 300~$km$^{2}$~s$^{-2}$, while a spacecraft with twice New Horizons' mass could achieve C3 of $200~$km$^{2}$~s$^{-2}$ \citep{smith2019interstellar}. SLS Block 2 rockets would enable even higher C3 values \citep{stough2021sls}. As detailed CPS spacecraft design and mass budgets require further study, we examine trajectory scenarios that assume all spacecraft are launched with the New Horizons C3 value of $157~$km$^2$~s$^{-2}$ as well as another scenario that takes the larger value of $220~$km$^2$~s$^{-2}$.

We consider a three-launch scenario in which a pair of spacecraft is launched toward Jupiter in the first year, one spacecraft is launched in approximately the opposite direction six months later, and a final pair is launched about three years later when Jupiter's orbital position has advanced by $\approx 90^\circ$.  Using gravitational scattering off Jupiter as described below, this scenario disperses the spacecraft from one another, as this is the objective that optimizes CPS cosmic expansion science, while potentially being achievable with just three launches.  However, the spacecraft could also be launched in separate vehicles rather than being paired to maximize the C3 of individual spacecraft.  (Launching all spacecraft at the same time may be possible if some spacecraft initially orbit the inner Solar System before employing gravitational assists from Venus or Earth.)

For each pair of spacecraft launched towards Jupiter in the considered scenario, one spacecraft undergoes significant gravitational scattering off Jupiter, while the other experiences minimal deflection. This differentiation can be achieved through post-deployment thrusts. For example, with C3 = $157~$km$^2$~s$^{-2}$, we calculate that the additional thrust resulting in a velocity increase of 100~m~s$^{-1}$ shortly after deployment from the final rocket stage would lead to that spacecraft reaching Jupiter $30~R_J$ earlier than the trailing spacecraft, where $R_J$ is the radius of Jupiter, such that if the leading spacecraft is deflected by $70^\circ$, the second would experience only a $15^\circ$ deflection. Course adjustments of $\sim 100~$m~s$^{-1}$ are within standard spacecraft maneuvering capabilities.

The spacecraft experiencing significant Jupiter scattering must do so at high inclination angles relative to the ecliptic to avoid Jupiter's equatorial radiation belts (which extend to approximately $30^\circ$ from Jupiter's equator, with Jupiter's obliquity being $3^\circ$). The high inclination angle also satisfies the mission requirement for a spacecraft outside of the ecliptic plane to enable three-dimensional spacecraft-spacecraft (and spacecraft-FRB) trilateration, with the second spacecraft out of the ecliptic using our nominal launch specifications to provide redundancy. Our ensuing calculations choose a Jupiter deflection angle of $70^\circ$ for the two spacecraft that scatter significantly off Jupiter, comparable to the $60$-$90^\circ$ deflections experienced by the Voyager, Pioneer 11, and Cassini missions.

A  $70^\circ$ deflection angle requires approaching Jupiter within distances of 1.1 million kilometers (15~$R_J$) for C3 = $157~$km$^2$~s$^{-2}$ or 0.6 million kilometers (9~$R_J$) for C3 = $220~$km$^2$~s$^{-2}$. While closer than New Horizons' 2.3 million kilometer approach, which results in relatively more radiation exposure, the high-inclination trajectories with respect to Jupiter's equator do help compensate. We estimate radiation doses of 15-20~krad during the Jupiter flyby, assuming the very approximate radiation intensity scaling of $1000~(R/R_J)^{-2}~$krad~hr$^{-1}$ \citep{1134214}. This exposure is comparable to the $\sim 10~$krad~year$^{-1}$ typically expected for interplanetary missions \citep{interplanetaryradiation}, although we caution that these numbers are order-of-magnitude estimates.

Figure~\ref{fig:trajectories} illustrates the resulting trajectories for the proposed deployment scenarios. The top panels assume C3 = $157~$km$^2$~s$^{-2}$ (matching New Horizons), while the bottom panels show trajectories with C3 = $220~$km$^2$~s$^{-2}$.  The left panels show a bird's-eye view of the ecliptic plane ($x-y$), while the middle panels show the spacecraft position in the dimension orthogonal to the ecliptic ($z$) with the radial coordinate in the ecliptic ($r$). The rightmost panels show the separations between all the spacecraft pairs. The separations for the C3 = $157~$km$^2$~s$^{-2}$ scenario range from 20-60~AU after ten years, increasing to 35-85~AU after fifteen years. The launch scenario with C3 = $220~$km$^2$~s$^{-2}$ yields even greater separations, with some exceeding 100~AU after twelve years. We use these spacecraft trajectories in \S~\ref{sec:science} to estimate the mission's sensitivity to extragalactic distance measurements and to the clumpiness of the dark matter.

\subsubsection{FRB distance sensitivity for nominal spacecraft trajectories}
\label{sss:trajectories_beam}

\begin{figure}[h]
\centering
\begin{tabular}{@{}c@{\hspace{0.05\textwidth}}c@{}}
\underline{C3=$157$ km$^2$ s$^{-2}$} &
\underline{C3=$220$ km$^2$ s$^{-2}$} \\
\includegraphics[width=.32\textwidth, trim=0 0 0 100, clip]{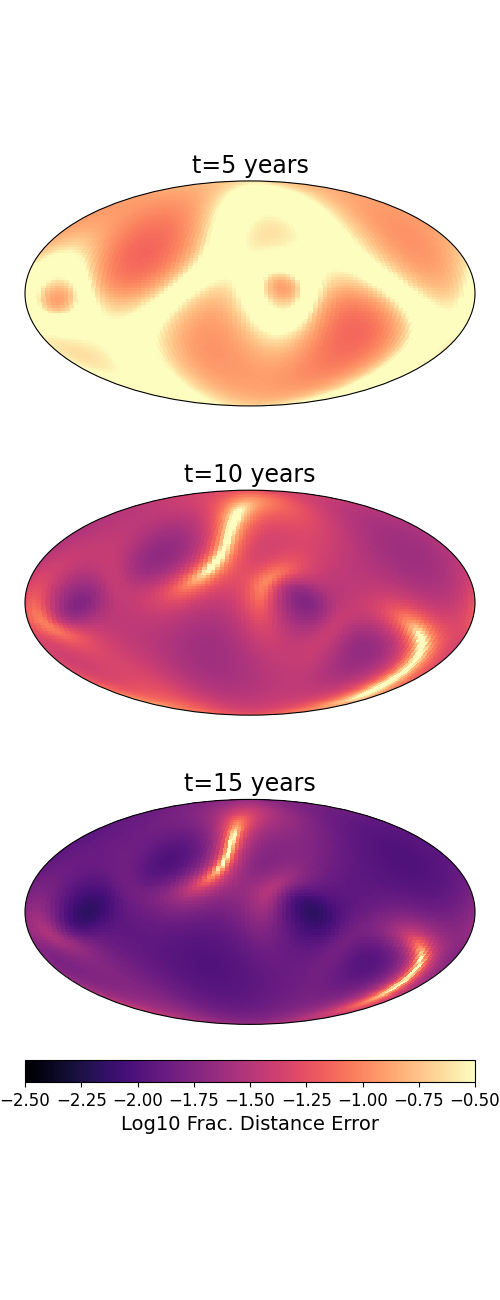} &
\includegraphics[width=.32\textwidth, trim=0 0 0 100, clip]{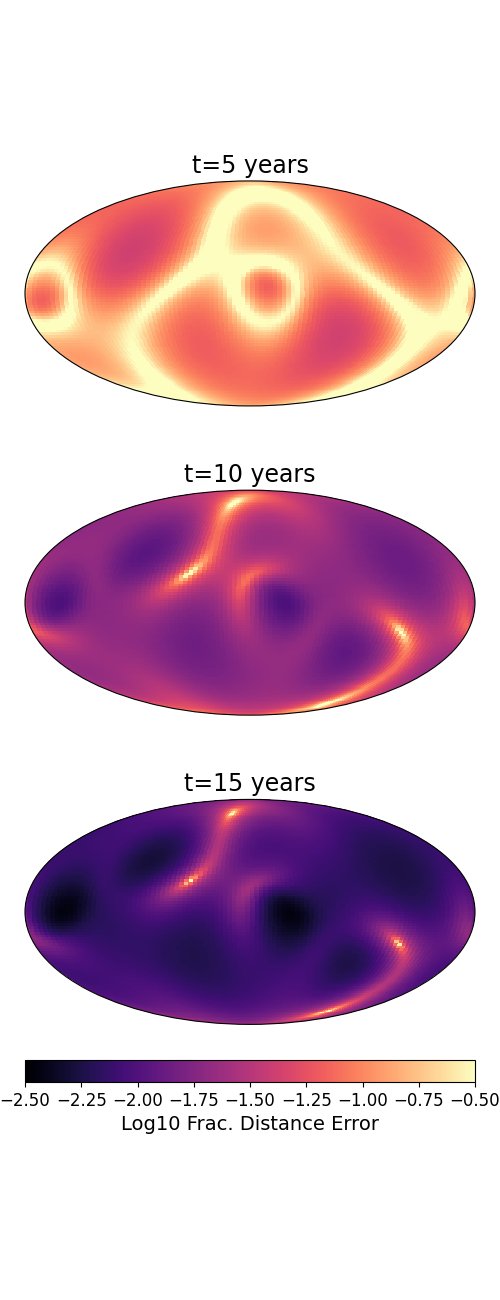}
\end{tabular}
    \vspace{-1.5cm}
  \caption{\small Sky pattern of fractional distance error for the mission with C3=$157$ km$^2$ s$^{-2}$ (left panels) and $220$ km$^2$ s$^{-2}$ (right panels) at three times for the nominal mission with five CPS spacecraft.  All Mollweide projections are shown in the ecliptic coordinate system with arbitrary longitudinal coordinate, as described further in the text. Fractional distance errors assume spacecraft positional errors of $5$ cm in each space-time coordinate for all the CPS spacecraft and a source at $d=100$ Mpc.  The fractional distance error scales proportionally with $d$.}
  \label{fig:beampattern}
\end{figure}

We can understand the sensitivity with direction that results from the nominal spacecraft trajectories. Figure~\ref{fig:beampattern} shows the fractional sensitivity at $t=5$, $10$, and $15$ years with a Mollweide projection in the ecliptic coordinate system. These fractional errors assume a source at $d=100$ Mpc, but this can be easily scaled to other distances using their scaling $\propto d$. 
The left panels are for C3=$157$ km$^2$ s$^{-2}$, while the right panels are for $220$ km$^2$ s$^{-2}$. This is computed using the fact that the time delay between detectors $0$ and $m>0$ is 

\begin{equation}
c \Delta t_{m} = -\boldsymbol{x}_{m} \cdot \boldsymbol{\widehat{d}} + \frac{x_{\perp, m}^2}{2d} + \mathcal{O}(x_{m}^3/d^2),
\label{eqn:deltat}
\end{equation}
where $\boldsymbol{x}_{\perp, m} = \boldsymbol{x}_{m} - (\boldsymbol{x}_{m} \cdot \boldsymbol{\widehat{d}}) \, \boldsymbol{\widehat{d}}$, and $\boldsymbol{x}_{m}$ is the baseline between detector 0 and $m$.  To calculate the sensitivity, we compute the Fisher information in wavefront curvature from the $\Delta t_{m}$ measurements:
\begin{equation}
F_{i j} = c^2 \sum_{m, n>0} \frac{\partial \Delta t_{m}}{\partial\lambda_i} [\mathbf{C}^{-1}]_{mn} \frac{\partial \Delta t_{n}}{\partial\lambda_j},
\end{equation}
where $\lambda_i$ are the set of parameters for each source: the two angles that specify direction ($\boldsymbol{\widehat{d}}$) and the source distance ($d$). We further assume the covariance is set by the nominal $\sigma_x = \sigma_t = 5$ cm space and time errors. For simplicity, the error on the index $0$ detector is taken to be zero, as this makes the covariance diagonal with $\mathbf{C}_{mn} \approx (\sigma_x^2 + \sigma_t^2) \delta^{\rm K}_{mn}$.  This covariance assumes that the error in timing the FRBs wavefront with respect to the onboard clock is smaller than the detector space-time positional uncertainty, which is likely a good approximation for our nominal specifications for FRBs that exceed our ${\rm SNR} >10$ detection threshold. This approximation of excluding the clock noise in detector $0$ will underestimate the distance error from the full covariance by up to a factor of $\sqrt{2/3} \approx 0.8$, but we expect the underestimate to be even less different from unity than this bound because the detector 0 noise has covariance between all of the $\Delta t_{m}$ measurements.\footnote{In practice, this expression for the Fisher matrix also means that our results will depend on which CPS spacecraft is selected as the detector 0 reference.  We find that this dependence is very slight by varying the reference detector in our calculations.}

We can calculate the fractional error on the source distance from the Fisher Matrix via $\sigma_d/d = \sqrt{[\mathbf{F}^{-1}]_{00}}/d$, where we take the distance parameter to be specified by the zeroth index.    
Figure~\ref{fig:beampattern} shows that this distance error is relatively uniform over much of the sky, especially at later times when the detectors are farther apart.  Let us orient to the $x-y-z$ coordinates in Figure~\ref{fig:trajectories}:  The direction aligned with the $x$-axis is the leftmost part of the Mollweide projection.  Continuing along the equator to the right, a quarter of the way in longitude corresponds to the $y$-axis in Figure~\ref{fig:trajectories}.  Varying latitudes in the Mollweide projections correspond to the out-of-the-ecliptic-plane $z$ direction. The regions of lower sensitivity, which are more prominent at earlier times in the mission, are due to sky directions where the projection of at least one baseline in the source direction, $x_\perp$, becomes small. This projection has a larger angular footprint for the CPS spacecraft with the shortest separations than for those with the largest.\footnote{The small ring of reduced sensitivity in Figure~\ref{fig:beampattern}  (at $b = 20^\circ$ and $\ell = -160^\circ$ near the ecliptic) is due to where $x_\perp$ from the spacecraft launched initially and the pair initially scattered off of Jupiter come into alignment. The larger ring is from the projection of the shortest spacecraft baselines.  
At later times, the regions of lower sensitivity in Figure~\ref{fig:beampattern} lift away from the ecliptic as one of the spacecraft in these pairs moves to higher latitudes, becoming less pronounced as the spacecraft pair further separates.}

  After some experimentation, we find that a good estimate for the sensitivity observed over much of the sky area in Figure~\ref{fig:beampattern} is given by the approximate distance sensitivity formula $\delta t = C_{00}/[x_\perp/(2cd)]$ from  \citet{2022arXiv221007159B}, where the value for $x_\perp$ is taken to be the full length of the second shortest baseline.  The reason that this formula is plausible is that, of the four $\Delta t_n$ that the 5-spacecraft CPS concept constrains, the one from the second shortest baseline provides the second largest error bar.  Since the system is over-constrained by one $\Delta t_n$  to solve for the three source parameters, the measurement from the shortest baseline (which provides the largest error bar) is not as important for shaping the constellations sensitivity to distance.

\begin{figure}[h]
  \centering
  \begin{center}
    \includegraphics[width=1\textwidth]{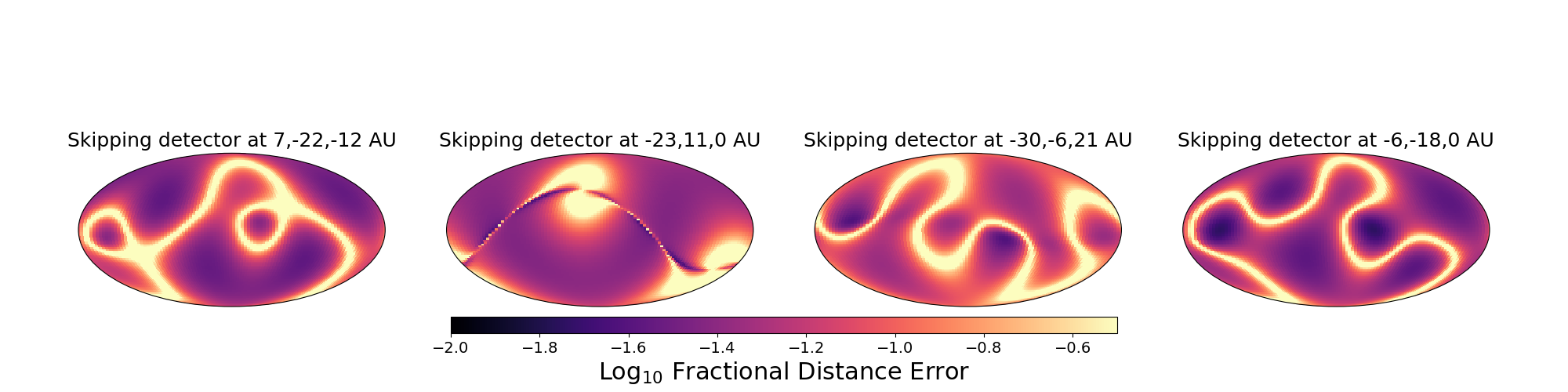}
    \end{center}
      \vspace{-.5cm}
  \caption{\small The same as Figure \ref{fig:beampattern}, but showing the impact of not using the detector at the referenced position and only considering $t=10~$yr and C3=$157$ km$^2$ s$^{-2}$. Since omitting one spacecraft results in only the critical number of four detectors being used for timing, the error is large in directions where the projected baseline vector for any antenna pair ($x_\perp$) is small.}
  \label{fig:beampatternskippped}
\end{figure}

The sensitivity of removing one spacecraft, so that there are four spacecraft in total contributing to the distance measurement, is shown in Figure~\ref{fig:beampatternskippped}. 
While even with four detectors the sensitivity is similar over all but a band of the sky, adding a fifth detector makes the sensitivity much more uniform across the sky.


\subsection{Telemetry and data processing}
\label{ss:signalacquisition}

After an FRB is detected, each spacecraft would reorient its main dish toward Earth to transmit the recorded data, likely to a Deep Space Network (DSN) station.  This FRB may occur days or even months before this data recovery step. Since repeating FRBs have known dispersion measures, most of the dispersion can be removed by de-convolution, compressing the FRB voltage data to occupy only the $\tau \sim 1\,$ms intrinsic width of the burst. With a bandwidth of $\Delta \nu = 1.6\,$GHz and two polarizations, a $\tau = 1\,$ms burst requires transmitting approximately ten million bits, assuming 2-bit quantization. A one-hour downlink to Earth would then require data rates of only $3\,$kbps. For comparison, the New Horizons spacecraft achieved $1$-$4\,$kbps data rates beyond Pluto's orbit at $r\sim 35~$AU \citep{2008SSRv..140...23F}; the CPS spacecraft would have a substantially larger main antenna than New Horizons, which would support higher data rates.  Assuming that CPS telemeters back three $1~$ms FRB voltage timeseries per month, consistent with the rates estimated in \S~\ref{ss:rates}, taking the conservative value of $3~$kbps for the downlinks would require three hours a month of DSN time per spacecraft.  Since the limited data that would be transmitted back can be easily stored on each spacecraft, the downlink can be scheduled for times of lower DSN utilization.

Once data is transmitted to Earth, the signals from different CPS spacecraft must be correlated to measure the precise arrival time differences between spacecraft. Here we assess the signal-to-noise ratio (SNR) required to reliably identify the correct geometric delay between signals arriving at different elements. If a spacecraft successfully triggers on an FRB, then the FRB arrival time is constrained to within approximately the burst duration of 1~ms. This timing window yields $\sim 2/(1{\rm~ ms}\times 5 ~\text{GHz}) = 4\times 10^7$ potential delays to search when correlating signals.  We adopt for our FRB rate estimates in \S~\ref{ss:rates} a 10$\, \sigma$ detection threshold, formally corresponding to a one-in-a-million chance for a spurious correlation when searching over $10^7$ delay values.\footnote{In the mode where CPS self-triggers (as will be described in \S~\ref{sec:science}), there are $N_{\rm sat} (N_{\rm sat} - 1)/2$ combinations of spacecraft pairs, the significance each delay is detected would be higher than our nominal detection requirement of 10$\, \sigma$ on a single baseline.}

This false-positive estimate, however, does not account for the additional phase from intervening plasma, which must be incorporated into the search alongside geometric delays. At 5 GHz, the wavefront advances by approximately 5~m from Earth's ionosphere \citep{misraenge}. Additionally, \citet{2022arXiv221007159B} find that the FRB radiation at two different spacecraft will have wavefronts that differ by $\sim 50~ (x/30 {\rm AU})^{5/6}$ meters at $5$~GHz due to the varying columns of interstellar plasma along the sightline to each spacecraft, where $x$ is the separation between the two spacecraft.  (This finding is based both on the inferred turbulent spectrum of the ISM and, more directly, on dispersion measure variations to millisecond pulsars as they move by tens of AU.)  When observing across a bandwidth $B$, the phase variation from the center to the edges of the band is reduced by a factor of $B/(2\nu)$ in the limit where $B \ll \nu$. Most FRBs exhibit emission covering a band of width $\Delta \nu \sim 500$ MHz, reducing phase variations by a factor of $\sim 1/20$. At $x\approx 30$~AU, this suggests a factor of $10^2$ increase in the search space compared to the limit of no differential dispersion, meaning our 10$\, \sigma$ threshold would correspond to a one-in-ten-thousand chance of spurious correlation once accounting for the number of trials.

In scenarios where the FRB signal is swamped by noise, as may occur when coordinating with large ground-based antennas (\S~\ref{sec:science}), the spacecraft itself may not be able to select the interval during which the FRB occurred. In such cases, ground-based instruments must first obtain precise angular coordinates of the FRB source and then communicate the specific intervals in the voltage timeseries that the spacecraft should telemeter back. The best angular localizations employ VLBI with continent-scale baselines. Assuming VLBI with an effective bandwidth of 200 MHz and SNR$=10$, the group delay timing precision would be $\delta t = 1/(2 \pi \, B\, {\rm SNR}) \approx 0.1$~ns. For a VLBI baseline $x_{\rm VLBI}=5000$~km, this yields an angular uncertainty of $\delta \theta = c /x_{\rm VLBI} \approx 1~$ milli-arcsecond, which is typical for terrestrial VLBI \citep{reidmicroarcsecond}.  Indeed, milli-arcsecond localizations have already been demonstrated for several repeating FRBs, including FRB 180916.J0158+65 \citep{VLBIFRB} and FRB 20121102A \citep{10.1093/mnras/stae632}.  VLBI localization constrains the arrival time at a CPS spacecraft to approximately $\delta t \approx \delta \theta \times r/c \approx 10^{-4}~$s for $r=20~$AU and $\delta \theta = 1~$milli-arcsecond, which is less than the typical duration of an FRB. Thus, Earth-based VLBI can predict FRB arrival times at the CPS spacecraft with sub-millisecond precision.


\section{Science forecasts}
\label{sec:science}

We envision several operational modes to achieve the key science objectives:
\begin{description}
    \item[Self-triggering mode:] The spacecraft independently triggers on bright FRBs and saves the voltage timeseries. Following detection, the system must reorient to transmit notification signals for clock calibration or distance measurements. The required response time depends on the stability of the onboard atomic clock, as described in \S~\ref{ss:atomic_clocks}.
    
    \item[Terrestrial coordination mode:] Large radio observatories on Earth operate in conjunction with CPS to achieve enhanced sensitivity. The spacecraft must continuously buffer voltage data, and a notification signal must be sent from Earth promptly to the CPS spacecraft after an FRB is detected to prevent overwriting the SSD storage. 
    Earth-based antennas may also need to send signals to calibrate onboard clocks soon after an FRB occurs, with the specifics of how soon thereafter a signal must be communicated depending on the precision of the onboard clock.
    
    \item[Gravitational wave science mode:] Pairs of spacecraft orient toward each other, transmitting and receiving a carrier wave to measure distance changes between them and constrain micro-Hertz gravitational wave signals.
\end{description}

The mass-distribution of the Solar System can potentially be studied concurrently with the two FRB modes by analyzing the series of space-time positions. However, it may be advantageous to minimize spacecraft reorientation to better model accelerations, potentially warranting an additional dedicated observation mode (\S~\ref{ss:science_solarsystem}).

In subsequent sections, we estimate the detection rate of FRBs in the first two science modes (\S~\ref{ss:rates}), use these estimates to project the constraints on FRB distances that CPS would achieve (\S~\ref{ss:science_distances}), consider potential constraints on dark matter clumpiness (\S~\ref{ss:science_distances}), and analyze gravitational wave detection capabilities (\S~\ref{ss:science_grav_waves}).

\subsection{CPS detection rates for FRBs}
\label{ss:rates}
The system will operate in two distinct modes for FRB science. In the self-triggering mode, each CPS spacecraft independently detects FRBs with millisecond durations above a fluence threshold of
\begin{eqnarray}
    F_{\nu} &=& 15\,\textrm{Jy-ms} ~~\eta_Q^{-1}
    \left(\frac{D_{\rm eff}}{8~\textrm{m}}\right)^{-2} 
    \left(\frac{\Delta \nu_{\rm FRB}}{500~\textrm{MHz}}\right)^{-1/2}
    \left(\frac{\tau}{1~\textrm{ms}}\right)^{1/2} \left(\frac{T_\text{sys}}{20~\text{K}} \right) \left(\frac{\text{SNR}}{10} \right).
    \label{eqn:Fnuself}
\end{eqnarray}
In the terrestrial coordination mode, a large ground-based telescope operates in conjunction with the CPS spacecraft, and the fluence sensitivity depends on the geometric mean of the respective collecting areas and system temperatures (and a fortuitous $1/\sqrt{2}$):
\begin{eqnarray}
    F_{\nu} &=& 1.5\,\textrm{Jy-ms}
     ~~\eta_Q^{-1/2} \left(\frac{D_{\rm eff}^{\rm CPS}}{8~\textrm{m}}\right)^{-1} \left(\frac{D_{\rm eff}^{\rm Earth}}{70~\textrm{m}}\right)^{-1} 
    \left(\frac{\Delta \nu_{\rm FRB}}{500~\textrm{MHz}}\right)^{-1/2}
    \left(\frac{\tau}{1~\textrm{ms}}\right)^{1/2} \nonumber \\
   & & \times  \left(\frac{T_\text{sys}^{\rm CPS}}{20~\text{K}}  \right)^{1/2}  \left(\frac{T_\text{sys}^{\rm Earth}}{30~\text{K}}  \right)^{1/2}  \left(\frac{\text{SNR}}{10} \right).
   \label{eqn:Fnuter}
\end{eqnarray}
Here, $\eta_Q$ represents the quantization loss at the CPS spacecraft, which, for our nominal 2-bit quantization, is $\eta_Q = 0.88$. As described in \S~\ref{ss:signalacquisition}, SNR $\approx 10$ is generally required to reliably detect the signal when searching across a grid of delays and dispersions, though somewhat lower SNR thresholds may be feasible in the self-triggering mode due to the larger number of spacecraft pairs than independent delays to an FRB source ($\Delta t_n$). The parameters $\Delta \nu_{\rm FRB} =500~$MHz and $\tau=1~$ms reflect typical values for FRBs, which are justified below.  

Potential ground-based observatories for the terrestrial coordination mode include the 70~m DSN Goldstone antenna, the $D=64$~m Parkes telescope, the $D=100$~m Green Bank Telescope, the $D=500$~m FAST telescope (though currently limited to $\nu <3$\;GHz), as well as interferometers such as the VLA (equivalent to a $D=120$~m), MeerKAT (equivalent to $D=100$~m), and eventually the Square Kilometer Array (SKA).\footnote{These telescopes achieve aperture efficiencies of $\epsilon \approx 50-70$\%, and $D_{\rm eff} = \epsilon^{1/2} D$.}

\subsubsection{Empirical constraints on FRB rates}
\label{sss:ratesemprical}

One of the primary uncertainties in the CPS concept is the rate of repeating FRBs at frequencies above 3 GHz that would be detectable by our system. Population models that characterize the statistics of the bulk population of FRBs (observed at $0.4-1.5~$GHz) suggest that the number density of FRBs at a given fluence decreases with frequency as approximately $\nu^{-\{1-2\}}$ \citep{james22, shin23}. This frequency dependence is inferred indirectly from the shape of the FRB dispersion measure (DM) distribution observed by CHIME at 400-800~MHz, constrained using that a decreasing power-law in frequency implies fewer high-redshift (and consequently higher-DM) bursts. Additionally, a spectral scaling of $\nu^{-1.5\pm0.3}$ has been measured between 1.1 and 1.4 GHz for the average spectrum from an ensemble of FRBs \citep{macquart2019}. 

There have been a limited number of direct observations of FRBs at 4-8 GHz. The first known repeating FRB source, FRB121102, has been monitored at these frequencies, with its fluences at 6-8~GHz found to be comparable to those observed below 3~GHz. FRBs from this source have been detected with fluences of several Jy-ms, within reach of CPS \citep{gajjar18}. In another campaign observing FRB121102 at $3~$GHz, \citet{2017ApJ...850...76L} detected nine bursts during 34 hours of observation, each with an approximate spectral width of 500 MHz.  Although these bursts were relatively faint, they had fluences of a few tenths of Jy-ms per burst.  Since FRB121102 is far away at $800~$Mpc relative to the ideal distance for CPS for distance constraints of $d\sim 200~$Mpc (\S~\ref{ss:science_distances}), this suggests that there are closer FRB sources that would be brighter. However, only a few other repeating FRBs have been studied at the target frequencies, generally during periods when the source is more active at lower frequencies. \citet{2022arXiv220211112A} reported 13 bursts from FRB 20190520B at 3-6~GHz in twenty hours of observation, with fluences ranging from 0.2-5~Jy-ms and no clear frequency trends. \citet{2022arXiv220713669B} detected eight bursts with fluences of 0.2-2~Jy-ms from FRB 20180916B at 4-5 GHz in 34 hours. Burst durations tend to be somewhat shorter at higher frequencies than the canonical $\tau = 1~$ms, with \citet{gajjar18} finding an average duration of 0.6 ms for FRB121102, while \citet{2022arXiv220713669B} reported that many bursts were temporally unresolved, with durations $<0.3~$ms. Using $\tau =0.3~$ms rather than $\tau =1~$ms would lower the fluence sensitivity of CPS (eqns.~\ref{eqn:Fnuself} and \ref{eqn:Fnuter}).

 While nominally CPS targets $\nu > 3$~GHz, since the mean scattering time at high Galactic latitudes becomes nanoseconds or smaller (\citep{2002astro.ph..7156C}; \S~\ref{ss:science_distances}), it is possible that CPS could operate at lower frequencies if there is a population of FRBs that have scattering delays that are well below the mean. This seems possible if scattering tails are driven by the chance alignment of current sheets, a leading theory \citep[e.g.][]{10.1093/mnras/stu1020}.  A repetition of FRB 20201124 exhibited an extraordinary fluence exceeding 180 Jy-ms at 2.2~GHz \citep{2023PASJ...75..199I}. Another remarkable burst observed between 2.2-2.3 GHz registered approximately 10 Jy over an extremely brief 30 µs duration \citep{Majid_2021}. At even lower frequencies around 1.5 GHz, three of the most actively repeating sources (located at redshifts z = 0.1, 0.2, and 0.3) all have recorded bursts with fluences in excess of $10$~Jy-ms \citep{2023ApJS..269...17H}.


A large sample of repeating FRBs has been found with the CHIME instrument in its $400-800~$MHz band \citep{2023arXiv230108762T}.  The fields of view of other telescopes that target FRBs are too small to reliably find repeaters, although this is likely to change in the coming decade with DSA-2000  \citep{2019BAAS...51g.255H}.  Thus,  CHIME provides the best reference for our rate models, which will consequently suffer from the extrapolation to the higher frequencies of CPS.  Owing to its configuration of North-South cylindrical elements, the CHIME field of view encompasses declinations of $0-90^\circ$ in the Northern Hemisphere as well as a couple degrees in right ascension.  Because of this geometry, CHIME has a significant integration time at declination $> 70$ degrees, with an average
of $9$ days of integration.\footnote{This integration time is for their `upper transit'.  CHIME observes each
source at high declinations twice in a day, but are less sensitive to one of these transits, their lower-transit.} 
Declinations $> 70^\circ$ constitute $3.0\%$ of the sky area.  At these declinations, CHIME has detected 4 repeaters, and they report another 10 potential repeaters, but are less confident because the chance of confusion with the non-repeating population becomes high due to its several arcminute instrumental beam.  They estimate that 3 of the potential repeaters are likely spurious. Thus, their results suggest that there are $\approx 11\pm \sqrt{11}$ repeaters in $3.0\%$ of the sky. Correcting for the sky area, this suggests $360 \pm 110$ repeaters over the entire sky that show two repetitions within $9$ days for a survey with the frequency and sensitivity of CHIME.

Only a fraction of our estimate of $360 \pm 110$ on the sky has been detected. In its first two years of operation, CHIME has detected 60 repeaters.  However, with other low-frequency FRB surveys like HIRAX, BURSTT and CHORD \citep{2016SPIE.9906E..5XN, 2022PASP..134i4106L, 2019clrp.2020...28V}, it is likely that most of the FRB sources that repeat on timescales relevant for CPS will be detected in the next two decades.  We will assume that all of these bright repeaters have been discovered before the start of the CPS mission for our estimates.

The distance distribution of FRBs is also important for our estimates.
At least six repeating FRBs have measured redshifts, with their redshifts translating in the concordance cosmological model to being at comoving light-travel distances of $0.8$, $15$, $140$, $410$, $780$, and $1300$~Mpc \citep[the comoving light travel distance is the cosmological distance to which CPS is sensitive]{VLBIFRB,Tendulkar2017, Heintz2020, 2020Natur.581..391M, 2023MNRAS.520.2281N}.  All but one of these repeaters were discovered by CHIME. The distance distribution suggests weak scaling per log distance, at least out to a gigaparsec.  The distance distribution of FRBs turns out to be nicely matched to where CPS is sensitive to distances (\S~\ref{ss:science_distances}).

\subsubsection{FRB rates model}
\label{sss:ratesmodel}

We now develop a model to extrapolate from the CHIME observations to the higher frequencies of interest.  This model parameterizes the significant uncertainty in our understanding of the rates of FRB repeaters with a flexible model.  While the predictions of the model are sensitive to inputs that need better empirical calibration, this exercise allows us to understand the potential rates for CPS and provides a framework for improving the empirical calibration.

As repeaters are observed to come from relatively low redshifts, with all known to have $z\lesssim 0.3$, we approximate spacetime as flat and the spatial geometry as Euclidean.  The repeat rate of an FRB source above an FRB energy ($E$) at a given frequency in a fixed bandwidth ($B$) is assumed to be of the form 
\begin{eqnarray}
R_{\rm obs}(R_{\rm int}', E, \nu, B) &=& R_{\rm int}' \left(\frac{E}{10^{40} \, \text{erg}}\right)^{-\beta} \left(\frac{\nu}{1 \, \text{GHz}}\right)^{-\alpha} \left( \frac{B}{1~\text{GHz}} \right).
\label{eqn:RobsE}
\end{eqnarray}
We can use  $F_\nu = {E} \Delta \nu_{\rm FRB}^{-1}/(4\pi d^2)$ to rewrite this rate in terms of the fluence threshold, $F_\nu$, and distance $d$ rather than the burst energy $E_\nu$, such that equation~(\ref{eqn:RobsE}) becomes
\begin{eqnarray}
R_{\rm obs}(R_{\rm int}, F_\nu, d, \nu, B)  &=& R_{\rm int} \left(\frac{d}{100 \, \text{Mpc}}\right)^{-2\beta} \left(\frac{F_\nu}{5 \, \text{Jy-ms}}\right)^{-\beta} \left(\frac{\nu}{1 \, \text{GHz}}\right)^{-\alpha} \left( \frac{B}{1\,\text{GHz}} \right),\label{eqn:Robs}
\end{eqnarray}
and $R_{\rm int}'$ and $R_{\rm int}$ are constants that set the intrinsic overall FRB rate. The latter constant, $R_{\rm int}$, is the repeat rate of an FRB source with fluences of $F > 5$~Jy-ms at distances of $100~$Mpc within a bandwidth of $B=1~$GHz centered on $\nu = 1$~GHz.   We assume all FRBs have the same spectral width in this model, $\Delta \nu_{\rm FRB}$, and that $\Delta \nu_{\rm FRB} \ll B$ so that the rate does not depend on the spectral width of FRBs.  This is justified since many bursts are found to have $\Delta \nu/\nu \sim 0.1$. 
Thus, the rate of repetition is proportional to the observation bandwidth $B$, which seems sensible.

We assume $\alpha = 1-2$ with a fiducial value of $1.5$, a range motivated empirically in \S~\ref{sss:ratesemprical}.  We use the observed distribution of repeating FRBs to constrain the power-law index in fluence, $\beta$, as described below. This power-law model in $F_\nu$ conflicts with
observations of the most rapidly repeating FRBs that suggest a lognormal distribution \citep{2023ApJS..269...17H}. However, since CPS is sensitive to the brightest and hence rarest FRBs, it is plausible that the repeat rate above some fluence for the brightest tail of bursts can be modeled with a power-law. Furthermore, even if a single FRB source has a lognormal fluence distribution, it is the aggregate of all sources at a given repeat rate that equation~(\ref{eqn:Robs}) needs to approximate, which could still be described by a power-law. 

The final ingredient necessary for predictions is a distribution of $R_{\rm int}$ for FRB sources, since different sources are observed to repeat at different rates.  Two plausible distributions are a power-law, capped at some maximum value for how often an FRB can repeat, or a lognormal distribution.  The form we primarily study is of the former class and is given by
\begin{equation}
\frac{{\rm d}n}{{\rm d}R_{\rm int}} = n_0 \left(\frac{R_{\rm int}} {1~ \text{day}}\right)^{-\gamma} \exp\left( -\frac{R_{\rm int}}{\Gamma} \right),
\end{equation}
where $n_0$ is a constant that sets the number density per $R_{\rm int}$ at 1~day.
Note that this is the famous Schechter function that describes the number distribution with luminosity for many astrophysical sources.  We have experimented with both this Schechter form and a lognormal distribution in which the standard deviation above the lognormal's median is similar to $\Gamma$.  The major difference is that the lognormal model tends to have fewer close-by bursts.  The models with smaller $\gamma$ in our Schechter function parameterization mimic this behavior.   




We now want to use CHIME observations to infer the distribution of $R_{\rm int}$. The number
of FRB sources above a fluence $F_\nu$ in an observation time of $T$ that have at least two observed FRBs is \citep{2023PASA...40...57J}
\begin{equation}
N_2(F_\nu) = \int {\rm d}\ln d \left( 4\pi d^3\right)\times  \int {\rm d}R_{\rm obs} \frac{{\rm d}R_{\rm int}}{{\rm d}R_{\rm obs}} \frac{{\rm d} n}{{\rm d}R_{\rm int}}  \left(1 - e^{-R_{\rm obs} T} - R_{\rm obs} T e^{-R_{\rm obs} T}\right),
\label{eqn:N2}
\end{equation}
where the function in the parentheses is just $1 - P(0| R_{\rm obs} T) - P(1| R_{\rm obs} T)$, where $P(n|\bar n)$ is the Poisson distribution function for the probability of integer value $n$ given a mean $\bar n$. We calibrate our model to reproduce the CHIME detection rate of FRBs, where their more complete observations at high latitudes suggest $360\pm 110$ repeating FRBs over the full sky in an observational window of $T=9$~days at a characteristic frequency of $\nu =500~$MHz and bandwidth of $B=400~$MHz (\S~\ref{sss:ratesemprical}).  We use that the CHIME instrument's average sensitivity threshold at these latitudes is approximately $F_\nu= 3~$Jy-ms to evaluate $R_{\rm obs}$.  With these specifications and equation~(\ref{eqn:N2}), if we specify $\alpha$, $\beta$ and $\gamma$ in the model -- and we motivate values for these parameters in \S~\ref{sss:ratesestimates} -- then the CHIME constraint for the full sky of $N_2 = 360$ fixes $n_0$. 

Once $n_0$ is set, we can calculate the number of CHIME FRB sources that CPS would detect in an observation time of $T_{\rm CPS}$ and per log distance:
\begin{eqnarray}
\frac{{\rm d} N_{\rm CPS}(F_{\nu, \rm CPS})}{{\rm d} \ln d} &=& \left(4\pi d^3 \right) \times  \int_{R_{\rm min}}^\infty {\rm d}R_{\rm obs,CHI} \frac{{\rm d}R_{\rm int}}{{\rm d} R_{\rm obs, CHI}} \frac{{\rm d} n}{{\rm d}R_{\rm int}}  \left(1- e^{-R_{\rm obs, CPS}  T_{\rm CPS}} \right) \nonumber\\
&&\times  \left(1 - e^{-R_{\rm obs, CHI} T_{\rm CHI}} - R_{\rm obs, CHI} T_{\rm CHI} e^{-R_{\rm obs,CHI} T_{\rm CHI}}\right), \label{eqn:NCPS}
\end{eqnarray}
where the subscripts `CHI' and `CPS' indicate that the values are appropriate for CHIME and CPS, respectively.  Equation~(\ref{eqn:NCPS}) yields the number of FRB sources that CPS would detect: We can also calculate the total number of bursts that CPS would detect with the replacement $\left[1- \exp\left(-R_{\rm obs, CPS}  T_{\rm CPS} \right) \right] \rightarrow R_{\rm obs, CPS}  T_{\rm CPS}$. i.e. this counts the total number of FRBs detected, even if many come from the same sources.

For our calculations, we choose an observing strategy in which only sources that are expected to appear at least once in a time $T_{\rm CPS}$ are observed by CPS -- the system does not waste time observing bursts that are unlikely to repeat in the observation window. Thus, we only include sources in the above integral with rates above $R_{\rm min} \equiv f/T_{\rm CPS}$, noting that $R_{\rm obs, CPS} \equiv R_{\rm obs, CHI}/f$, where we defined 
\begin{equation}
f \equiv (F_{\nu, \rm CPS}/3 ~\text{Jy-ms})^\beta \,(\nu_{\rm CPS}/500 ~\text{MHz})^\alpha.
\end{equation}
However, there are likely more optimized choices for CPS science.  For example, rather than observe all sources for a time $T_{\rm CPS}$, a program that targets the FRBs that repeat with the greatest frequency would enhance the number of detections over our estimate at the cost of detecting fewer distinct sources.  

This FRB rate model predicts a cumulative distribution of detected fluences that scales as $F_\nu^{-\beta}$. Thus, for our fiducial choice of $\beta=1.5$, a third of the bursts will exceed the flux at the threshold for detection by a factor of $2$. Thus, if CPS detects $N$ bursts, the brightest it detects will be $N^{1/\beta}$ above the sensitivity threshold, or a factor of $\approx 20$ for $N=100$ and again $\beta=1.5$.\footnote{For the distribution of bursts that CHIME detects, this model does not have the typical Euclidean space scaling $N(F_\nu) \sim F_\nu^{-1.5}$ that holds for any luminosity function of sources that have a maximum and minimum $F_\nu$.  This curious aspect is due to our power-law fluence distribution in equation~(\ref{eqn:Robs}).  If we capped this to have a maximum energy for bursts at $E_*$, then the scaling $N(F_\nu) \sim F_\nu^{-1.5}$ occurs once $4\pi d_{ *}^2 F_\nu \Delta \nu_{\rm FRB} > E_*$ is satisfied, where $d_{ *}$ is the $d$ that satisfies $R_{\rm obs}(\Gamma, F_\nu, d, \nu_{\rm CHI}) T_{\rm obs, CHI} = 1$. This scaling may hold only at extremely high FRB fluences, as the population of non-repeating FRBs in our fluence range comes from significantly further distances \citep{2023PASA...40...57J}, suggesting that the maximum FRB energy can be much larger than the typical energy of our repeating bursts. However, if we impose a sufficiently small $E_*$ so that $N(F_\nu) \sim F_\nu^{-1.5}$ applies, this would also have the effect of capping the distances of our observed FRBs (acting similarly to $\Gamma$ in our fiducial model). \label{footnote:maxdistance}}

\subsubsection{FRB rates estimates}
\label{sss:ratesestimates}

\begin{figure}[htbp]
\centering
\vspace{-2cm}
\includegraphics[width=\textwidth]{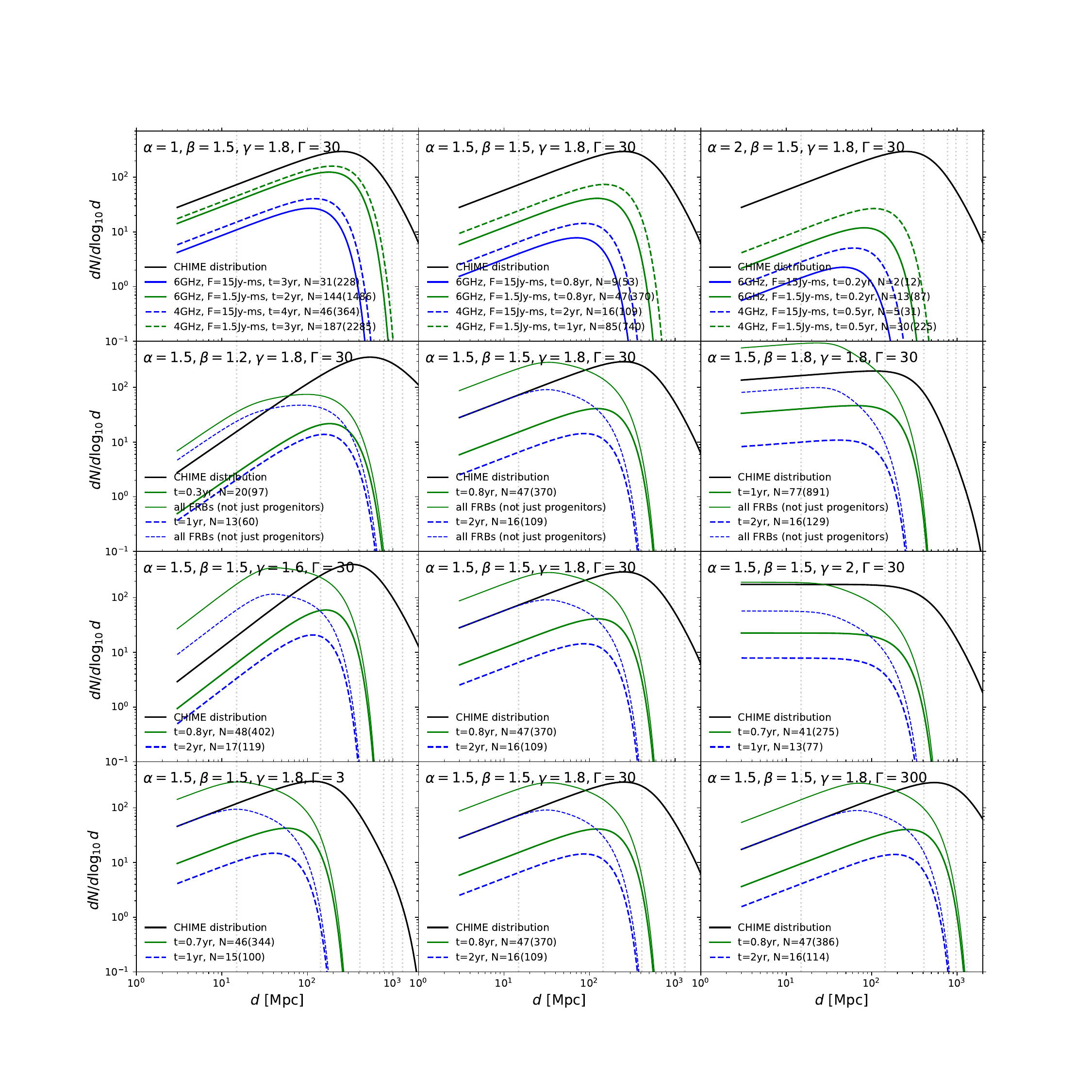}
\vspace{-2cm}
\caption{\small Forecasts for the number of FRB sources that CPS would detect per $\log_{10}$ distance as a function of our four model parameters: the power-law index of a sources repeat rate with frequency, $\alpha$; the power-law index of a source repeat rate with FRB fluence, $\beta$; the power-law index that describes the repeat rate distribution (at a given frequency and fluence threshold) for the population of FRBs, $\gamma$; and the cutoff in the Schechter function that describes the maximum repeat rate of this distribution in days, $\Gamma$.  Different panels show how the predictions vary with model parameters, with the range spanned for each parameter motivated by empirical constraints.  The black curves show model predictions for the distances of CHIME repeating FRBs normalized to the $360$ that we estimate should be present over the full sky (\S~\ref{sss:ratesemprical}).  The other curves show the distribution of repeating FRB sources CPS would detect if all CHIME FRBs are followed up that are anticipated to have at least one repetition that is detectable by CPS in the specified observation time.  
Each FRB is assumed to be observed in a bandwidth of $B=1.6~$GHz for $30$ days in self-triggering mode that reaches $F_\nu=15\;$Jy-ms, and $5$ days for the ground based monitoring that reaches $F_\nu=1.5\;$Jy-ms, with the total observation time for each campaign reported in the legend as well as the central frequency of each observation.  For each model, the total number of FRB sources as well as the total number of FRBs (as an FRB source can be observed multiple times) is annotated, with the latter in parentheses.  The total number of FRBs detected is also shown by the thin curves in the bottom three rows.   The vertical dotted translucent lines show the comoving light travel distances to the six repeating FRBs that have measured redshifts. }
\label{fig:rate_estimates}
\end{figure}

Figure~\ref{fig:rate_estimates} shows the predictions of this model for the number of FRB sources that CPS detects, assuming the fiducial bandwidth of $B=1.6~$GHz centered at $\nu=4$ (solid curves) and $6~$GHz (dashed curves). The different panels vary the four parameters in the model, while keeping the other parameters fixed at the fiducial values of $\alpha=1.5$, $\beta=1.5$, $\gamma=1.8$, and $\Gamma = 30$ day$^{-1}$.  (The middle panels repeat the same fiducial model for reference, so that there are nine models in total.)  The fiducial value of $\Gamma=30$ day$^{-1}$, which sets the exponential cutoff scale in the repeat rate, is similar to the repeat rate observed for the most active bursts at the sensitivity and frequencies of CHIME. The fiducial values of $\gamma$ and $\beta$, as well as the ranges considered, are chosen to roughly match the observed distances of repeating FRBs, which are indicated by the vertical lines (all but one of these, the original repeater, were found by CHIME).   The value of $\alpha$, the power-law scaling of the repeat rate with frequency, is varied from $\alpha=1$ to $2$, a range motivated by the observations (as discussed in \S~\ref{sss:ratesemprical}).  The solid curve represents the distribution of distances to repeating FRB sources that the model predicts would be detected by an instrument with the specifications of CHIME over the full sky in a 9-day observation on each field.  

The green curves in this figure represent the predictions for the mode in which CPS operates in conjunction with a large radio dish on Earth.  These curves are computed assuming a 5-day total follow-up of all CHIME bursts that are anticipated to produce an FRB bright enough to detect VLBI correlations confidently down to fluences of $1.5~$Jy-ms.  The value $1.5~$Jy-ms is motivated by the fluence sensitivity achieved on  a $1~$ms burst with a spectral width of $\Delta_{\rm FRB} =500~$MHz, a $D_{\rm eff} = 80~$m terrestrial antenna, and the nominal CPS specifications (c.f. eqn.~\ref{eqn:Fnuter}).  Across the nine models shown, CPS would detect FRB sources in number of $N = 13-187$ over the period indicated in the figure annotations (on the order of 1-3 years).  These are shown by the thicker curves, and the $N$ indicated in the panel legends.  
The value of $N$ likely needs to be at least ten to justify the CPS mission, and thus, a future observational program to pin down the model parameters and make better predictions for the FRB rates is motivated.  If the rates are too low, then a larger primary antenna on CPS or a lower system temperature is required (or, alternatively, a larger antenna on the ground).

The blue curves in Figure~\ref{fig:rate_estimates} represent the rates for the self-triggering mode, which assumes a $30$-day stare towards each FRB source detected by CHIME that is anticipated to produce a fluence above the system's fluence sensitivity of $15~$Jy-ms during this $30$-day observation (c.f. eqn.~\ref{eqn:RobsE}).  This sensitivity makes the same assumptions about the CPS sensitivity and the FRB properties as we made for the terrestrial observing mode. The self-triggering system detects somewhat fewer FRB sources in our models, with a number in the range of $N= 2-46$ among the models considered. 

 The thinner curves in the bottom three rows in Figure~\ref{fig:rate_estimates} show the total number of FRBs per log distance that CPS would detect (rather than the number of FRB sources per log distance shown by the thicker curves).  The total number of FRBs is also given by the parentheses in the panel legends.  The total number of FRBs CPS would detect is significantly higher than the number of FRB sources, with many coming from FRB sources at shorter distances.

Figure~\ref{fig:rate_estimates} also annotates the durations of the observing campaign  to achieve these numbers of FRB sources.  Since the assumed observing strategy is to target each FRB for $t_{\rm obs} =5$~days or 30 days, depending on whether a terrestrial antenna is used, the total time is the total number of targeted FRB sources multiplied by $t_{\rm obs}$.   As the nominal CPS mission is a decade long, and many of these durations are $\sim 1~$yr, the number of FRBs detected from each FRB progenitor could potentially be a factor of several times larger than the numbers given in these panels.\\

Let us now examine the dependence of our rates model:
\begin{description}
    \item[frequency index of FRB repeat rate:]  The top row of panels in Figure~\ref{fig:rate_estimates} shows how the predicted rates scale with $\alpha$, the power-law scaling of the repeat rate with frequency.  This parameter can have a large effect, as we are extrapolating from the CHIME observations at an order of magnitude lower frequencies than the nominal CPS band. The rates are reduced by a factor of three for our fiducial model of $\alpha=1.5$ (top-left panel) compared to $\alpha=1$ (top-middle panel), and by a further factor of $\approx 3$ for a softer scaling of $\alpha=2$ (top-right panel).  Observing with CPS at $4~$GHz (dashed curves) rather than $6~$GHz (solid curves) mildly changes the number of sources detected for $\alpha=1$, and the differences are more significant for $\alpha=2$.
    \item [energy index of FRB repeat rate and number of FRBs at fixed repeat rate:]  The middle two rows of panels in Figure~\ref{fig:rate_estimates} vary the model parameters $\beta$ and $\gamma$.  The former describes how the rate of FRB repetitions scales with their energy for a single source, and the latter describes how, at fixed energy, the number of FRB sources scales with their repetition rate.  We find that the predictions are roughly invariant if $\beta (\gamma-1)$ is fixed.  Smaller $\beta$ and $\gamma$ than our fiducial values result in ${{\rm d} N_{\rm CPS}(F_{\nu, \rm CPS})}/{{\rm d} \ln d}$ scaling more steeply with $d$ prior to the distances where it cuts off, and $\beta = 1.5$ and $\gamma =2$ result in equal numbers per $\log d$ out to several hundred megaparsecs.  The distribution of distances to the known repeaters with measured distances suggests that repeaters are weakly decreasing in number at closer distances, motivating our fiducial choices.  
    \item[maximum repeat rate of an FRB source:] 
The parameter most important for setting the maximum distances at which FRBs are detected by CPS is the maximum repeat rate in the Schechter function, $\Gamma$.  (We commented in footnote~\ref{footnote:maxdistance} that there could also be a maximum energy for FRBs, which would have a similar effect.)  The fiducial value of $\Gamma=30~$day$^{-1}$ is motivated by the rates of the most active repeaters for CHIME-like fluences, and the bottom panels show that increasing $\Gamma$ by an order of magnitude shifts the distances to which CPS detects repeaters by a factor of a few.  However, note that $\Gamma \lesssim 30~$day$^{-1}$ struggles to produce the observed distribution of repeaters, which is marked by the vertical lines.  Thus, it is likely that many of our models are underestimating the distances to bursts that CPS would detect.  This conclusion is supported by the fact that the observed fluences of FRB121102 at $800~$Mpc would be detectable by CPS in conjunction with ground based facilities.

\end{description}

These estimates ignore possible optimizations.  For example, FRBs are known to go through active periods when they can be targeted by CPS to increase detection rates. Furthermore, these estimates assume a program that observes each FRB that repeats at a rate greater than or equal to the $5$ or $30$ day CPS observations. These choices are certainly not optimized to maximize scientific returns.  Indeed, the thin curves in the bottom three rows show the total number of FRBs detected:  These models' observational strategy would result in the nearest repeaters yielding the most detections; yet, these nearby sources would be less useful for cosmic expansion science.

\subsection{Cosmic distances}
\label{ss:science_distances}

The CPS distance sensitivity is primarily shaped by the detector space-time position errors and the refractive delays from propagating through the plasma of the Galaxy.  In~\S~\ref{ss:trajectories}, we discussed the first of these error sources.  Here, we summarize the space-time position error and then discuss the refractive delay.  We then use these to estimate how well CPS can be used to measure FRB distances and, thereby, cosmological parameters.

\paragraph{Detector space-time position errors:}
As motivated in \S~\ref{ss:trajectories}, we assume $\sigma_x = 5~$cm delays in both space and time.  As remarked there, the sensitivity of CPS to the wavefront curvature appears to depend mostly on the space-time error between the second shortest detector pair. (We find that the sensitivity was marginally degraded if we substantially increased the error on the longest arms).  While the CPS error on distances is proportional to $\sigma_x$, we soon show that the refractive timing error is comparable at the lower frequencies targeted by CPS.  Only higher frequencies substantially benefit from reducing the positional error.

\paragraph{Refractive delays:}
Knowledge of the density distribution of the Milky Way interstellar medium is required to infer the refractive delay. Since our nominal $\sigma_x = 5~$cm, we require $\lesssim 0.2~$ns delays from refraction in order for this to not be the dominant delay. 
Pulsar timing arrays have achieved $\sim 30\,$ns timing residuals for the best pulsars at $\sim 1~$GHz \citep{2016ApJ...817...16C}. 
Scaling the PTA timing residuals to CPS frequencies of $\nu> 3~$GHz as $\nu^{-4}$, the expected scaling, as justified below, suggests that the CPS requirement of timing below $\sigma_x$ should be possible . 

The RMS diffractive delay $\sigma_\tau$ --  also known as the scattering time -- is measured at low frequencies using FRBs and pulsars.  It scales as a power-law in frequency, at least when the density can be approximated as Gaussian random with a power-law power spectrum given by $P(k) \propto k^{-\alpha}$  \citep{1989MNRAS.238..995G, narayan}. 
At frequencies that satisfy $\sigma_\tau \lesssim \nu^{-1}$ -- when the many images at lower frequencies merge into a single image --, the scattering time in these calculations becomes the refractive delay that limits geometric inference for CPS.  The scattering time is well measured using Galactic pulsars.  We use the $1~$GHz scattering delay times in \citet{2002astro.ph..7156C} to estimate the delay from refraction.  We then scale the scattering time to the higher frequencies targeted by CPS as $\nu^{-4}$.  This scaling would hold if structures larger than the Fresnel scale were responsible for the refraction (power spectrum indices of $\alpha>4$).  A different motivated assumption is that the spectrum of density fluctuations is given by Shridhar-Goldreich magnetohydrodynamic turbulence in which a Kolmogorov-like $\alpha = 11/3$ is achieved; in this case, the frequency-scaling is even steeper with $\nu^{-4.4}$.  

Observations find a factor of several scatter about the mean $\sigma_\tau$ in \citet{2002astro.ph..7156C}, such that for some FRBs the refractive delay will be smaller and for others larger than assumed.  Each CPS spacecraft should experience a largely distinct $\sigma_\tau$ from its sightline to an FRB, as the Fresnel scale of interstellar structures is $\sim 10^{-2}\,$pc.  As the Fresnel scale is traversed in days for typical interstellar velocities, multiple FRB observations from the same source should allow for an assessment of the scattering time distribution towards that source.

At $2.7~$GHz, sources undergo extreme scattering events about 0.1\% of the time \citep{1987Natur.326..675F}.  These events refract light out to a few tens of Fresnel scales.   As the CPS detectors are much further separated than the tenths of an AU size of the lenses that are responsible, one CPS spacecraft would likely be affected by an extreme scattering event, leading to a discrepancy between its $\Delta t_n$ and the $\Delta t_n$ measured with the other CPS spacecraft.  This discrepancy can potentially be isolated due to the FRB timing being overconstrained in the nominal 5-spacecraft mission.  Even for strong scattering events, the brightest image is often the nearly-undeflected ray \citep{2010ApJ...708..232B, 2014MNRAS.442.3338P}, such that sub-nanosecond timing may still be possible even for sightlines with extreme scattering events.\\



\begin{figure}[h]
  \centering
    \begin{tabular}{@{}c@{\hspace{0.02\textwidth}}c@{\hspace{0.02\textwidth}}c@{}}
    \underline{position error only} & \underline{position+refraction error, $4$~GHz} & \underline{position+refraction error, $6$~GHz} \\
    \includegraphics[width=.28\textwidth]{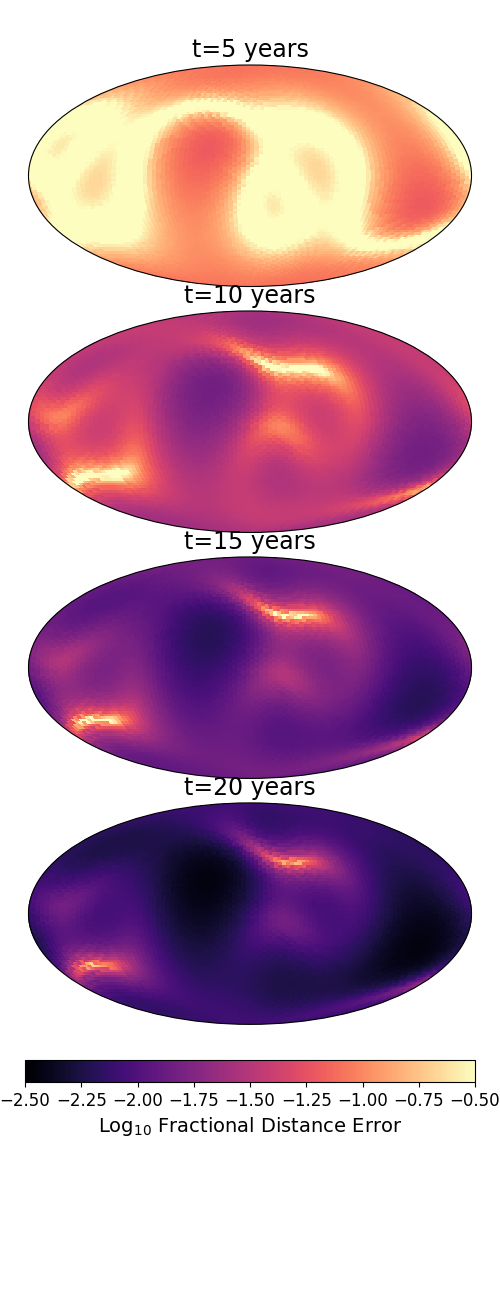} &
    \includegraphics[width=.28\textwidth]{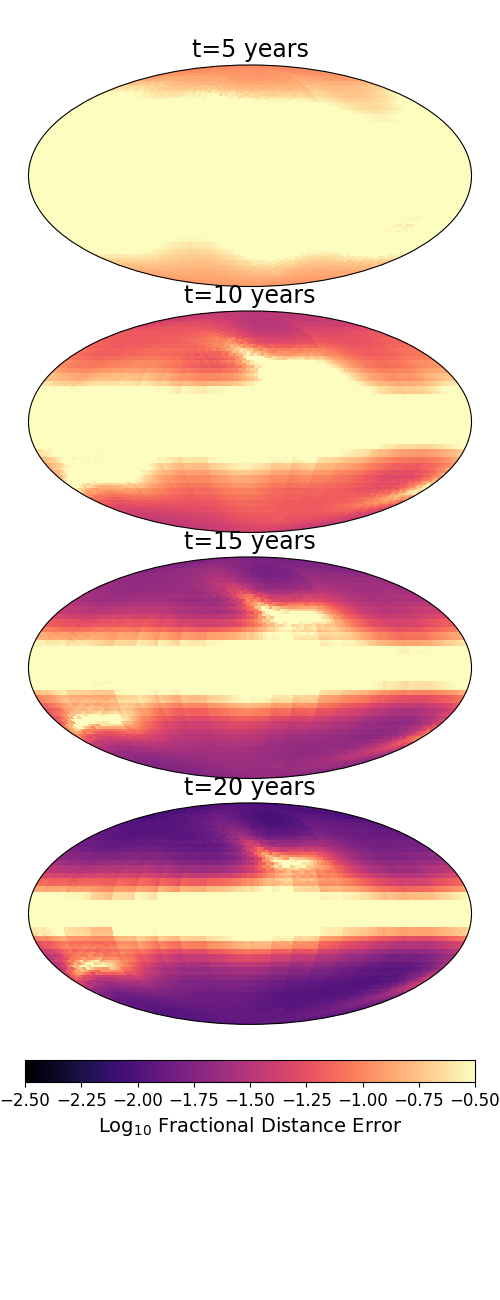} &
    \includegraphics[width=.28\textwidth]{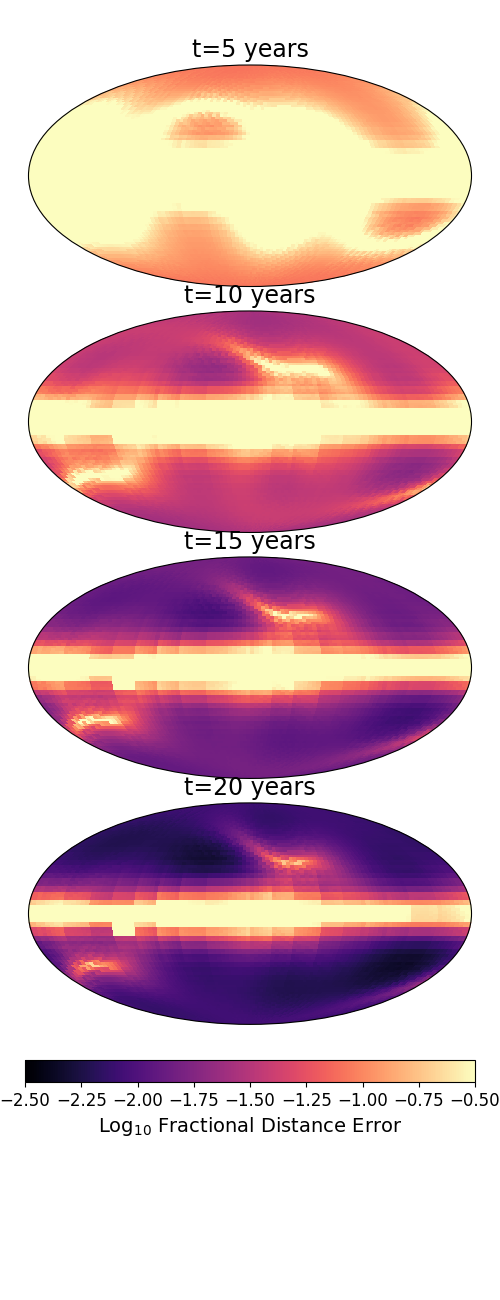}
    \end{tabular}
    \vspace{-1.5cm}
  \caption{\small Fractional distance error of CPS to a source at $d=100~$Mpc and the nominal launch velocity with C3=$157$ km$^2$ s$^{-2}$.  The panels on the left include just detector space-time errors of $\sigma_x = \sigma_t = 5~$cm, reproducing Figure~\ref{fig:beampattern} but now in Galactic coordinates.  The middle panels additionally include interstellar refraction for observations centered at $\nu =4$~GHz, and the right panels show the same but for $\nu =6$~GHz.  Position errors and interstellar refraction are likely the two dominant sources of error.  The refractive timing error is calculated from scaling the \citet{2002astro.ph..7156C} $\nu =1$~GHz scattering times to these frequencies, as described in the text.  The spacecraft trajectories are referenced to the orbital phase of Jupiter and this plot makes an arbitrary selection.
  }
  \label{fig:finalsensiviity}
\end{figure}

\paragraph{Distance sensitivity with interstellar refraction:}
Figure~\ref{fig:finalsensiviity} shows the sky-dependent fractional distance errors to an FRB at three mission epochs for the nominal C3 = $157$ km$^2$ s$^{-2}$ trajectories and for a burst at $d = 100$~Mpc.  The panels on the left include only detector space-time errors of $\sigma_x = \sigma_t = 5~$cm, reproducing Figure~\ref{fig:beampattern} but now in Galactic coordinates. The spacecraft trajectories are referenced to the orbital phase of Jupiter; this plot makes an arbitrary selection for this phase. The middle panels also include interstellar refraction assuming an FRB centered at $\nu =4$~GHz, and the right panels show the same but assuming $\nu =6$~GHz. The larger errors at low latitudes reflect the strong anisotropy of refractive delays from Galactic plasma: timing perturbations grow rapidly toward low Galactic latitudes, where the refractive time delays are largest.  
These refractive delays contribute a substantial fraction of the total error budget over $\sim 30$\% of the sky (primarily $|b| \lesssim 10^\circ$), particularly for the case considering a burst centered at $\nu =4$~GHz. 

\begin{figure}[h]
  \centering
    \begin{tabular}{@{}c@{\hspace{0.05\textwidth}}c@{}}
    \underline{C3=$157$ km$^2$ s$^{-2}$} & \underline{C3=$220$ km$^2$ s$^{-2}$} \\
    \includegraphics[width=.32\textwidth]{figures/beampattern_157_N5_galactic_rot90.0_with_scattering_nu6.png} &
    \includegraphics[width=.32\textwidth]{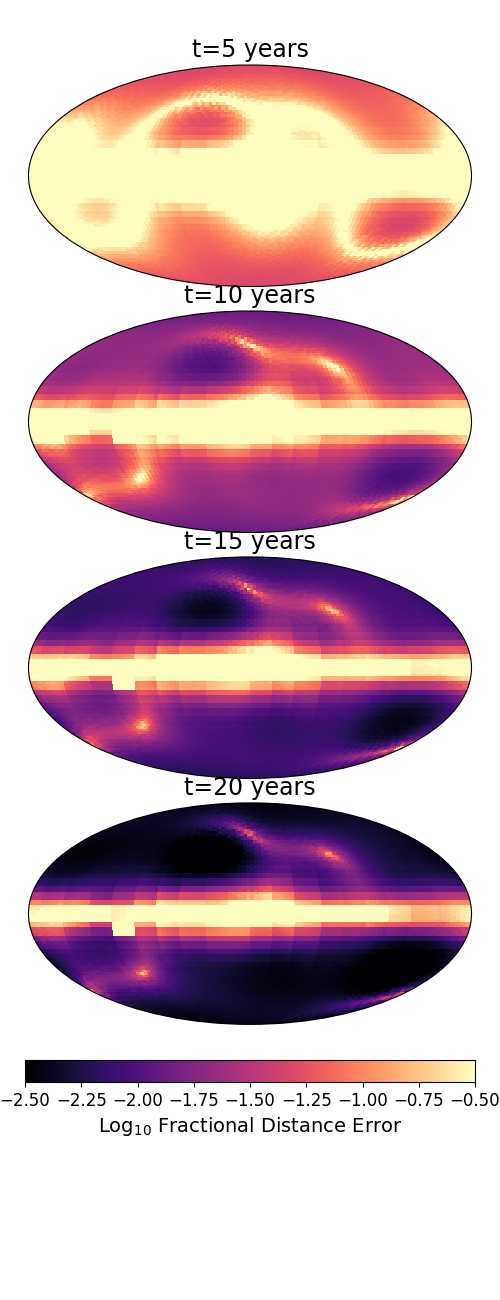}
    \end{tabular}
    \vspace{-1.5cm}
  \caption{\small The total (position+refractive) fractional distance error in Galactic coordinates, as also shown in Figure \ref{fig:finalsensiviity}, but specifically for a burst centered at $\nu=6~$GHz and for the nominal launch velocity with C3=$157$ km$^2$ s$^{-2}$ (left panels) and a more energetic launch with C3=$220~$km$^2$ s$^{-2}$ (right panels).}
  \label{fig:finalsensitivity3}
\end{figure}

Figure~\ref{fig:finalsensitivity3} compares the  total (position+refractive) fractional distance for the nominal C3 = $157$ km$^2$~s$^{-2}$ configuration with a higher-energy C3 = $220$ km$^2$~s$^{-2}$ launch for $\nu =6$~GHz. The higher C3 scenario produces larger baselines at all mission epochs, which improves the distance sensitivity as $\sigma_d/d \propto x^{-2}$. As a result, the median distance error improves by nearly a factor of two for the higher C3 value.

\begin{figure}[h]
  \centering
    \includegraphics[width=.45\textwidth]{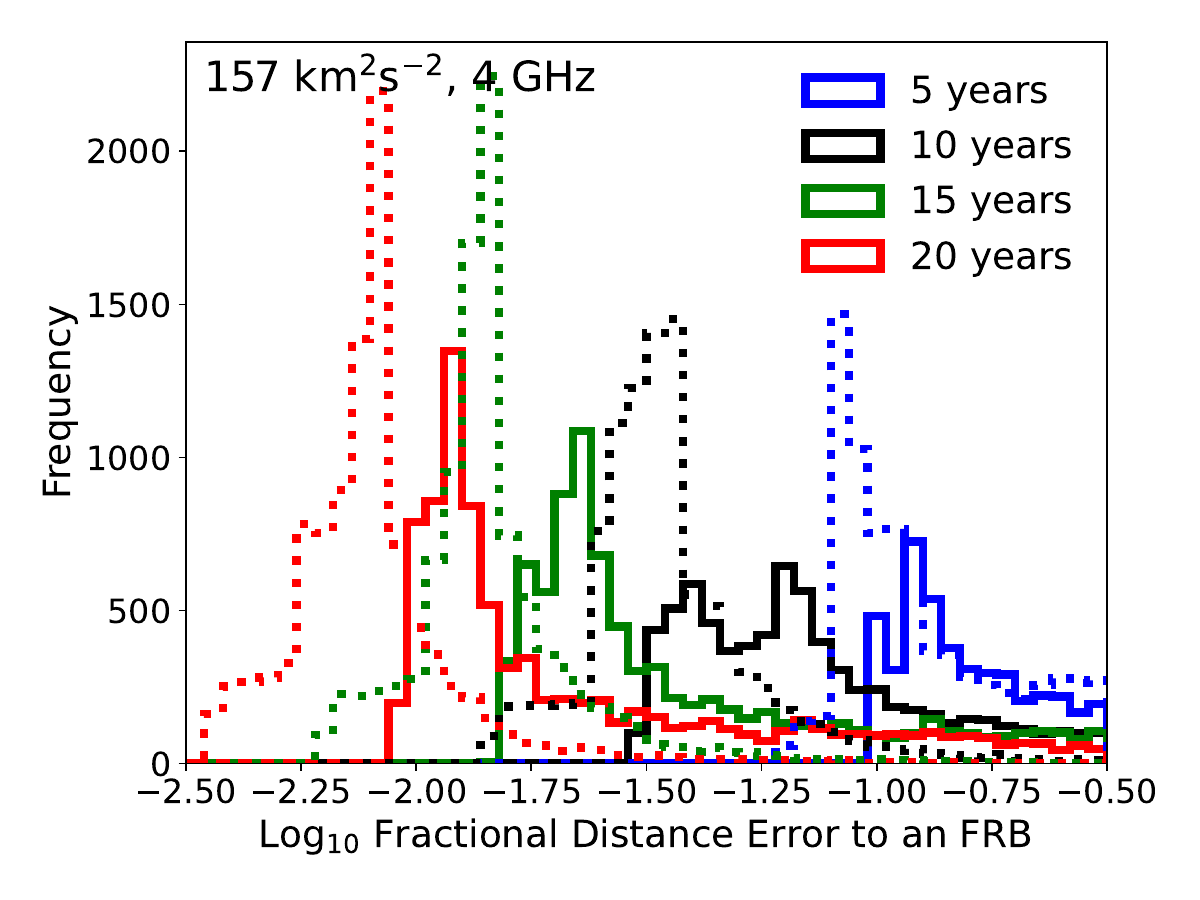}
    \hspace{-.5cm}
    \includegraphics[width=.45\textwidth]{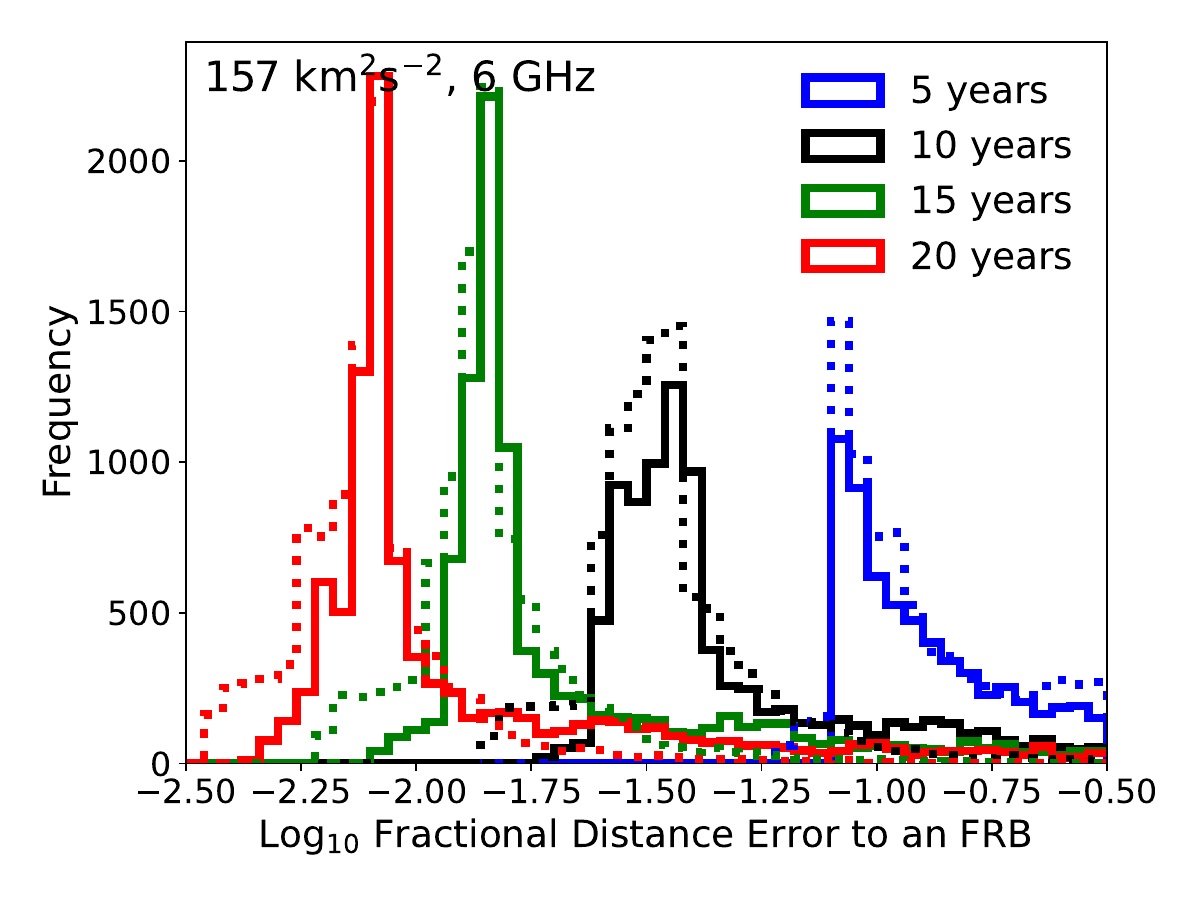}
        \includegraphics[width=.45\textwidth]{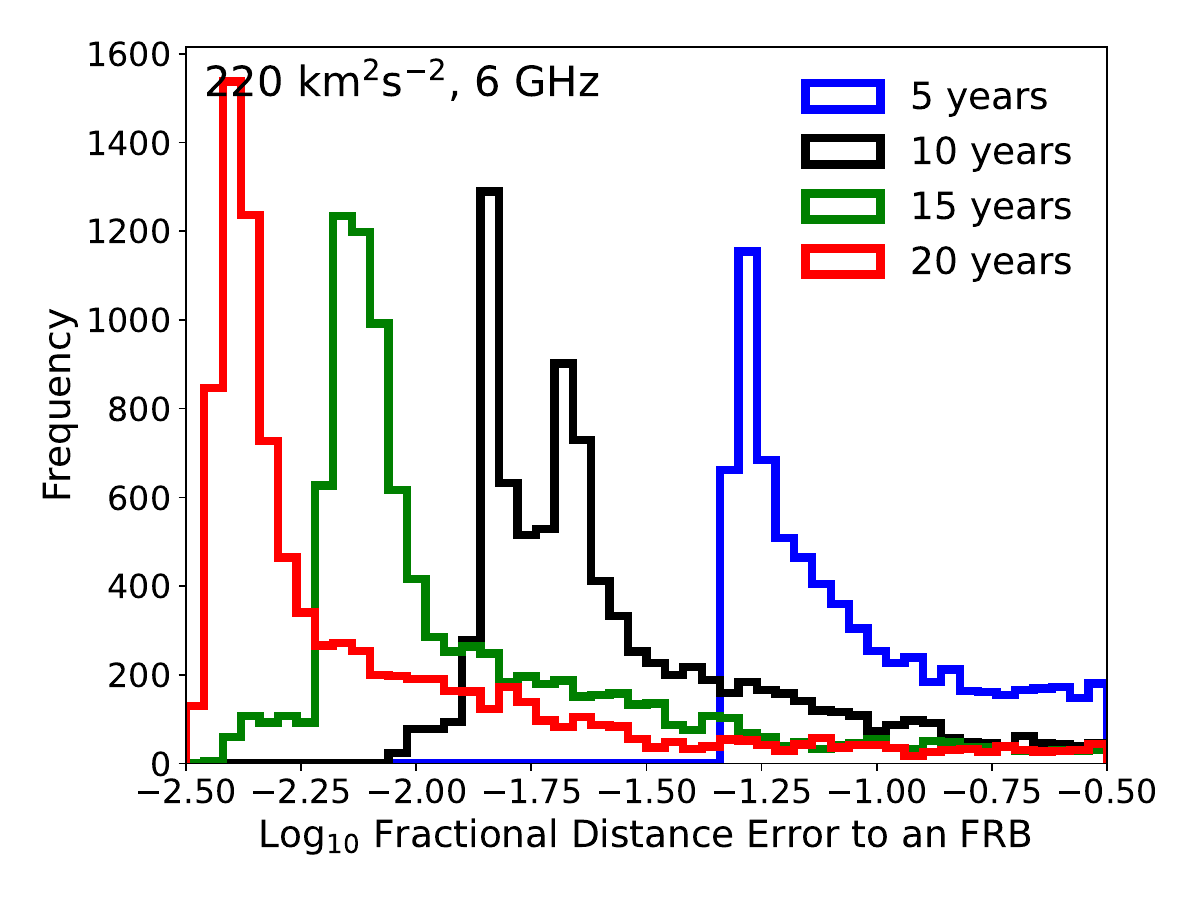}
     \hspace{-.5cm}
  \caption{\small Each panel shows the histogram of the fractional error on the distance to an FRB at $d=100~$Mpc at four different times during the CPS mission.  This error scales proportionally to the FRB distance $d$.  The top two panels are for C3=$157$ km$^2$ s$^{-2}$, with the two panels showing the constraints for a burst centered at $\nu=4~$and $\nu=6~$GHz, and the bottom panel is instead for C3=$220$ km$^2$ s$^{-2}$ and $\nu=6~$GHz.    The solid lines include the timing error from interstellar refraction, and the dashed (only shown in the top panels) exclude this contribution to the error. }
  \label{fig:finalsensivity_histogram}
\end{figure}

Finally, Figure~\ref{fig:finalsensivity_histogram} shows histograms over the full sky for the resulting total distance errors to an FRB for the cases just considered:  The top two panels are for C3=$157$~km$^2$ s$^{-2}$ and  $\nu=4~$GHz (left panel) and $\nu=6~$GHz (right panel), while the bottom panel represents the case with C3=$220$~km$^2$ s$^{-2}$ and $\nu=6~$GHz. The solid curves include both detector position and interstellar refraction contributions to the fractional FRB distance error, while the dashed curves reflect the performance when only spacecraft positional uncertainties are included. The comparison illustrates that refractive delays increase the median error by a factor of $\sim 2$ at 4~GHz, whereas at 6~GHz the histograms are only slightly shifted by including the refractive error.  For a burst at $d=100$~Mpc, the solid histograms show that median fractional distance sensitivity reaches 5\% for C3=$157$~km$^2$ s$^{-2}$ and $\nu=4~$GHz by the ten year nominal mission duration, with FRBs at ideal locations on the sky providing 3\% errors.  The median error in the FRB distance becomes 1\% for a $20~$yr extended mission. For C3=$157$~ km$^2$ s$^{-2}$ and $\nu=6~$GHz, the errors are similar for most bursts regardless of their sky position -- 2-3\% in year ten and below 1\% after twenty years. Instead, using C3=$220$~km$^2$ s$^{-2}$ results in all fractional distances improving by almost a factor of two.


\paragraph{sensitivity to $H_0$:} With these estimates for the distance sensitivity, we are now in a position to estimate how well CPS can constrain the Hubble constant $H_0$, as well as other parameters that pertain to the expansion of the Cosmos.  In \S~\ref{sss:ratesmodel}, we estimated the rate CPS would detect FRBs.  Our models tend to predict a power-law rate out to some distance that is sharply cutoff at some maximum distance, above which no FRBs are detected (c.f.~Fig.~\ref{fig:rate_estimates}).  To be more general than these models, we approximate the distance distribution of CPS-detected FRB sources as 
\begin{equation}
\frac{{\rm d}N_{\rm CPS}}{{\rm d} d} = N_{\rm FRB} \times \frac{d^{\gamma}}{\int_{d_{\rm min}}^{d_{\rm max}}{{\rm d} d} ~ d^{\gamma} }  ~~~~~~\text{if $d_{\rm min} < d < d_{\rm max}$},
\end{equation}
 and zero otherwise.  
 Motivated by the FRB source models in \S~\ref{sss:ratesmodel}, we consider $-1 \leq \gamma \leq 0$.  These $\gamma$ indices result in $N_{\rm CPS}$ as well as our resulting errors on $H_0$ being insensitive to the value of $d_{\rm min}$.  We further choose $d_{\rm max}=300-3000$~Mpc, again motivated by the source models in \S~\ref{sss:ratesmodel}, which show a sharp cutoff in the distance to which CPS is sensitive to FRBs, with the value of this cutoff depending on model parameters.  We can now use this rate and a cosmological model for $d(\boldsymbol{\lambda})$, where the vector $\boldsymbol{\lambda}$ represents the cosmological parameters of interest, to calculate these parameters' Fisher matrix:
\begin{equation}
F_{ij} =     \int {{\rm d} d}  \frac{{\rm d}N_{\rm CPS}}{{\rm d} d} \int {\rm d} \sigma_d P(\sigma_d)~ \sigma_d^{-2}~ \frac{\partial  d(\boldsymbol{\lambda}) }{\partial \lambda_i}  \frac{\partial d(\boldsymbol{\lambda})}{\partial \lambda_j},
\label{eqn:fisher}
\end{equation}
where $P(\sigma_d)$ is the probability of having a distance error $\sigma_d$.  The expression $\sigma_d = |\delta d/d|_{100{\rm Mpc}} \times d^2$ relates $\sigma_d$ to the $d=100~$Mpc fractional errors we have presented so far, and, if we include redshift uncertainties that owe to the peculiar velocities of the galaxies, which set the error at small $d$, then 
\begin{equation} \sigma_d^2 = \left[|\delta d/d|_{100{\rm Mpc}} d^2/(100 \;{\rm Mpc})\right]^2 + \sigma_v^2 H_0^{-2}.
\end{equation}
For the RMS peculiar velocity error, we use $\sigma_v = 300~$km~s$^{-1}$, somewhat larger than the 200~km~s$^{-1}$ that supernova studies find they can model the line-of-sight peculiar velocities \citep{Peterson_2022}.\footnote{The maximum of this effect is the typical line of sight peculiar velocity of  $650/\sqrt{3} \approx 380$km/s; although, especially for a survey that relies on nearby FRBs, there are bulk correlations that we are not modeling that may modestly increase the quoted error \citep{2024PhRvD.109l3029D}.}

Let us consider the case where the FRBs are in the Hubble flow  $v = H_0 \,d$.  This is a good approximation for the FRBs to which CPS is most sensitive; only for more ambitious realizations is CPS directly sensitive to the evolution of the dark energy (as discussed soon). In this limit, the vector $\boldsymbol{\lambda}$ becomes a scalar equal to the Hubble constant, and equation~(\ref{eqn:fisher}) reduces to:
\begin{equation}
\left(\frac{\delta H_0}{H_0} \right)^{-2} \equiv  H_0^{2} F_{H_0 H_0} =  \int {{\rm d} d} ~ \frac{{\rm d}N_{\rm CPS}}{{\rm d} d} d^2\times \int {\rm d}f P(f)~ \left(f^{2}~d^4/[100 ~{\rm  Mpc}]^2  +  \sigma_v^2 H_0^{-2} \right)^{-1},
\label{eqn:H0fishier}
\end{equation}
where $f \equiv  |\delta d/d|_{100{\rm Mpc}}$ and $\delta H_0$ is defined as the 1-$\sigma$ error on $H_0$. 

\begin{figure}[h]
  \centering
    \includegraphics[width=.45\textwidth]{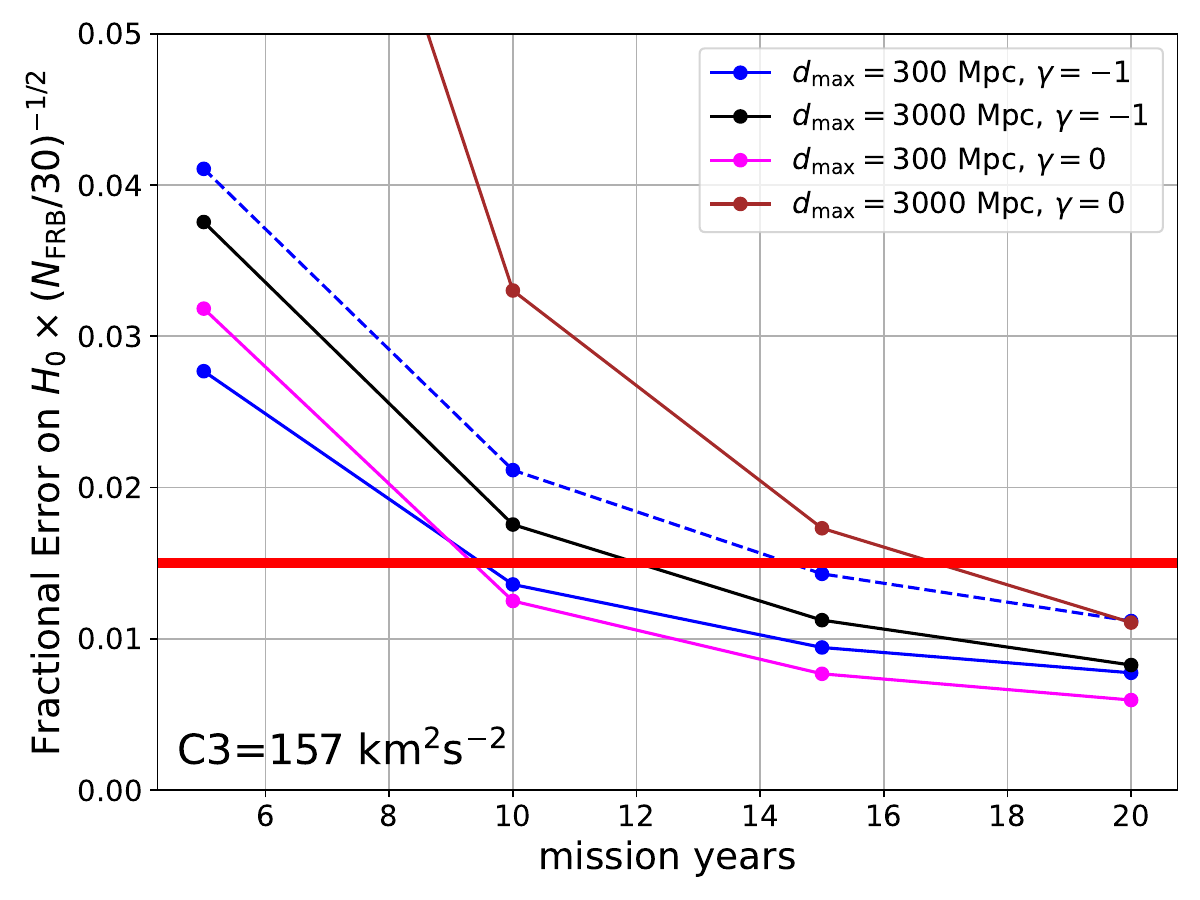}
        \includegraphics[width=.45\textwidth]{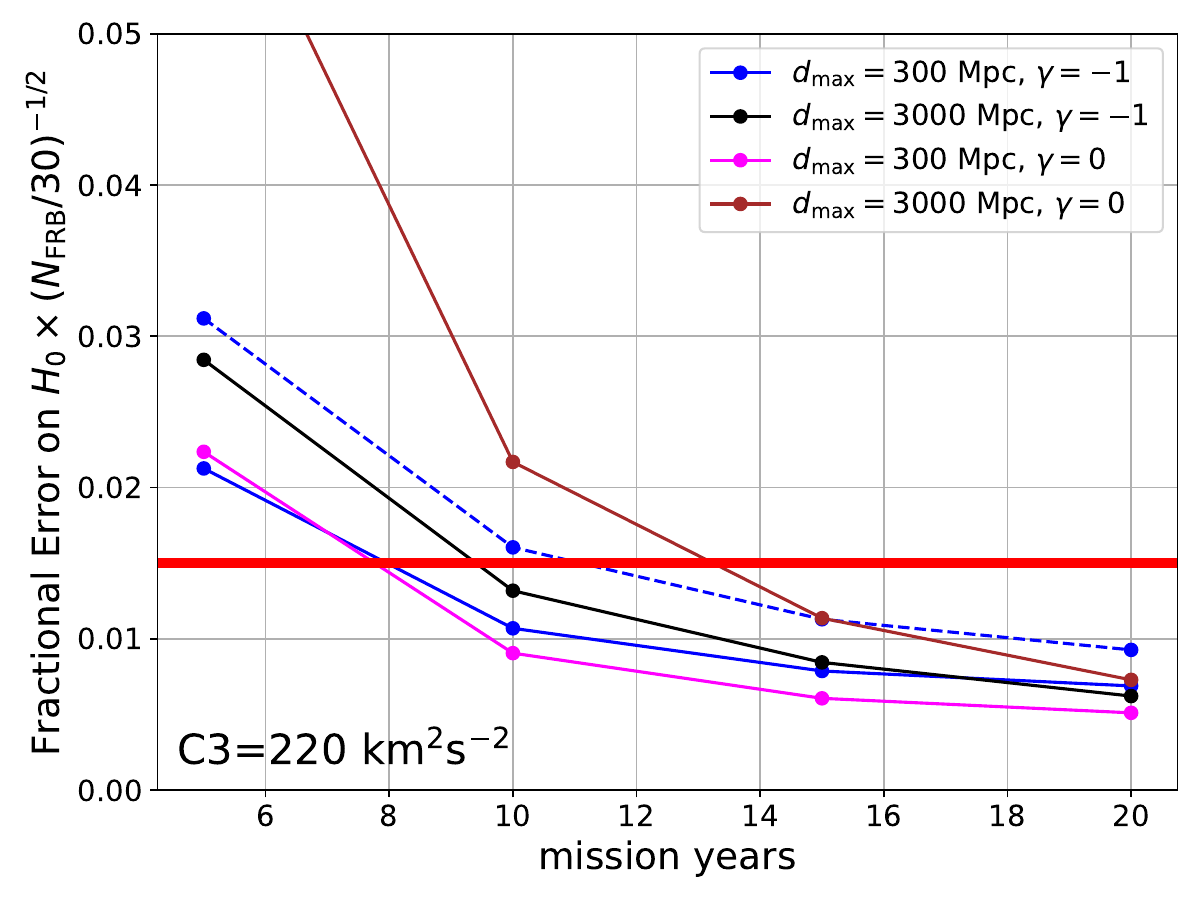}
  \caption{\small Projections for how well the Hubble constant, $H_0$, can be constrained by the nominal mission for the two C3 scenarios making different assumptions about the sources.  {\bf There are several optimizations that would improve the sensitivity for a given source population over what is projected, as discussed in the main text.}  The error is scaled to $N_{\rm FRB} =30$ sources, a number motivated by the 20-200 unique FRB sources that our ensemble of source models predicts (\S~\ref{sss:ratesmodel}).  These calculations assume detector position errors of $\delta x = 5~$cm in each space-time dimension and take the line-of-sight peculiar velocity error to be $300~$km~s$^{-1}$.  The dashed blue curves show the impact of including refractive delays for a burst centered at $\nu = 4$~GHz for one source case, which should then be compared to the solid blue solid that excludes it.  The solid red horizontal line at 1.5\% represents the middle of the range of the $1-3\%$ claimed error from Type 1a supernovae found by current measurements, albeit with larger discrepancies among these measurements \citep{Riess2019, 2025ApJ...985..203F, 2025arXiv251023823H}.}
  \label{fig:H0constraint}
\end{figure}

Figure~\ref{fig:H0constraint} shows this calculation for five models.  We see that CPS reaches a couple of percent error that is comparable to the current 1-3\% constraint from supernovae \citep{Riess2019, 2025ApJ...985..203F, 2025arXiv251023823H} around year ten in most of the source models, assuming CPS detects $N_{\rm CPS}= 30$ FRB sources, which is motivated by the range of 20-200 unique FRB sources predicted by our ensemble of source models presented in (\S~\ref{sss:ratesmodel}).  The errors improve progressively with the extent of the mission.  A twenty-year mission leads to errors for $N_{\rm CPS}= 30$ that approach half a percent in many of the source models.

\begin{figure}[h]
  \centering
    \includegraphics[width=.45\textwidth]{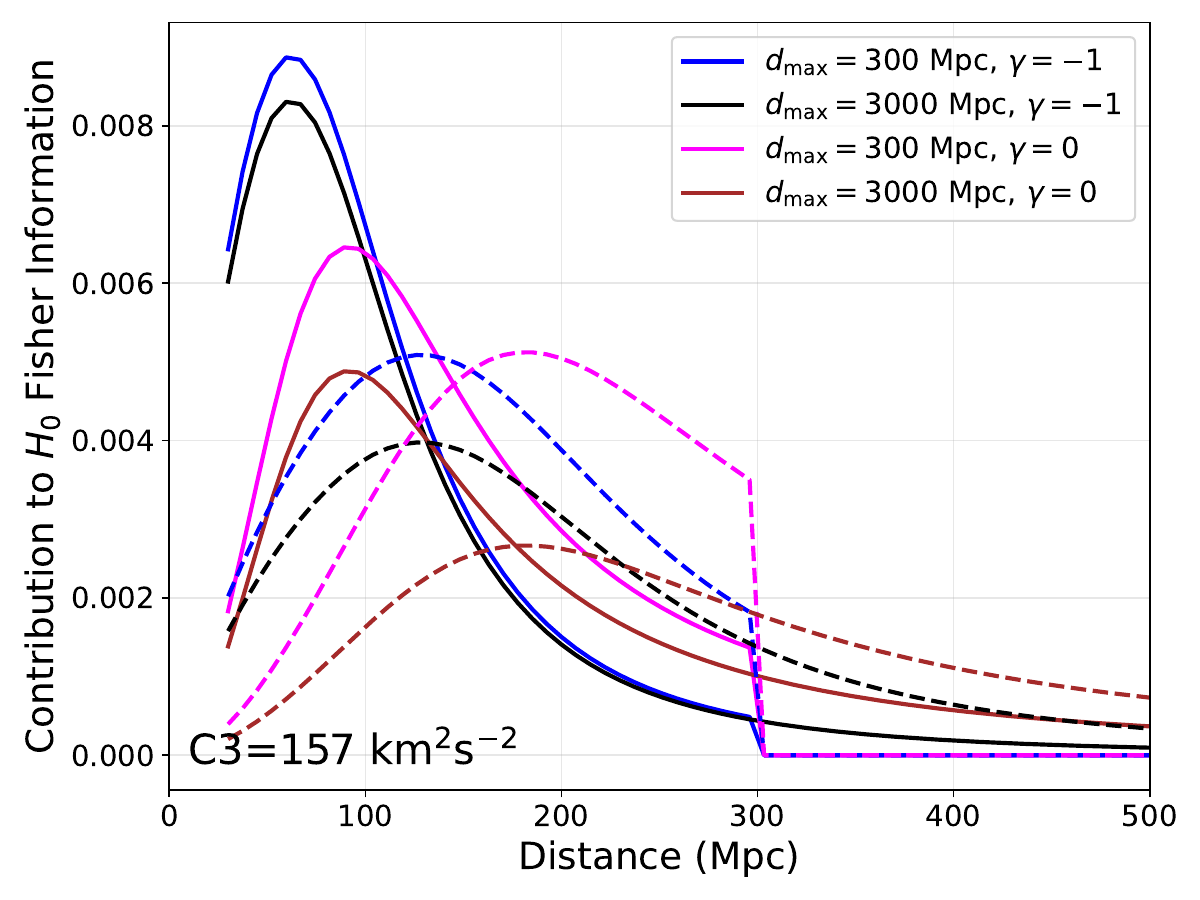}
        \includegraphics[width=.45\textwidth]{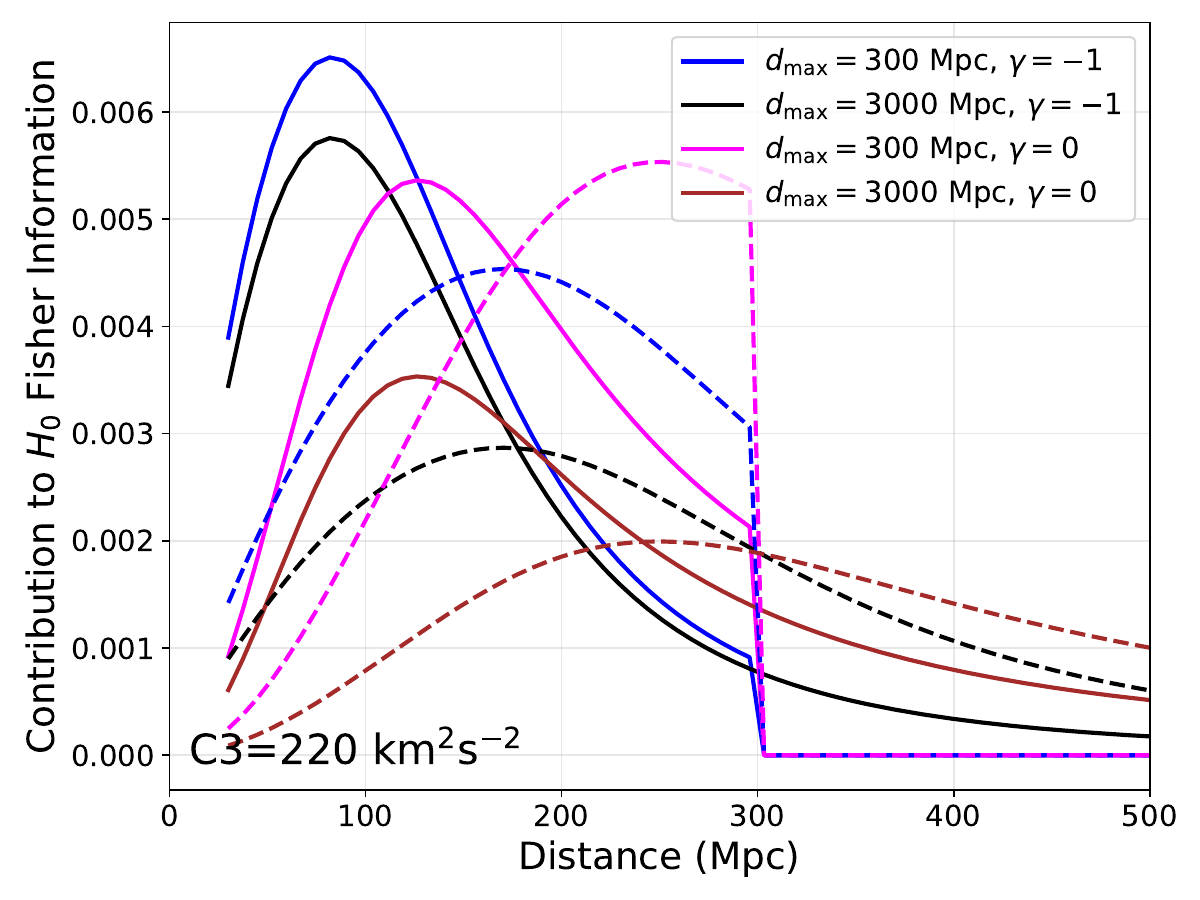}

  \caption{\small Plot of the differential Fisher information with distance (the integrand of eqn.~\ref{eqn:H0fishier}), which shows the distances that contribute most of the constraint in the nominal missions at 10~yr (solid) and 20~yr (dashed) for our different source models.  Two of our models have no sources at distances greater than $300~$Mpc, which is the source of the abrupt cutoff in some curves.}
  \label{fig:FRBdistances}
\end{figure}

The distance of the FRBs that shape the $H_0$ constraint increases as the spacecraft travels further out in the Solar System.  Figure~\ref{fig:FRBdistances} shows the contribution of FRBs at different distances to the Fisher information (the integrand of eqn.~\ref{eqn:H0fishier}). By ten years (solid curves), it is FRBs at $100~$Mpc that yield much of the constraint. 
In the source models in which the sources extend to distances of $d_{\rm max}= 3000~$Mpc, only the fraction of the FRBs at $\lesssim 200~$Mpc contributes to the $H_0$ constraint ten years into the mission.  By twenty years, the mode distance is closer to $d=200~$Mpc, with some of the Fisher information coming from distances that are even $2-3\times$ further.

We emphasize that these estimates were made before potential optimizations that may improve the constraints: 
\begin{itemize}
    \item  For many of our source models, only a couple of FRBs are contributing to the $H_0$ constraint, and therefore, the error could improve considerably with a more focused observational program that specifically targets the FRBs at optimal distances.  
    \item We have assumed that the errors do not improve with multiple FRBs from the same source.  Our results in \S~\ref{sss:ratesmodel} suggest that many FRB sources would be observed several times, indicating that multiple FRB measurements can be combined to statistically improve the distance error for each source.
   \item As the sensitivity of CPS is most limited by the shortest baselines, especially the second shortest for the nominal five spacecraft mission, dedicated ranging to improve the precision with which these baselines are measured would improve the error on $H_0$.  Similarly, prioritizing the calibration of the clocks at the nodes of the shortest baselines would have the same effect.
   \item Not launching spacecraft in pairs, as in the nominal mission, would enable an increase in the length of the shortest baselines. The separation between these spacecraft pairs is achieved by scattering off Jupiter in different directions; hence, the paired spacecraft achieve significant separations from one another only many years into the mission.  The second shortest baseline is $x_{\rm 2s} =25$ (40) AU after ten years in which the launch achieves C3=157 (220)~km$^2$ s$^{-2}$.  The CPS $H_0$ error scales as $x_{\rm 2s}^{-2}$.
\end{itemize}

\paragraph{Sensitivity to dark energy:}
CPS will be most sensitive to $H_0$ compared to other cosmological parameters.  However, pinning down $H_0$ complements other probes of the late-time cosmic expansion, especially in models where dark energy is more complicated than vacuum energy or where the universe is not flat.  Our most robust probes of the expansion history, the CMB and baryon acoustic oscillations (BAO), only yield precise distances to $z\gtrsim 0.5$ (as there is not enough volume at smaller distances for percent-level BAO measurements).  

 Figure~\ref{fig:H0constraint} shows how a reliable constraint on $H_0$ improves constraints on the parameters in a linearly-evolving-with-scale-factor dark energy cosmology, having a present-day dark energy density of $\Omega_w$ and an equation of state parameterized as $w = w_0 + w_a z/(1+z)$.  This plot uses current constraints from the Cosmic Microwave Background (CMB) from Planck \citep{2020A&A...641A...6P} and the baryonic acoustic oscillation measurements from the year-two Dark Energy Spectroscopic Instrument (DESI; \citealt{2025JCAP...04..012A}), assuming that the distance sound waves travel until recombination is exactly known.  See the footnote for additional details.\footnote{Implementing the CMB constraint is challenging, as the CMB does not merely measure a distance but a full spectrum of anisotropies that also constrain other parameters that affect the expansion history, especially the matter density.  To account for this, we use the two parameters related to the expansion history that the CMB best constrains: the angular size of the acoustic scale at the surface of last scattering, which Planck constrains to be $\theta_{*}/[10^{-2} \text{radian}] = 1.04124\pm 0.00028~$, and the matter density parameter for which Planck finds $\Omega_m h^2 = 0.1433 \pm 0.002$ \citep{2020A&A...641A...6P}. The latter constraint is their best fit for $\Omega_m h^2$ in a $w_0-w_a$ dark energy cosmology, which (owing to the additional parameters affecting the expansion in this cosmology) effectively removes geometric information from the $\Omega_m h^2$ constraint.  We further assume that the distance sound waves travel to recombination is exactly known so that the constraint on $\theta_{*}$ translates directly to a constraint on the distance to the surface of last scattering.  Thus, our treatment of existing constraints would underestimate the errors if there is additional physics in the early universe that changes the sound horizon.  We also do not account for the fact that the mass of neutrinos and their transition to non-relativistic behavior after recombination increases the matter fraction relative to what the CMB constrains, an effect that (owing to uncertainties in the neutrino mass) should slightly increase the CMB error on $\Omega_mh^2$ \citep{2024JCAP...12..048L}.}  This assumption means that the angular scales measured by the CMB and DESI BAO translate directly to comoving light travel distances at the corresponding redshifts (and, in the case of DESI, the Hubble parameter at that redshift via its line-of-sight BAO measurement).  The 5\% $H_0$ constraint shown in Figure~\ref{fig:H0constraint} is in line with current tensions, whereas the $1.5\%$ and $0.5\%$ $H_0$ constraints reflect how a trusted constraint on $H_0$, such as may be possible with CPS, translates to errors on dark energy: The errors on the $z=0$ dark energy density are most improved with an $H_0$ constraint (with the 1$\sigma$ error improving by a factor of $0.4$ or $0.2$ when going to $1.5\%$ or $0.5\%$ from $5\%$, respectively), followed secondly by the $z=0$ equation of state (by a factor of $0.5$ or $0.4$), and $w_a$ shows the least improvement but is still significant (by a factor of $0.7$ or $0.7$).  Generally, any model that adds additional degrees of freedom to the expansion history -- dark energy being one, neutrino mass and spatial curvature being others -- benefits from nailing down $H_0$.

CPS may also be directly sensitive to dark energy and not just indirectly via $H_0$.  CPS's direct sensitivity can be phrased in terms of how much different models for dark energy change the effective Hubble constant that relates distance to redshift.  A simple calculation that expands $d = \int_0^z c dz/H(z)$ shows that when expanding about a flat dark energy-matter cosmology in the limit $z\ll 1$, the effective Hubble constant defined as $d \equiv c z/H_{\rm eff}$ changes due to a variation in the dark energy density today of $\delta \Omega_w$, in the equation of state of dark energy $\delta w$, and in the spatial curvature density $\delta \Omega_k$ as 
\begin{equation}
\frac{\delta H}{H_0} = \frac{1}{4}z\left(3 \Omega_w \delta w + 3w \delta \Omega_w - \delta \Omega_k \right),
\end{equation}
where $\delta H \equiv H_{\rm eff}(z) - H_0$.\footnote{We can ignore the trigonometric function that alters $d = \int_0^z c dz/H(z)$ in non-flat cosmologies, as it only enters at cubic order in $z$.}  Since other cosmological surveys already constrain the curvature space density today, $\Omega_k$, and the dark energy density, $\Omega_w$, at the percent level (noting $ \Omega_w = 1 - \Omega_k -\Omega_m$), CPS by itself would be most interesting for constraining the equation of state of the dark energy $w$ at low redshift, which is uncertain at the tens of percent level. 
Ten years into the CPS mission, Figure~\ref{fig:FRBdistances} suggests that the $H_{\rm eff}$ constraint would be attributed to bursts between d=$100$ and $300~$Mpc (z=0.02-0.08), and $100-700~$Mpc ($z=0.02-0.2$) at twenty years. To estimate the sensitivity using $\delta w \approx 4\delta H/(3 \Omega_w  H_{0} z_{\rm max})$, let us assume the $H_{\rm eff}$ constraint is $\sqrt{2}$ as large as the nominal constraint on $H_0$ at the maximum distances/redshifts from the previous sentence.  This suggests that a one percent constraint on $H_0$ should yield an equation of state constraint of $\delta w \approx 30\%$ at ten years and $\delta w \approx 10\%$ at twenty years. 

A low-$z$ measurement of $w$ with CPS would complement BAO surveys, our most reliable method, which provides precise measurements at $z\gtrsim 0.5$. Indeed, the DESI+CMB value of the dark energy equation of state at $z=0$ in an evolving dark energy model parameterized as $w = w_0 + w_a (1-a)$ is $w_0 = -0.48^{+0.35}_{-0.17}$ \citep{2025arXiv250314738D}.  As this is likely only to improve at the factor of two level with future BAO surveys, our above estimates suggest CPS could contribute to improving this error.  Further optimizing CPS specifications over our nominal measurement of $\delta w$ may be motivated, as CPS could present a reliable way to constrain $w$ at low redshifts. \\


\begin{figure}
\centering
\vspace{-.5cm}
\includegraphics[width=0.6\textwidth]{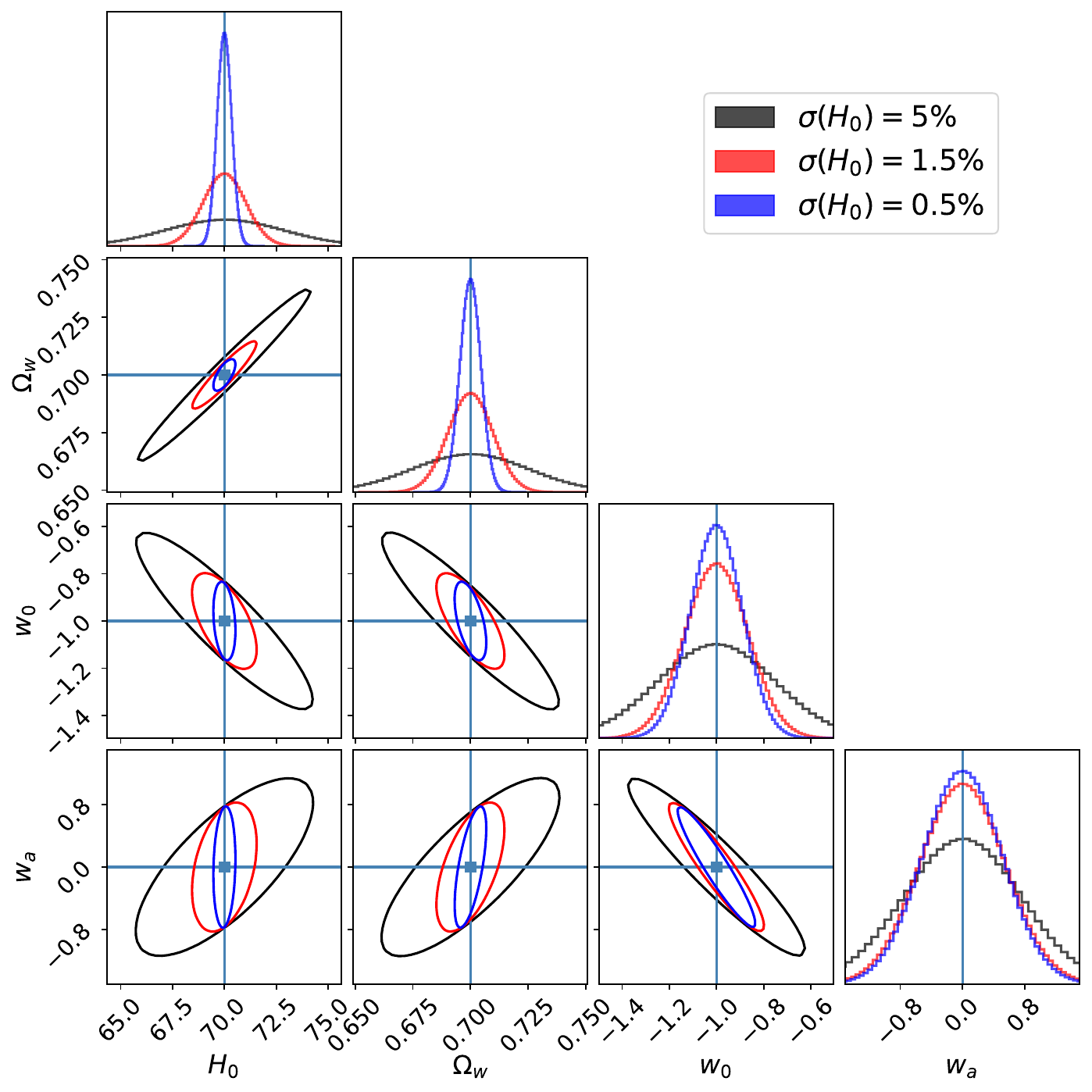}
\vspace{-.5cm}
\caption{\small The 1-$\sigma$ contours showing the potential gains for dark energy science with a trusted $H_0$ measurement with 1-$\sigma$ error indiated in the legend. This assumes a simplified evolving dark energy model with equation of state parameterized as $w = w_0 + w_a z/(1+z)$.  All contours take the constraints from DESI data release 2 and Planck CMB and then combine with a measurement of $H_0$ with the quoted precision. A trustworthy constraint on $H_0$ would also facilitate neutrino mass and spatial curvature constraints.}
    \label{fig:H0conenr}
\end{figure}

\subsection{Dark matter clumpiness}
\label{ss:science_darkmatter}

As discussed in \S~\ref{ss:darkmatter_preview} and in detail in \citet{xiao24}, CPS can probe the small-scale clumpiness of dark matter by monitoring the brightest, most active repeating FRBs and searching for month-to-year wobbles in their apparent positions.  With an accurate measurement, this wobble would likely be due to the gravitational time delays caused by sub-solar mass dark matter structures (anticipated by most leading dark matter models) pass in front of the line of sight \citep{xiao24}. 

Similar to distance-measurement science, measuring the varying angles to FRBs requires an accurate knowledge of the antennas’ relative space-time positions, which CPS determines via spacecraft-spacecraft trilateration. However, trilateration alone cannot constrain the overall rotation of the constellation, which must be known to measure the absolute position shifts of single sources. This limitation is overcome by observing at least two FRB sources and constructing a rotation-invariant observable -- the difference in apparent angle between the sources. The sensitivity of this observable to dark matter is likely set by the uncertainty in the overall rotation of the CPS constellation over the time between FRB detections from both FRB sources.

For our sensitivity estimates, we assume the two most active repeaters yield 30 (50) usable bursts over 5 (10) years in conservative (optimistic) scenarios, at distances of $d=200$ (1000)~Mpc. These values are consistent with the rates in \S~\ref{sss:ratesestimates} for low-redshift bursts. Given that many spacecraft-spacecraft separations exceed $50$~AU by year ten in our nominal C3 = 157~km$^2$~s$^{-2}$ trajectories, and exceed $100$~AU for an extended mission, we adopt 50~AU (100~AU) separations for the conservative (optimistic) forecasts. We further take the combined one-dimensional position plus clock error to be $\delta x = 30$ (10)~cm -- larger than our nominal $5\sqrt{2}$~cm -- because the common-mode constellation rotation between detections from the two FRBs cannot be removed via trilateration.  How the estimated sensitivity scales with these different parameters is detailed in \citet{xiao24}.

The value of $\delta x$ reflects the likely days-to-weeks gap between bursts from the two sources. Stochastic accelerations from solar irradiance variations and the Solar wind are small on these timescales (\S~\ref{ss:accelerations}), and so these errors would be set by the uncertainty introduced by reflections of solar light off the spacecraft, as well as spacecraft emissions from thermal radiation and outgassing. Determining a realistic value of $\delta x$ would require detailed modeling; therefore, we caution that the estimates in this section are merely intended to motivate this potential science.  

\begin{figure}
\centering
\vspace{-.5cm}
\includegraphics[width=0.6\textwidth]{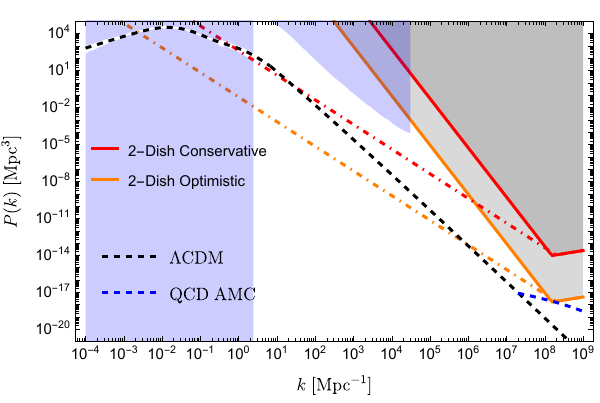}
\vspace{-.5cm}
\caption{\small Preliminary forecasts of the sensitivity of CPS to the nonlinear matter power spectrum $P(k)$ (given by the black dashed curve for the concordance $\Lambda$CDM cosmological model) from measurements of the wobble in angle of repeating FRBs making either conservative or optimistic assumptions about the CPS specifications and FRB properties (as described in the \S~\ref{ss:science_darkmatter}). The blue-shaded region has been excluded by current large-scale structure observations and cosmic microwave background spectral distortions. The dot-dashed curves represent the detection threshold for CPS when measuring the first derivative of time delay difference to FRB, and the solid curves represent the sensitivity by measuring instead its second derivative. The grayed-out regions would potentially be excluded by CPS second derivative measurements. Also shown is the estimated small-scale enhancement in $P(k)$ that may be present for the QCD axion in the post inflation scenario \citep{xiao24}. }
    \label{fig:shapirodelays}
\end{figure}

Figure~\ref{fig:shapirodelays} shows forecast CPS constraints on the matter power spectrum for these assumptions (see \citealt{xiao24} for calculation details). The shaded blue regions show current exclusions from large-scale structure and CMB spectral distortion measurements. The black dashed curve is the $\Lambda$CDM nonlinear $P(k)$ from the haloFit model \citep{2003MNRAS.341.1311S}, naively extrapolating this fit to much smaller scales than those where it is calibrated using its power-law scaling. Dot-dashed curves correspond to sensitivity from measuring the \emph{linear} evolution of the time-delay difference, which we find could detect $\Lambda$CDM-level nonlinear fluctuations at kiloparsec scales, enabling precise constraints on dark matter structure in a regime inaccessible to other techniques.  However, the contribution of kiloparsec scales to the linear time-delay observable means that it will be challenging to probe smaller scales; higher order changes in the time-delay difference are required.  The solid curves correspond to the \emph{quadratic} term in the time-delay difference, probing scales of $\sim 10^{-3}$~pc ($\sim 100$~AU) scales -- extremely small scales that are uniquely accessible by CPS. 

The dashed blue curve in Figure~\ref{fig:shapirodelays} shows a possible small-scale enhancement in the post-inflation QCD axion scenario, in which vacuum fluctuations produce $\sim10^{-3}$~pc miniclusters, from \citet{xiao24}. The QCD axion also solves the Strong CP problem and is thus one of the most theoretically compelling dark matter models. Less-motivated, lighter axions could produce even larger enhancements, while other scenarios -- such as an early matter-dominated era ending just prior to Big Bang Nucleosynthesis -- can yield similar excess power \citep{xiao24}. Remarkably, the optimistic CPS forecast approaches the sensitivity needed to detect the estimated QCD-axion enhancement. 

\subsection{Gravitational waves}
\label{ss:science_grav_waves}

\begin{figure}[htbp]
\centering
\includegraphics[
  width=0.95\textwidth,
  trim={0 0 0 1.0cm},
  clip
]{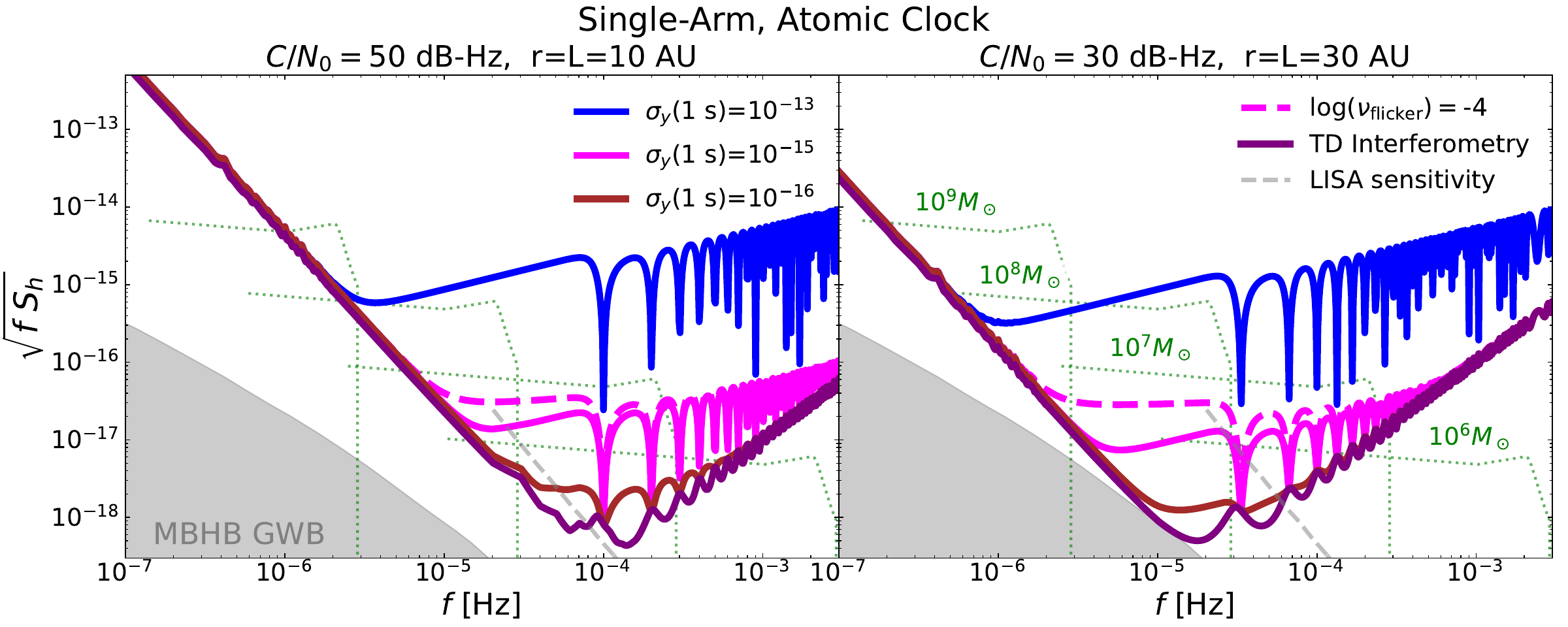}

\caption{\small Predictions from \citet{2024arXiv241115072M} for the gravitational wave sensitivity of a {\it single} arm plus the specified clock Allan deviations, and a `TD inteferometry' configuration where there is a mother spacecraft that is a node between two arms.  Allan deviations for $\tau=1\,$s of $\sigma_y = 10^{-13}$ are similar to the Deep Space Atomic Clock.   Phase measurements referenced to a local clock in each arm would achieve nearly the sensitivity of time-delay interferometry with two arms if the clock has $\sigma_y\lesssim 10^{-16}$.   The green dotted curves are the signal for $z=3$ merging equal-mass black holes with masses from left to right of $10^9$, $10^8$, $10^7$, and $10^6~M_\odot$, where the curves represent the characteristic strain for five year time-span before merger, and the grey shaded band in the lower left is a prediction for the stochastic background from merging black holes.}
\label{fig:clocksensitivity}
\end{figure}

As outlined in \S~\ref{ss:gw},  CPS has the potential to probe the $10^{-7}$ to $10^{-4}$~Hz `$\mu$Hz' gravitational waves by leveraging spacecraft situated in the naturally low-acceleration environment of the outer Solar System, thereby eliminating the need for complex test-mass isolation systems employed by missions like LISA. The ultra-long arms make CPS sensitive to low-frequency gravitational waves by broadcasting a phase across a baseline, transmitting it back, and then measuring the phase difference between its clock and the return signal. Gravitational waves would modulate this phase difference.

The solid curves in Figure~\ref{fig:clocksensitivity} illustrate potential CPS sensitivities (from \citealt{2024arXiv241115072M}; see that reference for details). These examples assume arm lengths of $L=10$~AU and $L=30$~AU, with the spacecraft located at heliocentric radii $r=L$. The gray region marks the stochastic gravitational wave background from merging massive black hole binaries, which constitutes an astrophysical noise floor. The green dotted curves show signals from $z=3$ equal-mass black hole mergers with component masses (left to right) of $10^9$, $10^8$, $10^7$, and $10^6~M_\odot$, with the characteristic strain over the five years prior to coalescence shown.  (The linear area between the characteristic strain power spectral density and the instruments sensitivity is proportional to the square of the signal to noise it would be detected.)  

The low-frequency sensitivity of CPS is limited by stochastic acceleration sources, primarily solar irradiance variations and the solar wind. At higher frequencies, the sensitivity is determined by the stability of the onboard atomic clock used to generate the reference phase -- or, for extremely stable clocks or in a time-delay interferometry (TDI) configuration discussed below, by the received signal strength $C/N_0$ (annotated in Figure~\ref{fig:clocksensitivity}).\footnote{As the curves are taken directly from \citet{2024arXiv241115072M}, they assume Ka-band broadcasts with $\lambda=1$~cm. This only affects the $f\gtrsim 10^{-4}$Hz sensitivity and only for the TDI and $\sigma_y=10^{-16}$ cases.  For these clock parameters, X-band -- as in our nominal mission -- would be a factor of a few worse for the same $C/N_0$.  Furthermore, plasma noise is assumed to be removed by broadcasting at two frequencies, although this is less important for the less sensitive configurations \citep{2024arXiv241115072M}}  

Most of the solid curves in Figure~\ref{fig:clocksensitivity} correspond to a single-arm mode that could require little or no CPS hardware augmentation. With a single-arm configuration, gravitational wave-induced phase shifts along a spacecraft-to-spacecraft arm are measured directly against an onboard atomic clock. The sensitivity scales with the clock’s Allan deviation, $\sigma_y$. For example, a system with $\sigma_y(\tau=1\,\mathrm{s})=10^{-13}$ -- similar to our nominal Deep Space Atomic Clock -- would achieve the blue solid curve sensitivity. This is sufficient to detect $10^9\,M_\odot$ mergers and some inspirals of $10^8\,M_\odot$ black holes for $L=30$~AU. Current rate estimates predict $\sim 1$–$10$ binaries with $M\gtrsim 10^8\,M_\odot$ within five years of merger \citep{2021FrASS...8....7S, 2024arXiv240602306Z}, suggesting that even this baseline configuration is likely to detect merging black holes.

There is also a different way to probe the stochastic background in $\mu$Hz gravitational waves by measuring the arrival times of FRB repeaters using the same setup employed for the cosmic expansion and dark matter experiment \citep{Lu:2024yuo}.  A stochastic gravitational wave signal introduces timing residuals between spacecraft of the size $c \, \delta t \sim \sqrt{f S_h} x $, where $x$ is the baseline, such that the anticipated $f \sim 1~\mu$Hz background of $\sqrt{f S_h} \sim 10^{-15}$ results in $c \, \delta  t\sim 1.5 (x/100{\rm AU})~$cm that can potentially be isolated with enough bursts (and the stochastic background is larger at even lower frequencies; Figure~\ref{fig:clocksensitivity}).  Just as in dark matter science, the noise for this gravitational wave detection method is likely set by the common mode array rotation, which likely has larger errors on $c \,\delta t$ than our nominal $5~$cm in each space-time dimension, as discussed in \S~\ref{ss:science_darkmatter}.

Improving clock stability expands the accessible mass range. The magenta and dark red curves show sensitivities for $\sigma_y(\tau=1\,\mathrm{s})=10^{-15}$ and $10^{-16}$, respectively.   (For $\sigma_y=10^{-16}$, clock noise becomes comparable to radiometer noise from measuring the broadcast phase; therefore, there are not significant gains from improving the clock beyond this specification.\footnote{The dashed pink curve in Figure~\ref{fig:clocksensitivity} shows the $\sigma_y=10^{-15}$ case if the clock enters the flicker noise regime at $f<10^{-4}$~Hz. Atomic clocks typically transition to flicker noise at $f\lesssim 10^{-5}$–$10^{-4}$~Hz; for DSAC, experimental data suggest $f\lesssim 10^{-5}$~Hz.})  Both $\sigma_y(\tau=1\,\mathrm{s})=10^{-15}$ and $10^{-16}$ would be sensitive to essentially all $\gtrsim 10^6\,M_\odot$ mergers in the observable Universe. The expected rate for $M \gtrsim 10^6\,M_\odot$ systems within five years of coalescence is $\sim 100$ \citep{2021FrASS...8....7S}. \citet{2021ExA....51.1333S} discusses several additional astrophysical sources to which these sensitivities would be sufficient for detection: these include a $10\,M_\odot$ stellar-mass black hole slowly inspiraling into the Milky Way’s central supermassive black hole and is destined for merger in $10^6$–$10^8$~yr  \citep{2021ExA....51.1333S}.

If a signal can be broadcast along two arms from a single “mother” spacecraft, clock noise can be canceled entirely via time delay interferometry (TDI; purple curve). This configuration synthetically forms a Michelson-like interferometer and removes spacecraft clock noise. An implementation of this using the nominal CPS spacecraft employs the trigger antenna to point toward a third CPS spacecraft, forming a second arm -- though this would reduce $C/N_0$ by $20\log_{10}(D_{\rm trigger}/D_{\rm main})$ compared to the main arm. As long as stable phase lock can be maintained, sensitivity below $\nu \lesssim 10^{-4}$~Hz is not $C/N_0$-limited.\footnote{A final alternative is that a laser ranging link could be used for the second synthetic arm, which would require frequency stability similar to GRACE-SO and $\gtrsim 10$~cm mirrors for 10~W lasers \citep{2024arXiv241115072M}.}  

A third possible configuration is Doppler tracking using CPS spacecraft and a terrestrial antenna, a technique employed for past outer Solar System missions \citep{dopplerranging}. This approach benefits from large Earth-based antennas and state-of-the-art terrestrial clocks, but it is limited by atmospheric phase noise and mechanical deformations of the ground dish. Both effects have historically limited sensitivity, but \citet{Armstrong_2021} showed that these noise sources can be largely eliminated if the spacecraft carries a stable atomic clock like the CPS spacecraft. If terrestrial noise sources can be eliminated, CPS's Doppler tracking sensitivity could match the single-arm atomic clock curves in Figure~\ref{fig:clocksensitivity}, and in combination with a large terrestrial antenna, gravitational wave detection over $\sim 100$~AU baselines may even be possible.

For all configurations -- particularly the atomic clock modes -- multiple CPS spacecraft could form several independent arms in different regions of the Solar System. This would improve both source localization and noise rejection. Longer arms also yield better angular resolution, with each arm localizing the source to within an angular error of $\delta\theta \sim \lambda_{\rm GW} / (\mathrm{SNR} \times L)$ radians, where $\theta$ is the angle towards the gravitational wave source with respect to the arm. For SNR~$=10$, $L=30$~AU, and $f=10^{-4}$~Hz, this gives $\delta\theta \sim 4^\circ$ -- comparable to the field of view of the Vera Rubin Observatory.

A few comments on implementing gravitational wave science with CPS.
This science system would impose more stringent specifications for CPS on acceleration noise from spacecraft outgassing \citep{2024arXiv241115072M}. Sensitivity could be improved if each spacecraft carried low-power, high-precision accelerometers optimized for the relevant frequency range. One example is the simplified gravitational reference sensor -- a CubeSat-class accelerometer about an order of magnitude less sensitive than the state-of-the-art acceleration control for the LISA mission \citep{2022JGeod..96...70D}. Because the target gravitational wave frequencies are low, downlink requirements are less demanding for gravitational wave science compared to the FRB distance science \citep{2024arXiv241115072M}.

\subsection{Solar System mass distribution}
\label{ss:science_solarsystem}
Section~\ref{ss:solarsystemmass} estimated that constraining spacecraft displacements to approximately $10^2$-$10^3$ meters over year-long timescales would enable the detection of major Solar System bodies such as Planet 9, as well as the characterization of the mass distribution within the Kuiper belt and among trans-Neptunian objects. We further argued in \S~\ref{ss:accelerations} that modeling non-gravitational effects -- including solar radiation reflections, thruster accelerations, and outgassing -- would likely limit the precision with which such gravitational displacements can be distinguished from other sources.

The self-ranging capability of CPS would enable the correction of thruster-induced accelerations by performing ranging measurements immediately before and after thruster firings. Gravitational accelerations operate on $\gg1$yr timescales, whereas reflections and outgassing effects are typically coupled to spacecraft orientation and thus vary on shorter timescales. This temporal separation suggests that self-ranging programs may also be designed to help distinguish between sources of acceleration. Additionally, the effects of solar radiation diminish quadratically with heliocentric distance, meaning that the multiple, nearly-identical CPS spacecraft positioned at different solar distances would provide additional leverage for separating spacecraft displacements caused by gravitational pulls in the Solar System.


\section{Conclusions and future directions for study}
\label{sec:future_directions}

This study investigated the feasibility of a mission deploying spacecraft to the outer Solar System to establish sub-percent constraints on the Hubble constant, $H_0$, and possibly competitive constraints on the dark-energy equation of state at $z\sim 0.1$.  This mission would work by creating a GPS-like system on the scale of the Solar System that provides microsecond timing accuracy (and equivalently $10^{-4}$ micro-arcsecond angular resolution, besting the state-of-the-art by a factor of $10^5$) to detect the curvature in the fast radio burst (FRB) wavefront and thereby measure its distance.  A robust measurement of the late-time expansion history would aid searches for dark-energy models that deviate from a cosmological constant.  It would also benefit cosmological measurements of the summed masses of neutrinos and of the spatial curvature.  Indeed, even aside from the current 10\% Hubble tension, current baryon acoustic oscillation and uncalibrated type 1a supernova measurements are discrepant between their inferred $\Omega_m$ and early time probes, which leave little room for neutrinos to have mass at the level anticipated by oscillation experiments \citep{2024JCAP...12..048L}, favor dark energy models in which the equation of state evolves dramatically with redshift \citep{2025arXiv250314738D}, or favor models where the geometry of the universe is (unexpectedly) closed \citep{2025arXiv250314738D}.  A precise and reliable low-redshift expansion measurement would aid progress by removing at least one critical degree of freedom in models of late-time cosmic expansion.\footnote{In the broader landscape of future $H_0$ measurements, gravitational waves from neutron star mergers detected by fourth-generation terrestrial gravitational wave detectors (such as the Einstein Telescope; \citealt{Punturo_2010} and Cosmic Explorer; \citealt{2019BAAS...51g..35R}, both of which are slated to begin operation a decade from now) represent another method with the potential for robust $H_0$ constraints.  CPS's expansion constraints would be most valuable if these missions prove unable to produce definitive improvements on $H_0$. Key challenges for gravitational waves to improve $H_0$ constraints include large uncertainty in the neutron star--neutron star merger rates, difficulty in identifying electromagnetic counterparts needed to obtain redshifts, modeling uncertainties related to tidal evolution and the neutron star equation of state, and the intrinsic degeneracy between luminosity distance and binary inclination in gravitational waveforms.}  

Additionally, our results suggest that a mission capable of providing sub-percent constraints on the cosmic expansion rate would probe dark matter clumpiness at Earth-mass scales with unprecedented sensitivity. We further showed that this mission could also detect gravitational waves from the mergers of the most massive supermassive black holes at gravitational wave frequencies of $10^{-7}-10^{-5}\,$Hz, which fall between pulsar timing arrays and the LISA mission.  This mission could also provide exceptional constraints on the mass distribution of outer Solar System bodies.

Perhaps the greatest uncertainty concerning this mission's viability relates to the flux distribution of FRBs at our targeted frequencies. While we have shown that detecting tens or even hundreds of FRB sources is plausible given existing empirical constraints on FRB rates, these estimates require verification through follow-up observations of known repeating FRB sources, as well as possibly surveys covering a substantial fraction of the sky to search for higher-frequency repeating sources. Establishing a sample of several well-characterized repeating FRB sources with sufficient fluxes at $\gtrsim 4$~GHz and at distances of $50-1000~$Mpc is a critical prerequisite before proceeding to mission development.

A second major challenge is that our representative instrument designs require power budgets several times larger than that of the New Horizons science system (Table~\ref{table:representativecomponents}). The atomic clock dominates this power budget with our nominal instrumentation, so this mission would benefit significantly from the development of less-power-intensive clocks capable of achieving comparable timing precision.\\

Specific research directions critical for establishing the viability of this concept include:

\begin{description}
\item[Understand FRB rates and fluences at $>3$ GHz:] Establishing a sample of at least ten well-characterized repeating FRB sources (with detectable fluxes at $\gtrsim 4$~GHz, sufficiently small scattering times, and some having optimal distances for expansion science) is essential for validating the CPS concept. This requires systematic follow-up of known repeaters and possibly dedicated higher-frequency surveys.

\item[Study time delays from interstellar medium refraction:] Further investigation of scattering and refraction in the Milky Way interstellar medium would inform the minimum operational frequency. Low-scattering sightlines might enable sufficiently precise timing at frequencies lower than the CPS target of $\nu=4-6~$GHz. This would benefit CPS as FRBs tend to be brighter at lower frequencies. 

\item[Develop low-power space clocks:] The atomic clock represents the largest power requirement in our nominal design, using specifications from the Deep Space Atomic Clock. These power demands likely exceed what can be supported by radioisotope sources with wattage comparable to that on the New Horizons spacecraft. The development of more efficient atomic clocks with similar Allan deviations is likely essential for mission feasibility.

\item[Develop solid-state storage for deep space:] Our analysis indicates that at least 30~TB of onboard storage per spacecraft is needed to store the voltage timeseries when coordinating with terrestrial observatories. This would require, for example, four Phison 8~TB SSDs (Table~\ref{table:representativecomponents}; TRL6).  This would allow CPS to coordinate with terrestrial observatories for an estimated cumulative time of two years before drive failure. Additional storage capacity or longer-lifetime solid-state drives would benefit the mission.

\item[Understand achievable C3 and orbit optimization:] The cosmic expansion and dark matter science benefits from higher launch velocities (C3). Development of more powerful third- and fourth-stage boosters would enable higher C3 values, as would minimizing spacecraft mass. CPS's distance sensitivity to cosmological sources is limited by the shortest baselines.  While our nominal mission envisions paired spacecraft launches, separate launches for each spacecraft to maximize separations would provide more sensitive distance measurements.

\item[Optimize receiver system temperature:] 
Our nominal 20~K system temperature may be achievable even at a $300~$K internal temperature for the spacecraft given recent advances in LNA technology \citep{weinreb2021low}, and this could be aided by putting the LNA on a cold plate or possibly even employing LED cooling.  Minimizing system temperature should be a priority as this allows proportionally smaller antennas on each spacecraft.


\item[Develop radioisotope power systems:] CPS likely requires radioisotope power sources comparable to the 10~kg of plutonium-238 (Pu-238) power source on New Horizons, which had 6\% electric conversion efficiency. NASA currently produces 1.5~kg/year of Pu-238 at Oak Ridge National Laboratory -- a rate requiring more than a decade to produce the plutonium for the five CPS spacecraft. Thus, CPS would require increased Pu-238 production and ideally improved conversion efficiency. Continuing research into thermoelectrics, which have the potential for $2-3\times$ improvements in energy-conversion efficiency, would benefit the mission.

\item[Study control of spacecraft accelerations and positions:]  CPS's dark matter clumpiness and Solar System mass distribution science improve proportionally to how well the accelerations and positions of the spacecraft can be measured. 
We showed that stochastic accelerations on the spacecraft are likely to be smaller than unmodeled reflections of Solar radiation and possibly outgassing. 
A detailed simulation of spacecraft performance would likely be necessary to understand what control over the spacecraft accelerations and positions is achievable.

\end{description}

\section{Acknowledgments}

We thank Eric Agol, Kenneth Carpenter, Holly Leopardi, Michael Kobayashi, and Liz Matson for useful conversations, and Kyle Boone for the early effort on the concept that led to \citet{2022arXiv221007159B}.  We especially thank JPL Team A for their work to prepare for a study, even though we were ultimately unsuccessful in transferring funds for this work.

CM acknowledges that materials presented here were based upon work supported by NASA under award number 80GSFC24M0006.

\bibliography{References}

\end{document}